\documentclass{article}
\usepackage{graphicx}
\usepackage{amsmath,amssymb,amsthm}
\usepackage{subcaption}
\usepackage{bm}

\usepackage{amsmath,amssymb,amsthm}
\usepackage{bm}
\usepackage{color}
\usepackage[normalem]{ulem}

\newcommand{\bit}{\begin{itemize}}
\newcommand{\eit}{\end{itemize}}
\newcommand{\benu}{\begin{enumerate}}
\newcommand{\eenu}{\end{enumerate}}
\newcommand{\be}{\begin{equation}}
\newcommand{\ee}{\end{equation}}
\newcommand{\bea}{\begin{eqnarray}}
\newcommand{\eea}{\end{eqnarray}}
\newcommand{\bean}{\begin{eqnarray*}}
\newcommand{\eean}{\end{eqnarray*}}
\newcommand{\ben}{\begin{equation*}}		
\newcommand{\een}{\end{equation*}}

\newlength{\figwidth}
\newlength{\figwidthtwo}
\newlength{\figwidththree}
\newcommand{\figref}[1]{Fig. \ref{#1}}

\newcommand{\fref}[1]{Fig.\,\ref{#1}}

\newcommand{\eref}[1]{Eq.\,(\ref{#1})}
\newcommand{\erefs}[2]{Eqs.\,(\ref{#1}--\ref{#2})}
\newcommand{\sref}[1]{Sec.\!~\ref{#1}}

\newcommand{\cref}[1]{Ref.\,\cite{#1}}
\newcommand{\crefs}[1]{Refs.\,\cite{#1}}
\newcommand{\cf}{{\it cf.}\! }

\newcommand{\ie}{{\it i.e.}\! }
\newcommand{\eg}{{\it e.g.}\! }

\newcommand{\etal}{{\it et al.}\! }
\newcommand{\apriori}{{\it a priori} }
\newcommand{\aposteriori}{{\it a posteriori} }

\newcommand{\prob}{p} 
\newcommand{\strain}{\varepsilon}
\newcommand{\young}{E}
\newcommand{\yield}{Y}
\newcommand{\hard}{H}
\newcommand{\satmod}{K}
\newcommand{\satexp}{B}
\newcommand{\stress}{{\sigma}}

\newcommand{\Dc}{\mathcal{D}}

\newcommand{\bb}{\mathbf{b}}

\renewcommand{\sb}{\mathbf{s}}
\newcommand{\alphab}{{\boldsymbol{\alpha}}}

\newcommand{\xib}{{\boldsymbol{\xi}}}
\newcommand{\thetab}{{\boldsymbol{\theta}}}
\newcommand{\etab}{{\boldsymbol{\eta}}}
\newcommand{\taub}{{\boldsymbol{\tau}}}

\newcommand{\Fb}{\mathbf{F}}

\newcommand{\Ib}{\mathbf{I}}

\newcommand{\tr}{\operatorname{tr}}

\newcommand{\dev}{\operatorname{dev}}

\begin{document}
 
\title
{
{\bf \Large Plasticity models of material variability based on uncertainty quantification techniques}
}

\author{
F. Rizzi \\
\footnotesize{\it Scalable Modeling and Analysis Department,} \\
\footnotesize{\it Sandia National Laboratories, P.O. Box 969, Livermore, CA 94551, USA} \\
R.E. Jones \\ 
\footnotesize{\it Mechanics of Materials Department,} \\
\footnotesize{\it Sandia National Laboratories, P.O. Box 969, Livermore, CA 94551, USA} \\
J.A. Templeton \\
\footnotesize{\it Thermal/Fluid Science and Engineering Department,} \\
\footnotesize{\it Sandia National Laboratories, P.O. Box 969, Livermore, CA 94551, USA} \\
J.T. Ostien \\
\footnotesize{\it Mechanics of Materials Department,} \\
\footnotesize{\it Sandia National Laboratories, P.O. Box 969, Livermore, CA 94551, USA} \\
B.L. Boyce \\
\footnotesize{\it Materials Mechanics and Tribology Department,} \\
\footnotesize{\it Sandia National Laboratories, P.O. Box 5800, Albuquerque, NM 87185, USA} \\
}

\date{}
\maketitle
\begin{abstract}
The advent of fabrication techniques like additive manufacturing has focused attention on the considerable variability of material response due to defects and other micro-structural aspects.
This variability motivates the development of an enhanced design methodology that incorporates inherent material variability to provide robust predictions of performance.
In this work, we develop plasticity models capable of representing the distribution of mechanical responses observed in experiments using traditional plasticity models of the mean response and recently developed uncertainty quantification (UQ) techniques.
We demonstrate that the new method provides predictive realizations that are superior to more traditional ones, and how these UQ techniques can be used in model selection and assessing the quality of calibrated physical parameters.
\end{abstract}

\section{Introduction} \label{sec:introduction}

Variability of material response due to defects and other micro-structural aspects has been well-known for some time \cite{hill1963elastic,nemat1999averaging,mcdowell2011representation,mandadapu2012homogenization}. 
In many engineering applications inherent material variability has been ignorable, and traditionally the design process is based on the mean or lower-bound response of the chosen materials.
Material failure is a notable exception since it is particularly sensitive to outliers in the distributions of micro-structural features \cite{dingreville2010effect,battaile2015crystal,emery2015predicting}.

Currently, additive manufacturing (AM) is of particular technological interest and provides strong motivation to not only model the mean response of materials but also their intrinsic variability.
Additive manufacturing has the distinct advantages of being able to fabricate complex geometries and accelerate the design-build-test cycle through rapid prototyping \cite{frazier2014metal}; however, currently, fabrication with this technique suffers from variability in mechanical response due to various sources, including defects imbued by the process, the formation of residual stresses, and geometric variation in the printed parts.
As an example, high throughput tensile data from Boyce \etal \cite{boyce2017extreme} clearly shows pronounced variability in the resultant yield and hardening.

Given this current state of the technology, the need to enhance design methodology to account for this variability in order to meet performance thresholds with high confidence is clear.
In this work, we leverage tools from uncertainty quantification (UQ) \cite{le2010spectral,xiu2010numerical,smith2013uncertainty} to provide material variability models, realizations, and, ultimately,  robust performance predictions.

It is well-known that any model is an approximation of the physical response of a real system. 
Typically, models are characterized by many parameters, and thus appropriately tuning them becomes a key step toward reliable predictions.
The most common approach to model calibration is least-squares regression which yields a deterministic result appropriate for design to the mean.
Bayesian inference methods provide a more general framework for model calibration and parameter estimation by providing a robust framework for handling multiple sources of calibration information as well as a full joint probability density on the target parameters.
Traditionally, Bayesian techniques have been applied in conjunction with additive noise models that are appropriate for modeling external, uncorrelated influences on observed responses.
Recently, a technique to embed the modeled stochasticity in distributions on the physical parameters of the model itself was developed by Sargsyan, Najm, and Ghanem \cite{sargsyan2015statistical}, and in this work we adapt it to model the inherent variability of an AM metal \cite{boyce2017extreme}.
This is not the only method available in this emerging field of probabilistic modeling of physical processes for engineering applications.
There are commonalities between many of the methods.
Notably, the work of Emery \etal \cite{emery2015predicting} applied the stochastic reduced order model (SROM) technique \cite{field2015efficacy} to weld failure.
The SROM technique has many of the basic components of embedded noise model: a surrogate model of the response to physical parameters, a means of propagating distributions of parameters with Monte Carlo (MC) sampling and computing realistic realizations of the predicted response.

In \sref{sec:experiment} of this work, we describe the selected experimental dataset \cite{boyce2017extreme} that motivates this effort and provides calibration data.
This deep dataset provides real-world relevance that a synthetic dataset would not; however, we apply some pre-processing and simplifying assumptions to facilitate the task of developing the methodology.
In \sref{sec:theory}, we review the basic plasticity theory that provides the basis for the material variability models developed in \sref{sec:method}.
In \sref{sec:method}, we develop the methods necessary to perform Bayesian calibration of the material parameters: selection of prior distributions to represent the state of knowledge prior to calibration, design of the likelihood function that determines how close the model response is to the calibration data, and the Markov chain Monte Carlo sampling needed to evaluate the posterior distribution of the parameters that quantifies their means and uncertainties.
In particular, we adapt both the traditional additive error \cite{kennedy2001bayesian} and the newer embedded error \cite{sargsyan2015statistical} UQ methods to the representation of the observed mechanical response; and we develop surrogate models of the full finite element simulation tailored to the elastic-plastic response of interest to facilitate efficient Monte Carlo sampling.
In \sref{sec:results}, we provide the results of the surrogate response building and calibration processes in order to compare the two methods in light of the selected data.
We also employ sensitivities provided by the surrogate in order to discuss model selection, and make assessments about the importance of the various parameters.
In \sref{sec:discussion}, we discuss the results in light of a simple analytic version of the representation problem that serves to illustrate the flow of the calibration process and emphasizes the attributes that make the embedded noise model particularly suitable to representing inherent material variability.
We also describe how the variability models can be used in an enhanced design process.
In \sref{sec:conclusion}, we emphasize the innovations of the proposed approach to modeling the mechanical response to microstructural material variability.

\section{Experimental Data} \label{sec:experiment}

\begin{figure}[!tb]
\centering
\subcaptionbox{}%
{\includegraphics[width=\figwidth]{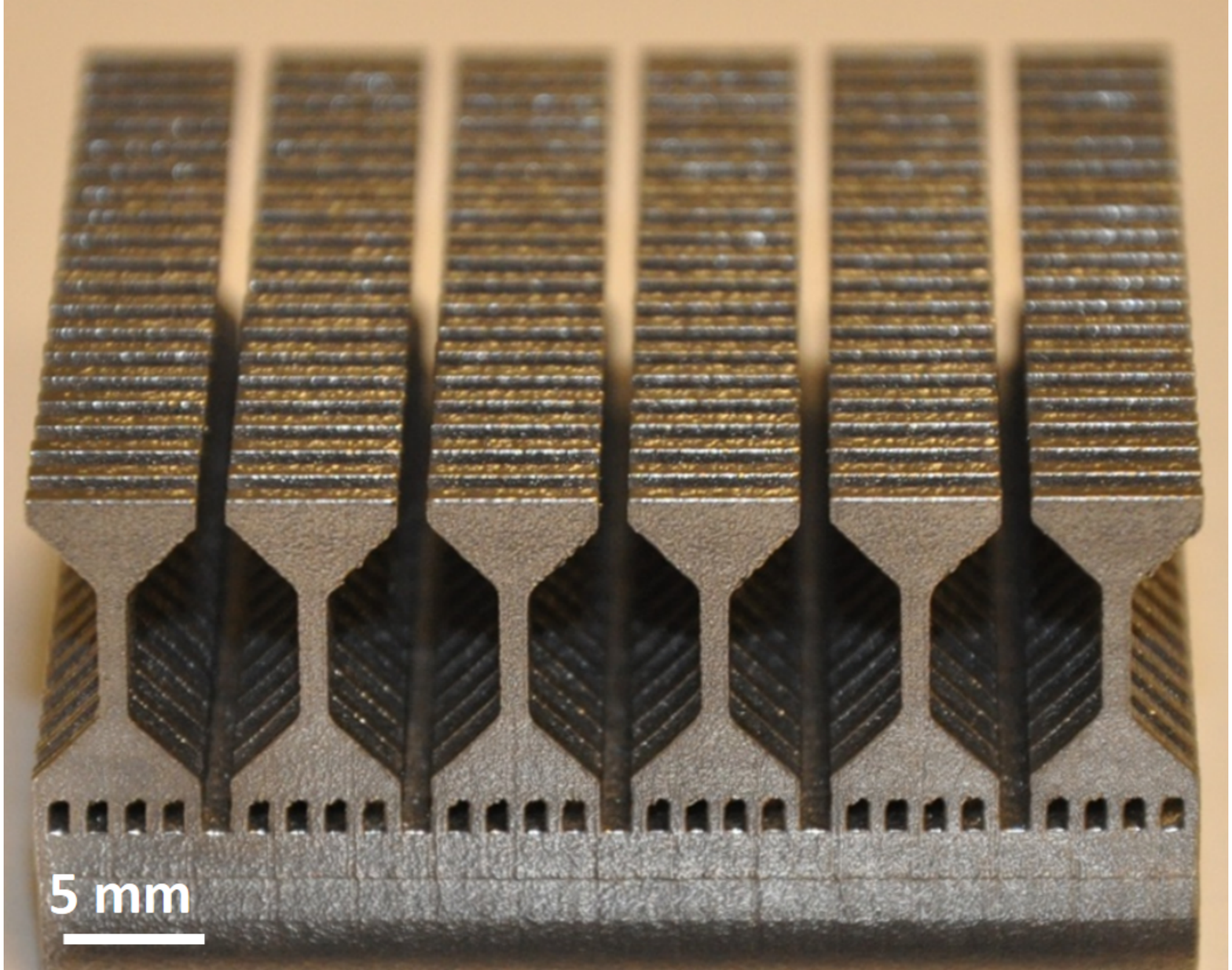}}
\hfill%
\subcaptionbox{}%
{\includegraphics[width=\figwidth]{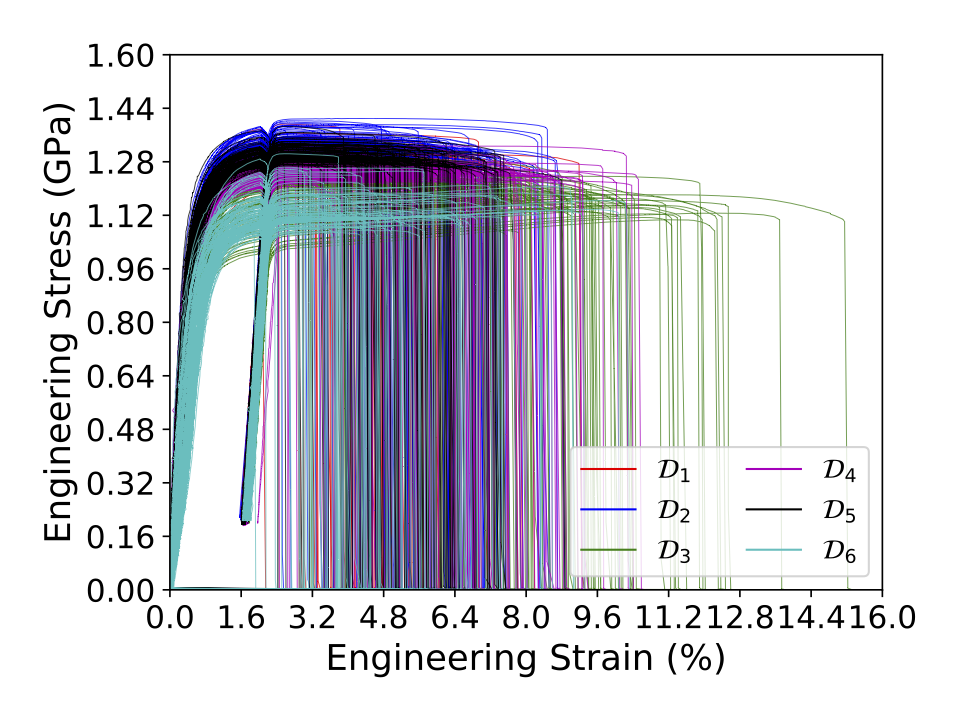}}
\hfill%
\subcaptionbox{}%
{\includegraphics[width=\figwidth]{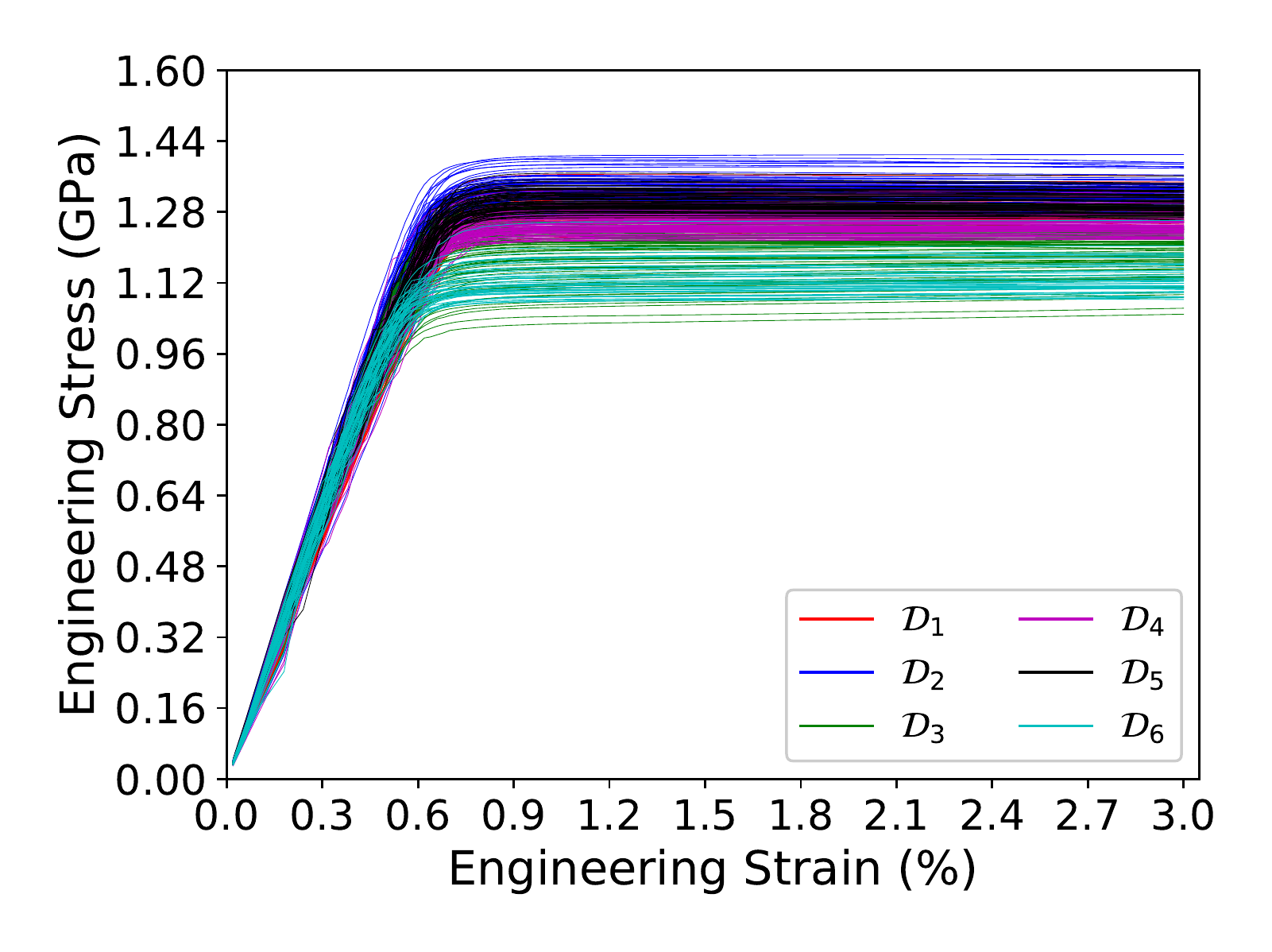}}
\caption{(a) An array of nominally identical micro-tension ``dogbone'' specimens, 
(b) experimental data from Boyce \etal \cite{boyce2017extreme} color-coded by batch, 
and (c) the reduced data set used in this work.}
\label{fig:experiment}
\end{figure}

We focus this work on the analysis of high-throughput, micro-tension experimental measurements of additively manufactured stainless steel. 
From the experiments of Boyce \etal \cite{boyce2017extreme}, we have six experimental datasets, each consisting of 120 stress-strain curves from the array of nominally identical dogbone-shaped specimens shown in \fref{fig:experiment}(a).
(The data from distinct builds of the array are referred to as \emph{batches} throughout the remainder of the manuscript.)
Each stress-strain curve \fref{fig:experiment}(b) is qualitatively similar and behaves in a classically elastic-plastic fashion; however, the material displays a range of yield strengths, hardening and failure strengths and some variability in its elastic properties.

To simplify the data and remove some of the uncertainties associated more with the loading apparatus than the material, we omit the pre-load cycle to approximately 0.2\% strain.
The remainder of the mechanical response is monotonic tensile loading at a constant strain rate, see \fref{fig:experiment}(c). 
We associate zero strain reference configuration with the zero-stress, mildly worked material resulting from the pre-load cycle.
The resulting stress, $\stress$, and strain, $\strain$, values are derived from the customary engineering stress and strain formulas.
We assume the measurement noise to be Gaussian with $\pm$~0.009\% standard deviation in the strain measurement and $\pm$~20.0 MPa in the stress measurement based on the analysis of the random variations for individual curves and the noise in their zero-stress/zero-strain intercepts.
Since we do not try to model failure in this effort, we discard tests that do not reach at least 3\% strain.
This threshold was chosen to be sufficiently large that each sample curve is well within the plastic regime (and near peak stress), and yet retain sufficient data to enable calibration.
This preprocessing yielded $N_b = 6$ batches of stress data with $N_i = \{ 64, 77, 91, 79, 64, 46 \}$ curves, respectively.
To make the data suitable for the inverse problem of parameter calibration, we interpolate each curve and extract $n_{\strain}=151$ points over the interval (0,3)\% to finally arrive at
\begin{equation}\label{eq:data}
\Dc = \{ \Dc_{i} \}_{i=1}^{N_b}, \quad 
\text{with} \quad \Dc_i = \{ \Dc_i^{(k)}\}_{k=1}^{N_i} \quad 
\text{and} \ \Dc_i^{(k)} = \{ \stress^{(i,k)}_{j} \}_{j=0}^{n_{\strain}-1}, 
\end{equation}
where $i$ enumerates the batches, $k$ enumerates the $N_i$ stress curves within the $i$-th batch, and $\stress^{(i,k)}_{j} = \stress^{(i,k)}(\strain_j)$ represents the stress measured at the $j$-th strain value, $\strain_j = 0.03 j/(n_\strain-1)$, for the $k$-th curve of the $i$-th batch. 
The resulting dataset is shown in \fref{fig:experiment}(c).

To expedite development of the appropriate analysis and modeling of materials with significant intrinsic variations, we assume all variability beyond the nearly negligible measurement noise stems from the underlying material response. 
This will lead to conservative estimates of material variability; however, given relevant data, variations in the as-built geometry could be included in the variability analysis or corrected for in pre-processing of the stress-strain data.

\section{Plasticity Theory} \label{sec:theory}

To model the observed behavior which resembles standard von Mises plastic response, we adopt a standard finite deformation framework \cite{simo1988framework} with a multiplicative decomposition of the deformation gradient into elastic and plastic parts
\begin{equation}\label{eq:W}
\Fb = \Fb_e \Fb_p \ ,
\end{equation}
where $\Fb_e$ is associated with lattice stretching and rotation, and $\Fb_p$ is associated with plastic flow.
Following \cref{Simo1998}, we assume an additive stored energy potential written in terms of the elastic deformation
\begin{equation}
W= \frac{\kappa}{2}\left(\frac{1}{2}(J_e^2-1)-\log(J_e)\right) + \frac{\mu}{2}\left(\tr[\bar{\bb}_e]-3\right).
\end{equation}
Here, the elastic volumetric deformation is given by $J_e=\det(\Fb_e)=\det(\Fb)$ since plastic flow is assumed to be isochoric, and the deviatoric elastic deformation is measured by $\bar{\bb}_e=J_e^{-2/3} \Fb_e\Fb_e^T$. 
We associate the elastic constants $\kappa$ and $\mu$ with the bulk modulus and shear modulus, respectively, and relate them to Young's modulus, $E$, and Poisson's ratio, $\nu$, via the linear elastic relations $\kappa=\frac{E}{3(1-2\nu)}$ and $\mu= \frac{E}{2(1+\nu)}$.
The Kirchhoff stress resulting from the derivative of the stored energy potential, $W$, is
\begin{equation} \label{eq:tau}
\taub = \frac{\kappa}{2}(J_e^2-1) \, \Ib + \sb \ \ \text{with} \ \
\sb=\mu \dev[\bar{\bb}_e].
\end{equation}

For the inelastic response, we employ a $J_2$ (von Mises) yield condition between an effective stress derived from $\sb = \dev[\taub]$ and an associated flow stress, $\Upsilon$, as
\begin{equation} \label{eq:yield}
f:=\sqrt{\frac{2}{3}}\|\sb\|-\Upsilon \le 0.
\end{equation}
The rate independent, associative flow rule is written in the current configuration as the Lie derivative of the elastic left Cauchy-Green tensor (\cf \cref{Simo1998})
\begin{equation}
L_v\bb_e=-\frac{2}{3}\gamma~\tr[\bb_e]\frac{\sb}{\|\sb\|}.
\end{equation}
The Lagrange multiplier $\gamma$ enforces consistency of the plastic flow with the yield surface, obeys the usual Kuhn-Tucker conditions, and can be interpreted as the rate of plastic slip.
Finally, we make the flow stress 
\begin{equation} \label{eq:Y}
\Upsilon(\bar{\epsilon}_p) = Y + H\bar{\epsilon}_p + K(1-\exp(-B\bar{\epsilon}_p)),
\end{equation}
a function of the equivalent plastic strain
\begin{equation} \label{eq:eqps}
\bar{\epsilon}_p= \sqrt{\frac{2}{3}} \int_0^t \gamma dt,
\end{equation}
and the following parameters: initial yield, $\yield$; linear hardening coefficient, $\hard$, and nonlinear exponential saturation modulus $\satmod$ and exponent $\satexp$.
In tension, the yield strength, $\yield$, determines the onset of plasticity; the hardening coefficient $\hard$ determines the linear trend of the post-yield behavior; and $\satmod$, $\satexp$ superpose a more gradual transition in stress-strain from the trend determined by Young's modulus $\young$ in the elastic regime to $\hard$ in the plastic regime.
These material parameters form the basis of our analysis of material variability. 
To be clear, this standard J$_2$ plasticity model is a coarse-grained representation of the microstructural variations that engender the variability in the mechanical response, with the plastic strain representating a wide variety of underlying inelastic mechanisms and the physical definitions of the material parameters shaping our interpretation of the underlying causes of the variable response.

We approximate the tensile test with a boundary value problem on a 
rectangular parallelepiped of the nominal gauge section with
prescribed displacements on two opposing faces and traction free
conditions on the remaining faces to effect pure tension. Finite
element simulations are performed in \textsc{Albany}
\cite{salinger2016albany} using the constitutive model described in
this section. The engineering stress $\stress$ and strain $\strain$
corresponding to that measured in the experiments are recovered from
the reaction forces, prescribed displacements, original
cross-sectional area and gauge length.

\section{Calibration formulation} \label{sec:method}

In general, a calibration problem involves searching for the parameters $\thetab$ of a given model that minimize the difference between model predictions and observed data. 
In this work, we adopt a Bayesian approach to the calibration problem \cite{kennedy2001bayesian,Sivia:1996,Rizzi:2012b,Rizzi:2013b,sargsyan2015statistical,MARZOUK2007}.
In contrast to least-squares fitting resulting in a single set of parameter values, in a Bayesian perspective the parameters are considered random variables with associated probability density functions (PDFs) that incorporate both prior knowledge and measured data.  
The choice of Bayesian methods is well motivated by the data which agree with the chosen model to a high degree, but uncertainty is present in the model parameters both within and across all batches.
Bayesian calibration results in a joint distribution of the parameters $\prob(\thetab|\Dc, M)$ that best fits the available observations $\Dc$ given the model choice $M$.
The width of the distribution depends on the consistency of the model with the data and the amount of data.
By using this probabilistic framework and physical interpretations of the parameters, we aim to quantify the material variability.

\subsection{Bayesian inference for parameter calibration} \label{sec:calibration}

Consider our model $M$ for stress $\stress=M(\strain;\thetab)$ comprised of \erefs{eq:tau}{eq:eqps}, where $\strain$ is the independent variable and $\thetab = \{ E, Y, H, K, B\}$ are the parameters of interest.
By setting to $\{H,K,B\}$ or $\{K,B\}$ zero we can form a nested sequence of models with 2, 3, or 5 parameters with perfect plastic, linear hardening, or saturation hardening phenomenology, respectively.
Given that we only have one dimensional tension data, we fix the Poisson's ratio $\nu=0.3$; however, we allow the Young's modulus, $E$, to vary, so that the locus of yield points is not constrained to a line.

Bayes rule relates the data and prior assumptions on the parameters into the posterior density of the target parameters as
\begin{equation} \label{eq:bayes}
\prob(\thetab|{\Dc}, M) 
= \frac{\prob({\Dc}|\thetab, M) \, \prob(\thetab|M)}{\prob(\Dc|M)} \ ,
\end{equation}
where $\prob({\Dc}|\thetab, M)$ is the likelihood of observing the data $\Dc$ given the parameters $\thetab$ and model $M$, $\prob(\thetab|M)$ is the prior density on the parameters reflecting our knowledge {\it before} incorporating the observations, and $\prob(\Dc|M)$ is the evidence. 
It is important to note that the denominator is typically ignored when sampling from the posterior since it is a normalizing factor, independent of $\thetab$, that ensures the posterior PDF to integrate to unity. 
Here, we will employ relatively uninformative uniform prior densities based on plausible ranges for each of the parameters (more details will be given in \sref{sec:results}).
Experimental data influences the resulting posterior probability only through the likelihood $\prob({\Dc}|\thetab, M)$, which is based on the difference between the data $\Dc$ and the model predictions $M(\strain; \thetab)$.
The likelihood plays an analogous role to the cost/objective function in traditional fitting/optimization in the sense that it defines what model predictions are close to the data.
Specific forms of the likelihood will be discussed in \sref{sec:likelihood}.
As \eref{eq:bayes} suggests, the outcome is conditioned on the model chosen which leads questions of model discrepancy, comparison and selection which will be briefly discussed in \sref{sec:likelihood} and \sref{sec:results}.
In general, given the complexities of the model $M$, the posterior density $\prob(\thetab|{\Dc}, M)$ is not known in closed form and one has to resort to numerical methods to evaluate it. 
Markov Chain Monte Carlo (MCMC) methods \cite{gamerman2006markov,berg2008markov} provide a suitable way to sample from the posterior density, and to estimate it using, \eg, kernel density estimation. 

\subsection{Surrogate Model} \label{sec:surrogate}

MCMC sampling of the posterior density involves many sequential evaluations of the model $M({\strain};\thetab)$. Since the finite-element based forward model $M({\strain};\thetab)$ is relatively expensive to query (each tension simulation takes approximately 1 cpu-hour), the inverse problem of determining the parameters $\thetab$ becomes infeasible. 
We overcome this by building an efficient, sufficiently accurate surrogate model $\tilde{M}({\strain};\thetab)$ of the physical response $M(\strain,\thetab)$ over the region of interest with a polynomial chaos expansion (PCe, see \ref{app:pce} for a brief review of polynomial chaos expansions).
The fidelity of the surrogate with the full model will be discussed in detail in \sref{sec:surrogate_construction}.

Since rough bounds of each of the parameters can be estimated from the data and knowledge of similar materials, we assume that the model parameters $\thetab$ follow a uniform prior density $\prob(\thetab)$. 
We construct a corresponding PC expansion of the random vector $\thetab$ via
\begin{equation}
\thetab(\xib) = \sum_{I=0}^{P_\theta} \thetab_{I} \Psi_I(\xib),
\label{eq:PCEpriorTheta}
\end{equation}
where $\xib$ is a vector of standard uniform random variables and $P_\theta$ defines the number of terms in the expansion.
A corresponding expansion of the model response can be written as
\begin{equation} \label{eq:PCEforM}
\tilde{M}({\strain};\xib) = \sum_{I=0}^{P_M} \stress_I({\strain}) \Psi_I(\xib),
\end{equation}
and serves as a suitable surrogate model which, given $\{ \stress_I \}$, can be evaluated by drawing samples from the distribution of $\xib$ and then evaluating the polynomial expansion, \eref{eq:PCEforM}.
Methods to obtain the coefficients $\{ \stress_I \}$ are discussed in \ref{app:pce}.
A by-product of this expansion is the direct access to parametric sensitivities over the range of interest. 
The sensitivities of the response to the input parameters can be used to identify the most influential parameters. 
This analysis is particularly useful for parameter elimination in problems with large dimensionality, as described in more detail in \sref{sec:results}.

\subsection{Likelihood Formulation} \label{sec:likelihood}

As mentioned before, the likelihood is the term in \eref{eq:bayes} that accounts for the data. 
To formulate the likelihood, one needs to reason about what data is available and its relationship with the model predictions. 
From the discussion in \sref{sec:experiment}, one can argue that the batches are independent.
Furthermore, within a given batch, we assume all the $N_i$ stress-strain curves are independent since each experiment is a self-contained test, performed on separate specimens, \ie the variability of each specimen is the result of its specific microstructure.

We consider two different formulations of the likelihood, which lead to different formulations of the inverse problem, and hence models of the material variability.
The formulations differ by how they account for measurement noise and other variability, and how they are affected by (systematic) model discrepancies. 
Since each formulation leads to qualitatively different predictions, interpretations, and realizations, we are interested in how each is able to discriminate material variability from other sources of randomness.
In this section and in the Results section we will discuss how, given that plastic strain is a coarse metric of the inelastic deformation in additively manufactured materials, discrepancies between the observed data and the model predictions can be interpreted physically.
The results in \sref{sec:results} will illustrate how the posterior responds to the quantity of data and its variability.

\subsubsection{Additive error formulation} \label{sec:additive_error}

Consider the $k$-th stress-strain curve from the $i$-th batch which
consists of a sequence of stress observations $\{\stress_j^{(i,k)}\}_{j=0}^{n_\strain-1}$ obtained at the strain locations $\{\strain_j\}_{j=0}^{n_\strain-1}$.
A widely-adopted approach is to express the discrepancy between an observation and surrogate model prediction using an additive noise model as
\begin{equation}
\stress_j^{(i,k)} = \tilde{M}(\strain_j;\thetab) + \eta_j^{(i,k)},
\end{equation}
where $\{\eta_j^{(i,k)}\}_{j=0}^{n_\strain-1}$ is the $(i,k)$ sample from the set of random variables $\{\eta_j\}_{j=0}^{n_\strain-1}$ capturing the discrepancy between observations and model predictions at a given $\strain_j$.
This formulation is predicated on the assumption that the model $\tilde{M}(\strain;\thetab)$ \emph{accurately} represents the true, physical process occurring with fixed, but unknown, parameters. 
This a strong assumption (and one of the main deficiencies of this approach) since models are, in general, only approximations of observed behavior.
Nevertheless, this is a commonly used method due to its simplicity.

In lieu of a completely characterized measurement error model (which is rarely obtained in practice), it is reasonable to assume the errors to be independent and identically distributed (i.i.d.) Gaussian random variables with zero mean, \ie $\eta_j \sim {\cal N}(0,\varsigma^2)$, where $\varsigma^2$ is the variance.
This yields the following likelihood
\begin{equation} \label{eq:additive_error}
\prob({\Dc_i^{(k)}}|\thetab, \tilde{M}) = 
\prod_{j=0}^{n_\strain-1} 
(2 \pi \varsigma^2)^{-1/2}
\exp\left( 
-\frac{(\stress_j^{(i,k)} - \tilde{M}(\strain_j;\thetab))^2}
{2 \varsigma^2}
\right),
\end{equation}
where we recall that $\Dc_i^{(k)}$ represents the stress observations collected from the $k$-th stress-strain curve of the $i$-th batch. 
By assuming that each curve is independent from another, we can write
\begin{equation} \label{eq:independent_additive_error}
\prob(\Dc_i|\thetab, \tilde{M}) = 
\prod_{j=0}^{n_\strain-1} \prod_{k=1}^{N_i}
(2 \pi \varsigma^2)^{-1/2}
\exp\left( 
-\frac{(\stress_j^{(i,k)} - \tilde{M}(\strain_j;\thetab))^2}
{2 \varsigma^2}
\right).
\end{equation}
for the full dataset of the $i$-th batch.
The standard deviation $\varsigma$ can either be fixed in advance, if some knowledge about the experimental process is available, 
or it can be inferred along with the target parameters $\thetab$. Moreover, it can be assumed to be either constant or varying with the data points. 

The basic additive error formulation can be enriched by adding a term capturing the discrepancy between the model prediction and truth represented by the physical data, leading to
\begin{equation}
\stress_j^{(i,k)} = \tilde{M}(\strain_j;\thetab) + 
\hat{\eta}_j^{(i,k)} + \eta_j^{(i,k)},
\end{equation}
where $\hat{\eta}$ represents the discrepancy between the model prediction and truth.
A structure for the model error is more difficult to prescribe than that for the data error.
In fact, the calibrated model now effectively becomes $\tilde{M}(\strain,\thetab)+{\hat{\eta}(\strain)}$, and not simply the original $\tilde{M}(\strain,\thetab)$. 
Given that this additional term is not physically associated with the presumed sources of non-measurement variability its applicability outside the training regime is delicate.
Lastly, this additive term can yield difficulties because it can lead to violations of physical laws \cite{salloum2014inference,salloum2014inferenceb}.

\subsubsection{Embedded model discrepancy} \label{sec:embedded_discrepancy}
A more suitable approach to representing variability embedded in the physical model involves adding the model discrepancy error \cite{Sargsyan2015,Safta:2017} to the parameters
\begin{equation}
\stress_j^{(i,k)} = \tilde{M}(\strain_j;\thetab+{\hat{\etab}}) + \eta_j^{(i,k)} , 
\end{equation}
where $\eta_j^{(i,k)}$ is an additive noise term akin to that in \eref{eq:additive_error}. 
In this case, $\thetab+{\hat{\etab}}$ is a random vector with density and moments to be estimated, whereas $\eta_j^{(i,k)}$ is determined by \apriori estimates of measurement noise.
The random vector $\thetab+{\hat{\etab}}$  can be represented with a PCe. 
For instance, for a single parameter $\theta$ we can write
\begin{equation} \label{eq:pcPerturbed}
\theta + \hat{\eta} = \sum_{I} \alpha_I \Psi(\xi). 
\end{equation}
The problem is thus transformed into a density estimation problem, where our objective is now to estimate $\alphab=\{\alpha_0, \alpha_1, \ldots\}$ that parametrize and define the density of $\theta + \hat{\eta}$.
This is in contrast to the conventional use of Bayesian inference for parameter estimation, \ie additive error formulations, in which one strictly infers the \emph{parameter} and not its density.  
Also, the data for our present calibration problem motivates the embedded approach since it suggests the uncertainties are aleatoric/irreducible rather than epistemic/reducible.

In the conventional case, as more data is taken into account, the width of the posterior density shrinks, tending to a Dirac delta function at the true parameter value assuming informative data and negligible model discrepancy.
On the other hand, in the present context of \emph{density estimation}, the objects of inference are the parameters $\alphab$, and the posterior density is thus on $\alphab$. 
Thus, the more data is taken into account, provided the data is sufficiently informative, the distribution on $\alphab$ narrows while the width of the distribution of the parameters $\thetab$ remains finite and conforms to the data.
For this embedded technique, the model calibration problem thus involves finding the posterior distribution on $\alphab$ via Bayes’ theorem  \eref{eq:bayes}
\begin{equation} \label{eq:bayes_embedded}
\prob(\alphab|{\Dc}, \tilde{M}) \sim \prob({\Dc}|\alphab, \tilde{M}) \, 
\prob(\alphab|\tilde{M}),
\end{equation}
where $\alphab$ has been substituted for $\thetab$ and, again, $\prob(\alphab|{\Dc}, \tilde{M})$ is the posterior, $\prob(\Dc|\alphab, \tilde{M})$ is the likelihood, and $\prob(\alphab|\tilde{M})$ is the prior. 
Once the posterior for the parameters is characterized, it can be propagated through the model to obtain the posterior predictive distributions for quantities of interest (namely stress in this case).
The key feature of these predictions is that their uncertainty is affected by both parameter and model uncertainties. 
For brevity, we leave the full mathematical details of this embedded approach, including the likelihood formulation, to \ref{app:embedded}.

\section{Results} \label{sec:results}
In this section, we present the details of the construction of the particular surrogate models from the full finite element plasticity model, their calibration to the experimental data, and the physical interpretation of the resulting predictions and parameter estimates. 
Most of the numerical results presented below are obtained using the UQ Toolkit \cite{uqtk:web} package.

\subsection{Surrogate Model} \label{sec:surrogate_construction}

To describe the material stress-strain behavior we analyze, calibrate and compare three nested plasticity models of increasingly complex phenomenology, namely perfect plasticity, linear hardening and saturation hardening.
As mentioned in \sref{sec:theory}, we focus on five parameters: Young's modulus, $\young$; yield strength, $\yield$; hardening modulus, $\hard$; saturation modulus, $\satmod$; and saturation exponent, $\satexp$, which control the elastic-plastic stress response.
We build Legendre-Uniform PC expansions of these parameters by assuming that they are uniformly distributed over a chosen range 
\begin{alignat}{5} 
\young  &=& 200  \  &+& 80     \,\xi_1  \quad &\text{[GPa]}, \nonumber \\
\yield  &=& 1.2  \  &+& 0.5    \,\xi_2  \quad &\text{[GPa]}, \nonumber \\
\hard   &=& 3.005\  &+& 2.995  \,\xi_3  \quad &\text{[GPa]}, \label{eq:inputPCE} \\
\satmod &=& 0.2  \  &+& 0.2    \,\xi_4  \quad &\text{[GPa]}, \nonumber \\
\satexp &=& 1750 \  &+& 1250   \,\xi_5  \quad &  \nonumber
\end{alignat}
where $\{\xi_1,\xi_2,\xi_3,\xi_4,\xi_5\} \sim {\cal U}(-1,1)$ are independent identically distributed (i.i.d.) standard uniform random variables.
We chose these parameters ranges to be large enough that the corresponding predictions can capture the variability of the experimental data shown in \fref{fig:experiment}b. 
Also, we remark that the expansion with i.i.d.~random variables is a common step to build the surrogate model. 
Any correlations between the parameters will then be discovered through the inverse problem, see \eg \crefs{Rizzi:2012b,Rizzi:2013b,MARZOUK2007}.
The priors for $\alphab$ are constructed such that the target physical parameters have their target priors, \eref{eq:inputPCE}.

\subsubsection{Two parameter perfect plasticity model} \label{sec:two_parameter}
\begin{figure}[!t]
\centering

\subcaptionbox{}%
{\includegraphics[width=\figwidthtwo]{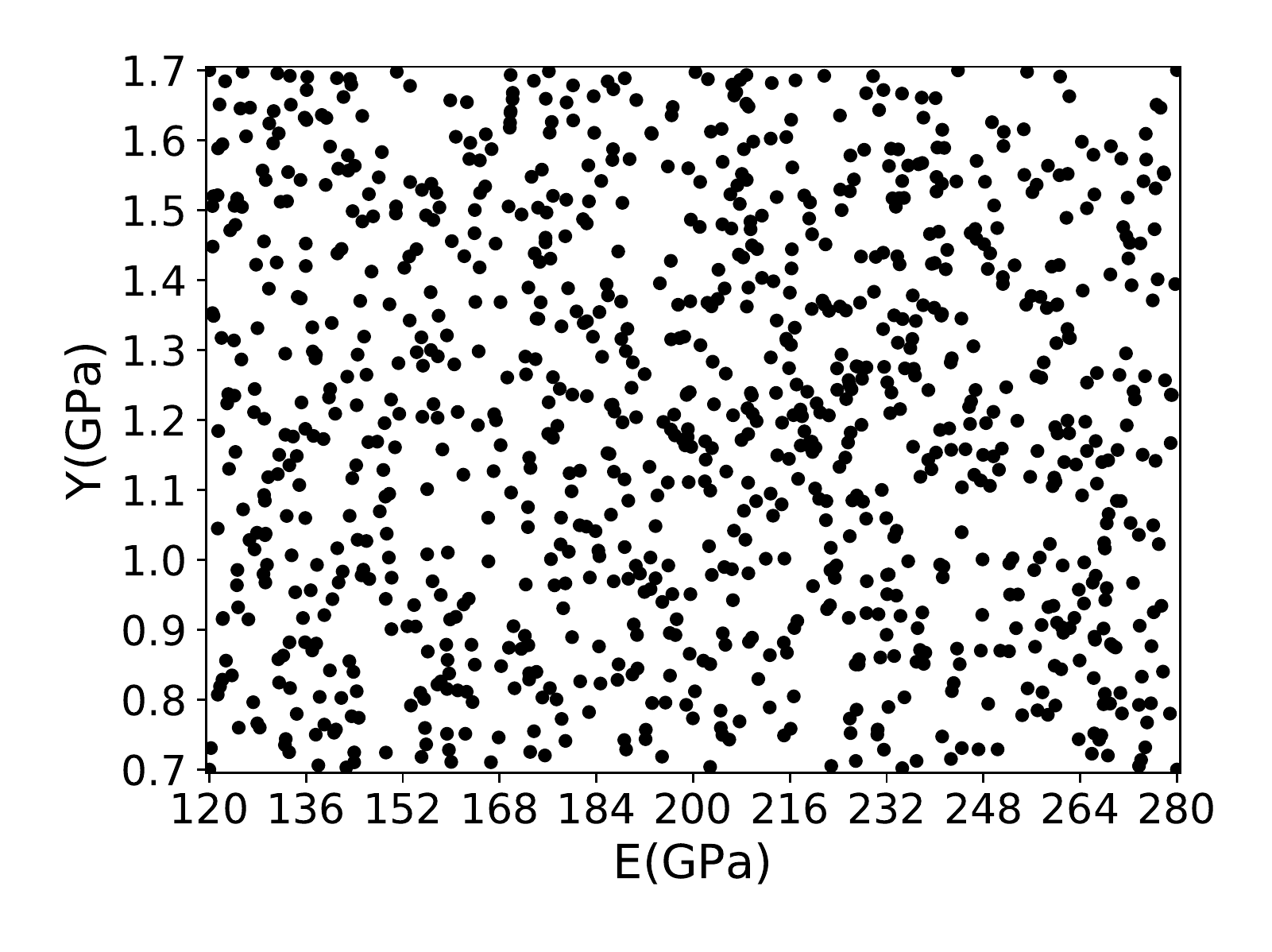}}%
\hfill%
\subcaptionbox{}%
{\includegraphics[width=\figwidthtwo]{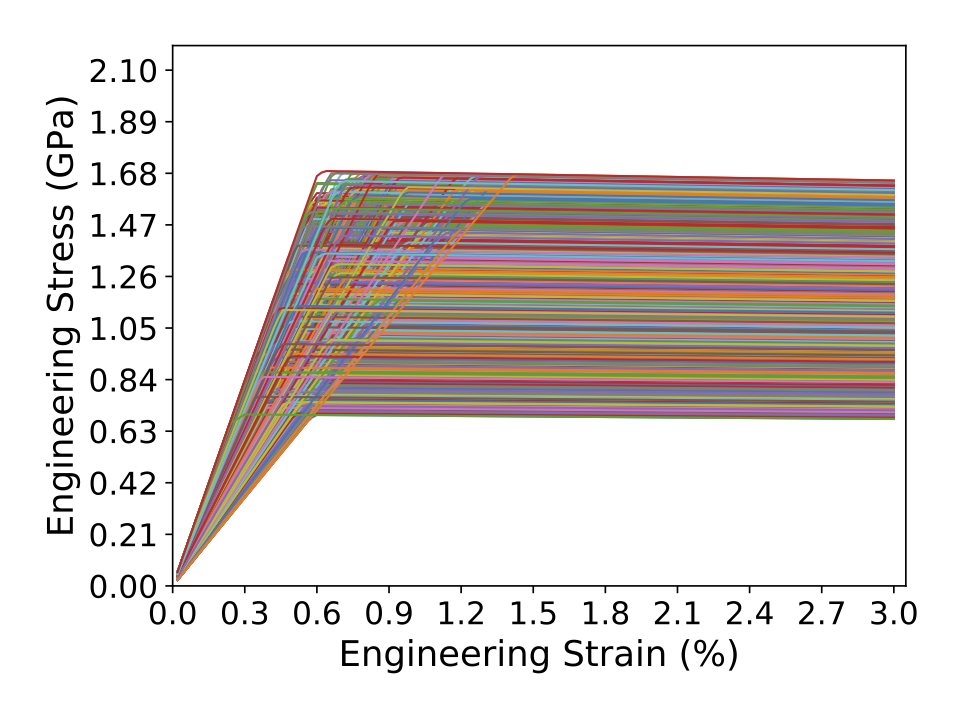}}

\caption{Two parameter surrogate data: (a) training samples in the space $(\young,\yield)$ and (b) corresponding stress-strain curves used to build the surrogate for the two parameter model.
}
\label{fig:2P_samples2d}
\end{figure}

For the two parameter model, the stress is expressed as a function of strain and 
Young's modulus, $\young$, and yield strength, $\yield$ according to
\begin{equation}
\stress(\strain) = M^{(2)}(\strain; \{ \young, \yield\}),
\end{equation}
where we use the superscript ``$(2)$'' to identify this as the two parameter model.
As mentioned before, the full tension simulation is expensive to evaluate, so we leverage the PC expansions of the inputs, \eref{eq:inputPCE}, to create PCes of the stress at each strain $\{\strain_j\}$ value
\begin{equation}
\stress(\strain_j) = \tilde{M}^{(2)}_j(\xi_1, \xi_2) 
\approx M^{(2)}(\strain_j, \young(\xi_1), \yield(\xi_2)),
\quad j = 0,\ldots,n_{\strain}-1.
\end{equation}
To build this sequence of PCes, $\{\tilde{M}^{(2)}_j, j=0,n_\strain-1\}$, we employ regression using uniform random samples in the two-dimensional space ($\xi_1,\xi_2)$.
In particular, we generate $1004$ training ($1000$ samples in the inner domain and four additional ones for the corners) to build the surrogate model, and $250$ validation samples to assess its accuracy and check for over-fitting.
(We chose regression over the more computationally efficient stochastic collocation via sparse quadrature grids given the superior results in constructing the piece-wise smooth surrogates to be introduced forthwith.)
\fref{fig:2P_samples2d} shows the training samples mapped back to the physical domain ($\young,\yield$) using \eref{eq:inputPCE}, and the corresponding plasticity simulations.
Given the number of samples, the regression approach is suitable for constructing PC expansions of up to ninth order. 
Computing a higher-order expansion would lead to an under-determined problem. 
The ensemble of responses, shown in \fref{fig:2P_samples2d}b, are clearly piece-wise linear, as expected, with a slight downward slope in the post-yield response due to finite deformation effects.

\begin{figure}[!t]
\centering
\subcaptionbox{}%
{\includegraphics[width=\figwidth]{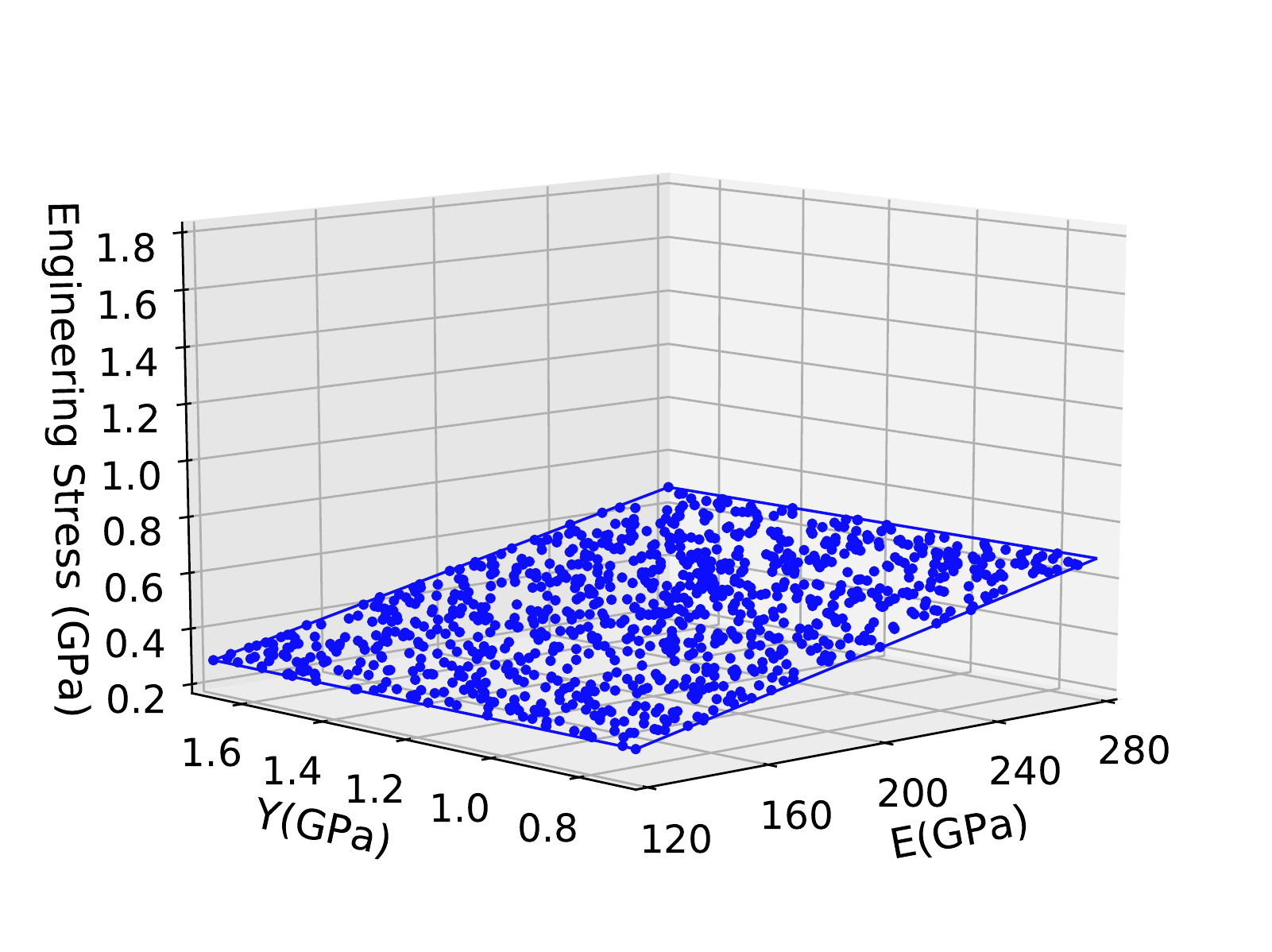}}%
\hfill%
\subcaptionbox{}%
{\includegraphics[width=\figwidth]{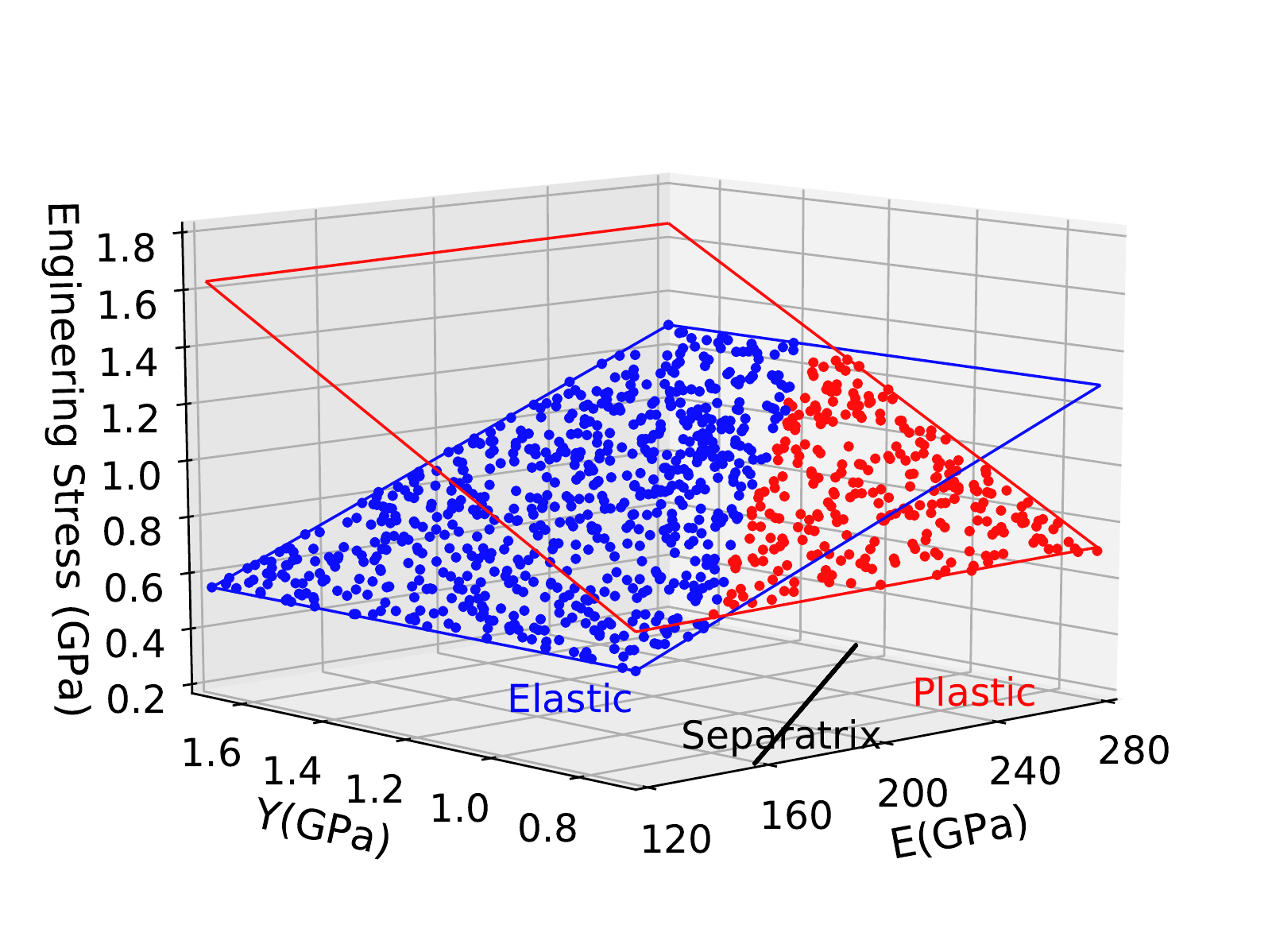}}%
\hfill%
\subcaptionbox{}%
{\includegraphics[width=\figwidth]{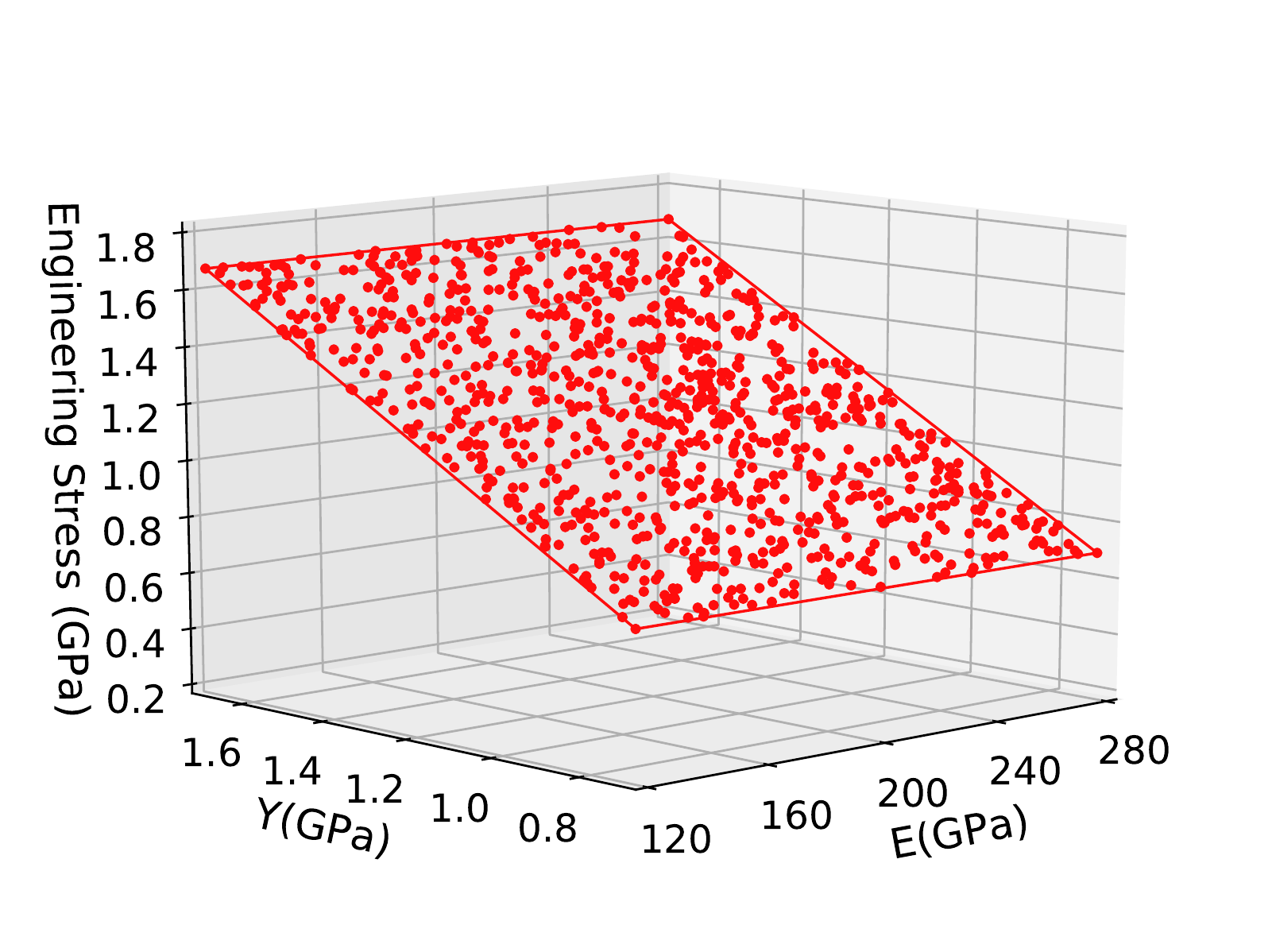}}%
\caption{Stress plotted as a function of elastic modulus ($\young$) and yield ($\yield$) at three different strains: (a) fully elastic,  $\strain=$0.12\%, (b) elastic-plastic, $0.39\%$, and (c) fully plastic $1.95\%$, over the domain of the two parameter model.}
\label{fig:2P_3dPlots}
\end{figure}

For the two parameter model, a key feature is that the stress-strain behavior has a discontinuity of the first derivative when the behavior switches from elastic to plastic as the stress, which depends on $\young$, exceeds the particular yield $\yield$ value.
This dependence on both parameters can be observed in \fref{fig:2P_3dPlots}, which shows the stress plotted as a function of $\young$ and $\yield$ at three different strain locations, $\strain = 0.12, 0.39, 1.95 \%$. 
These points have been chosen such that the first, $\strain=0.12\%$, is within a regime for which all points in the $(\young,\yield)$ space are fully elastic; the second value, $\strain=0.39\%$, is in a mixed elastic-plastic regime, and the third point, $\strain=1.95\%$, identifies a fully plastic regime. 
\fref{fig:2P_3dPlots} shows that within the elastic and plastic regimes, the behavior is linear, and the separator is a line.
We will exploit this observation, and use the plastic strain (which is zero in the elastic regime) as a classifier to sub-divide the training data. 

\begin{figure}[!t]
\centering
\subcaptionbox{}%
{\includegraphics[width=\figwidth]{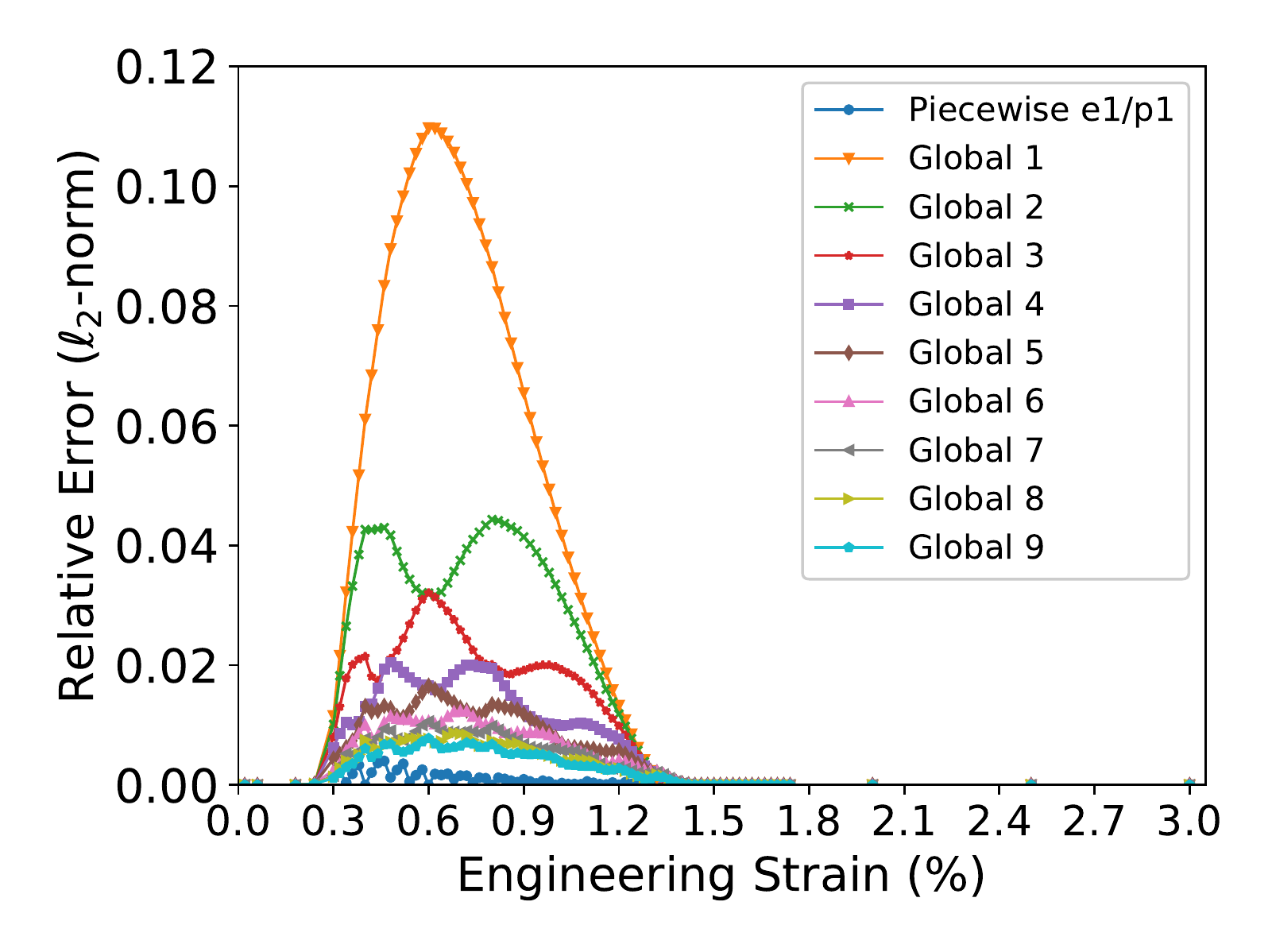}}%
\hfill%
\subcaptionbox{}%
{\includegraphics[width=\figwidth]{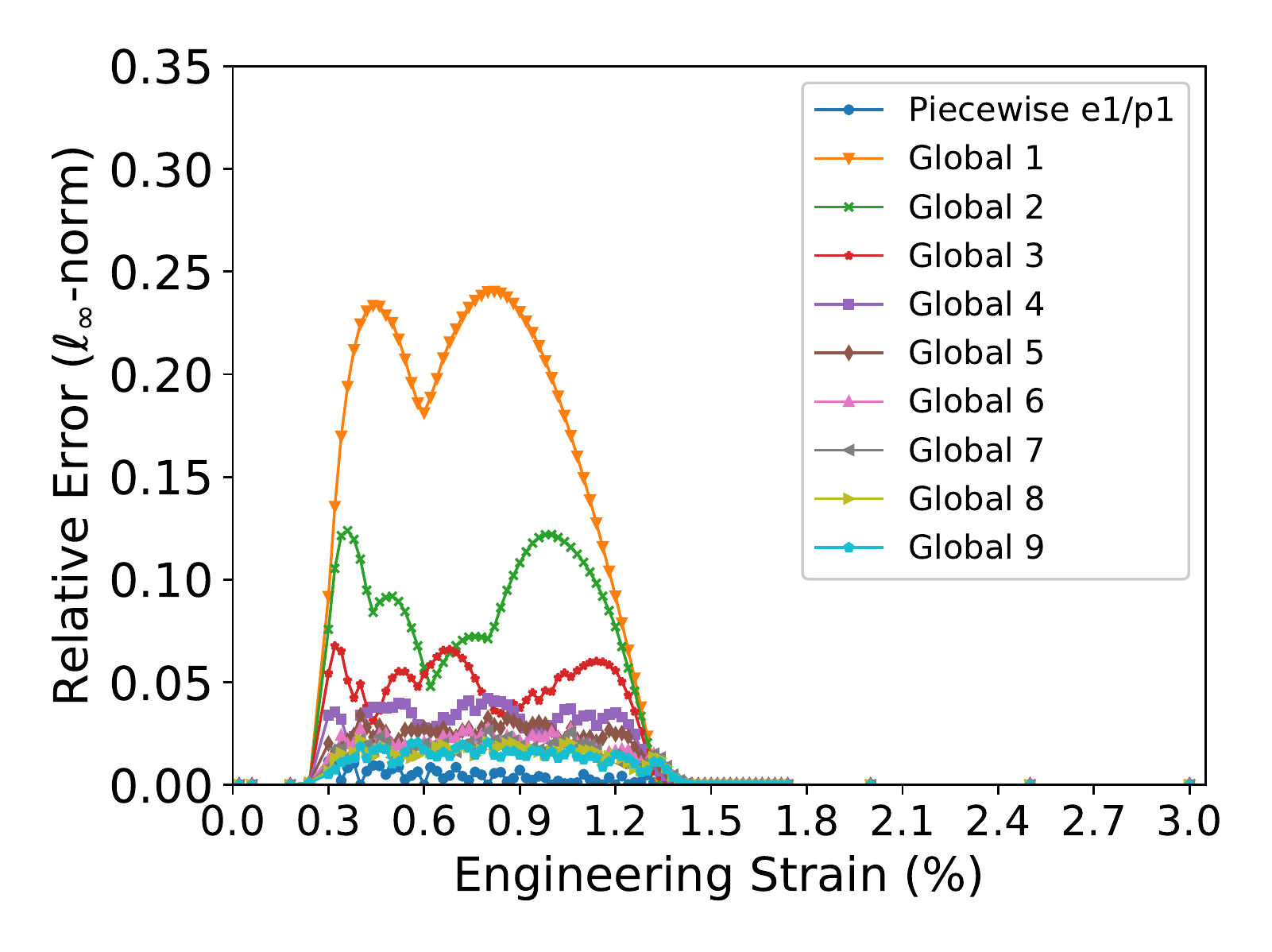}}
\caption{Surrogate error for the two parameter model at each target strain location based on (a) the $\ell_2$-norm~(a) and (b) the $\ell_{\infty}$-norm. 
Results are shown for global PC expansions of orders up to nine, as well as the 
piece-wise surrogate based on linear polynomials for both the elastic and plastic response.}
\label{fig:2P_errors}
\end{figure}

To compare the accuracy of global  PC expansions up to ninth order and a piecewise linear surrogate built over the elastic and plastic sub-domains, in \fref{fig:2P_errors} we show the relative error based on the $\ell_2$-norm and $\ell_{\infty}$-norm for the various PCes as a function of the strain. 
From \fref{fig:2P_errors} we observe that a \emph{global} linear PCe is accurate where the regime is either fully elastic or plastic, but inaccurate in the mixed region where the discontinuity in the response makes a global representation sub-optimal.
Also, as we increase the order of the PCe from first to fifth order, the results do not change within the elastic and plastic regions, but improve in the mixed region.
However, when the order of the expansion is at least fifth order, the errors do not decrease as rapidly which suggests over-fitting.
Lastly, the low-order piece-wise linear surrogate has the lowest error in both norms across the strain range and, hence, it is more suitable than a high order global surrogate for this model response.

\subsubsection{Three parameter linear hardening model} \label{sec:three_parameter}
Augmenting the two parameter model with the post-yield phenomenology controlled by the hardening modulus ($\hard$) results in the three parameter model, $\stress(\strain) = M^{(3)}(\strain; \{ \young, \yield, \hard \})$.
In this case, we use a total of $2008$ training samples ($2000$ in the inner domain and the $8$ additional ones for the corners), and $500$ validation points.

Again, we build global polynomial surrogates of increasing order and a piece-wise low order polynomial surrogate to represent the response surface.
To capture the additional complexity in the post-yield response we used a quadratic PC in the plastic regime of the mixed, piece-wise surrogate which is linear in the elastic regime.
The resulting $\ell_2$-norm and $\ell_{\infty}$-norm errors (not shown for brevity) have the same trends as those for the two parameter model shown in \fref{fig:2P_errors} due to similarity in the slope discontinuity of the response and, likewise, the piece-wise surrogate is the best representation of the full simulation response for linear hardening.

\subsubsection{Five parameter saturation hardening model} \label{sec:five_parameter}
\begin{figure}[!t]
\centering
{\includegraphics[width=0.55\textwidth]{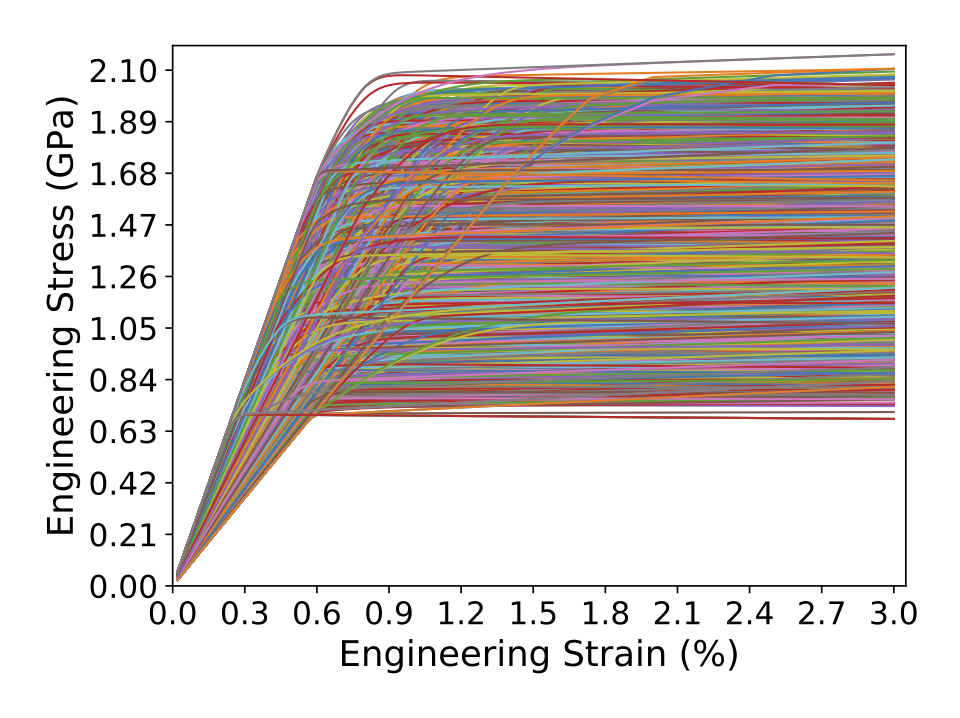}}
\caption{Stress-strain curves samples used to build the surrogate for the five parameter model.
}
\label{fig:5P_samples}
\end{figure}

The five parameter model adds the saturation modulus ($\satmod$) and saturation exponent ($\satexp$), yielding $\stress(\strain) = M^{(5)}(\strain; \{\young, \yield, \hard, \satmod, \satexp\})$.
To build and check this surrogate, we collected $5032$ training and $500$ validation samples. 
As shown in \fref{fig:5P_samples}, the five parameter model adds a smoother elastic-plastic transition through the presence of the saturation modulus and exponent. 

\begin{figure}[!t]
\centering
\subcaptionbox{}%
{\includegraphics[width=\figwidth]{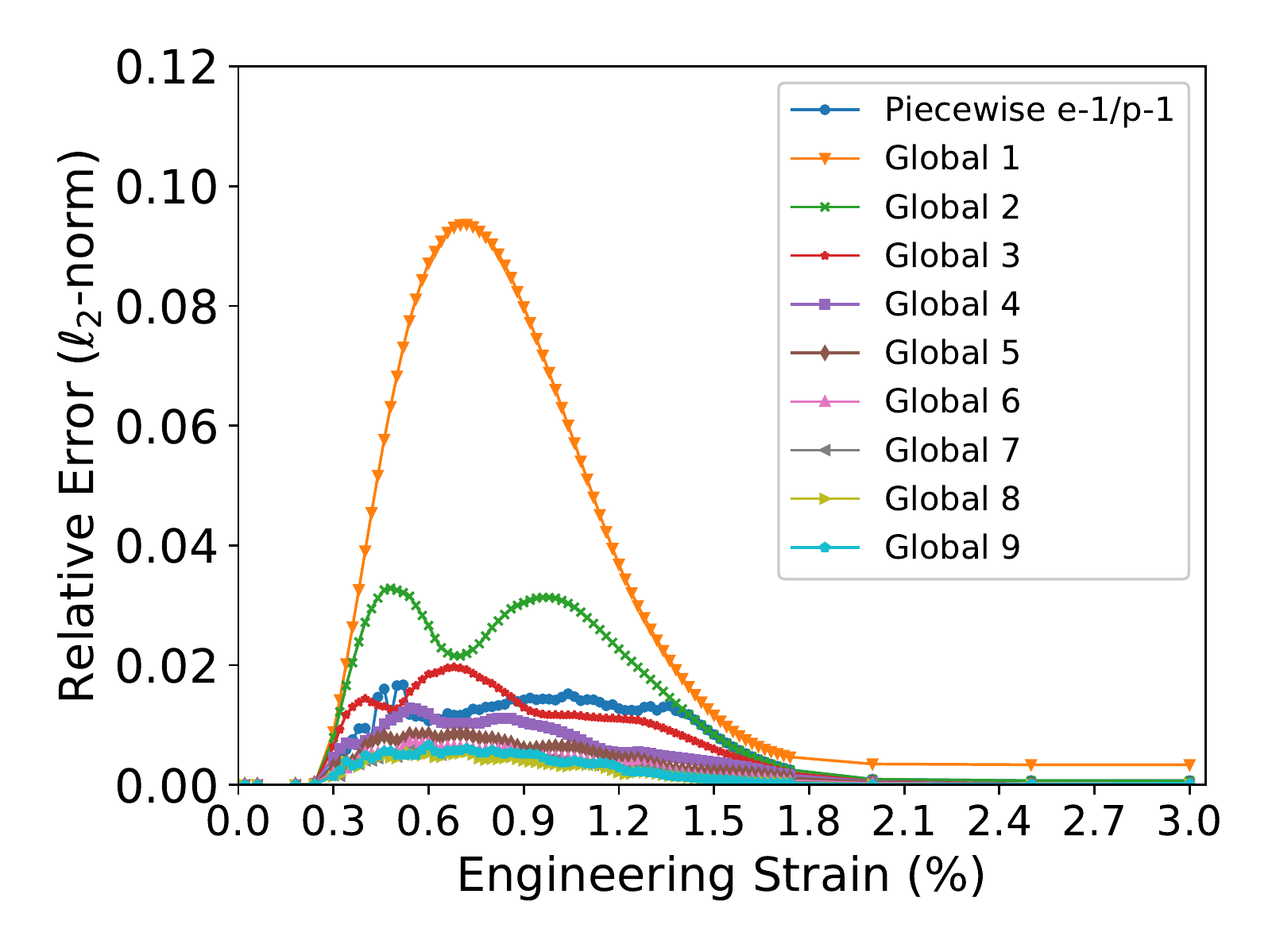}}
\hfill%
\subcaptionbox{}%
{\includegraphics[width=\figwidth]{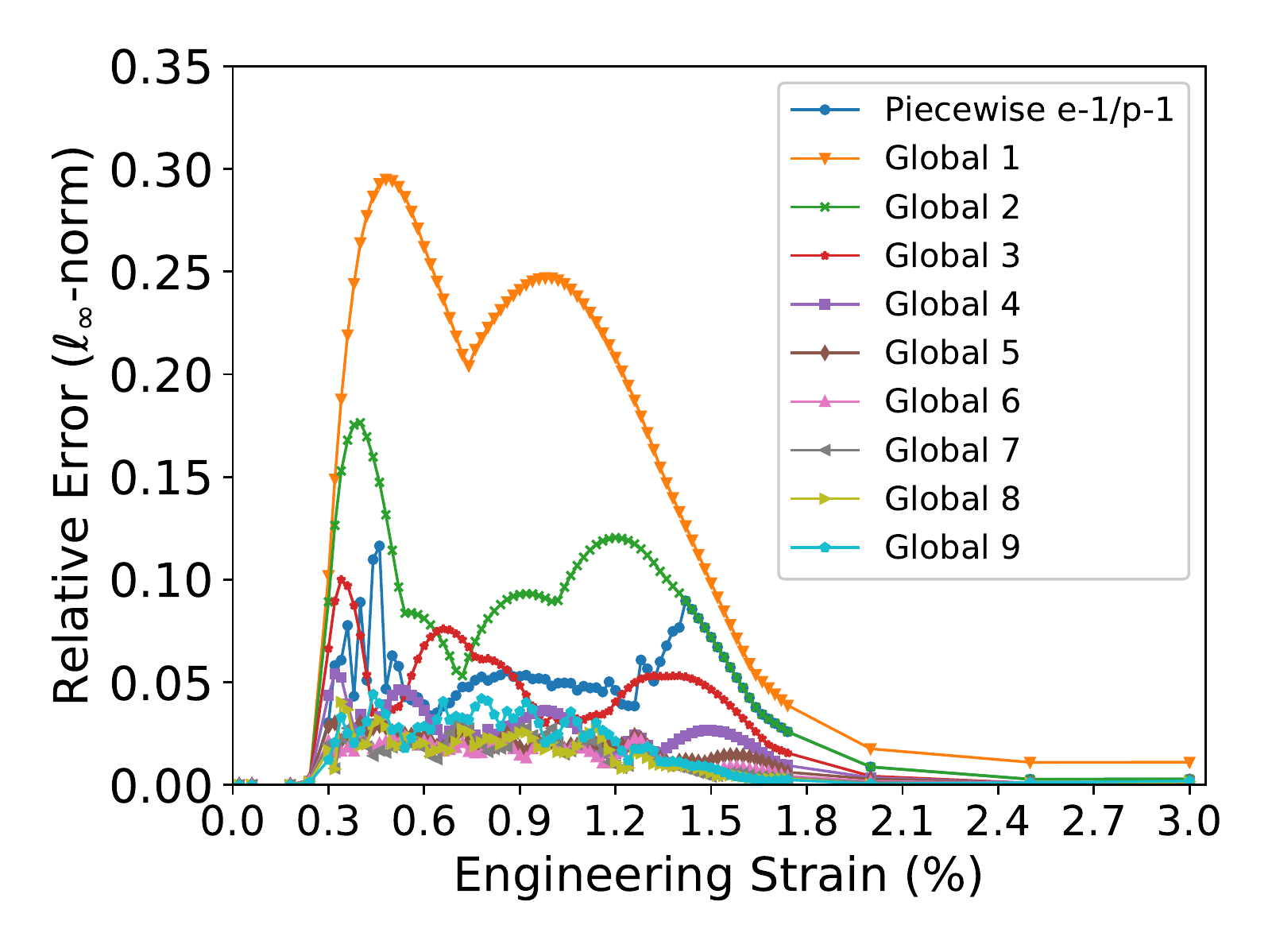}}
\caption{Surrogate error for the five parameter model at each target strain location based on (a) the $\ell_2$-norm and (b) the $\ell_{\infty}$-norm. 
Results are shown for global PC expansions of orders up to nine, as well as the piece-wise surrogate with a linear polynomial over the elastic regime, and a quadratic within the plastic regime.}
\label{fig:5P_errors}
\end{figure}

\fref{fig:5P_errors} shows the relative error based on the $\ell_2$-norm and $\ell_{\infty}$-norm for the various PCes as a function of the strain. 
As with the piece-wise surrogate for the three-parameter model, here we use a linear PCe for the elastic and a quadratic PCe for the plastic.
Unlike the results for the two simpler models, the piece-wise surrogate does not outperform the global surrogates.
In this case, a global PCe of order $\ge$6 gives the lowest errors. 
Since the errors are comparable for these polynomials, orders $>$6 are likely over-fitting the full simulation data.

\subsection{Sensitivity Analysis} \label{sec:sens_results}

\begin{figure}[!t]
\centering
\subcaptionbox{}%
{\includegraphics[width=\figwidth]{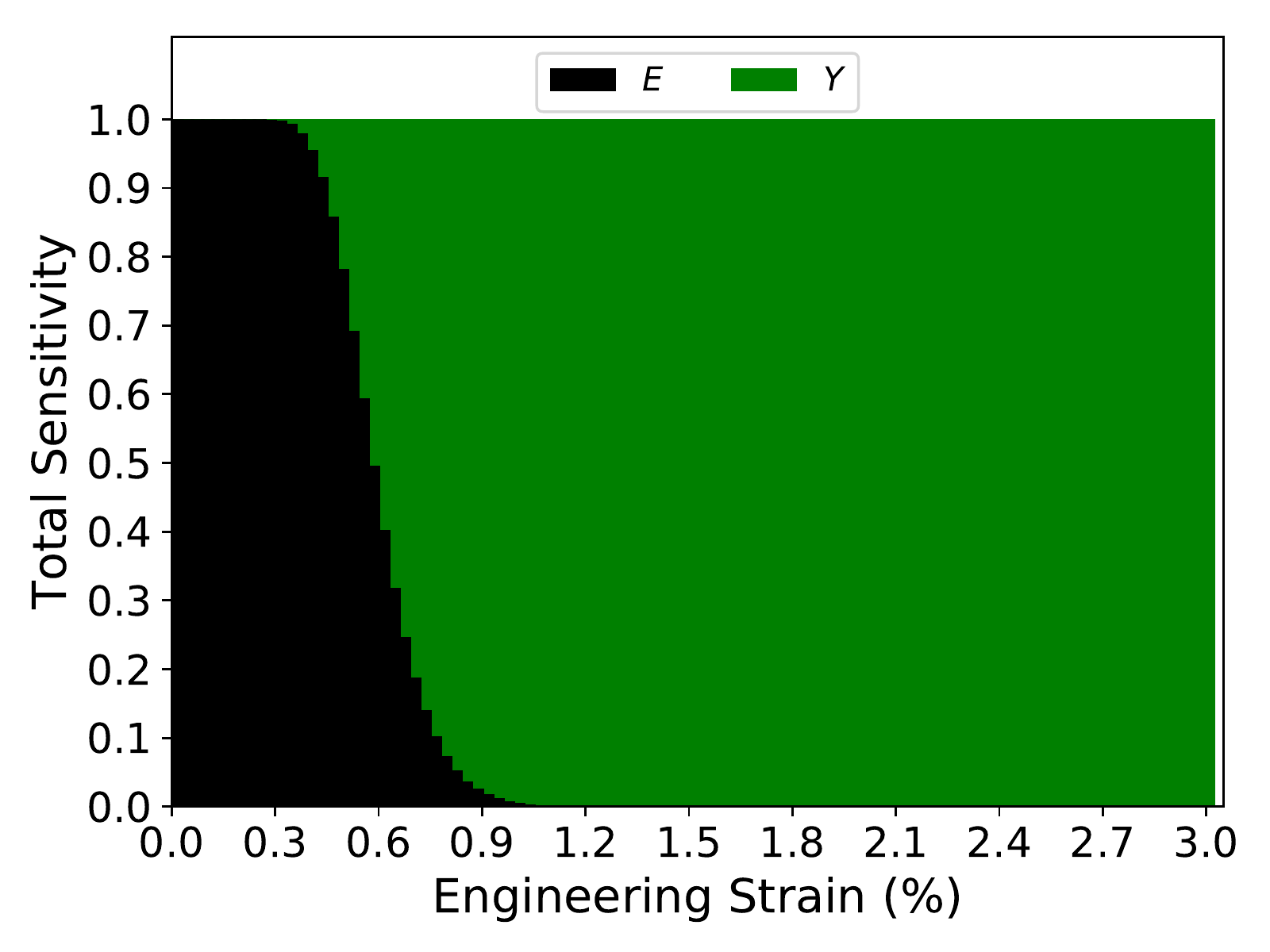}}%

\subcaptionbox{}%
{\includegraphics[width=\figwidth]{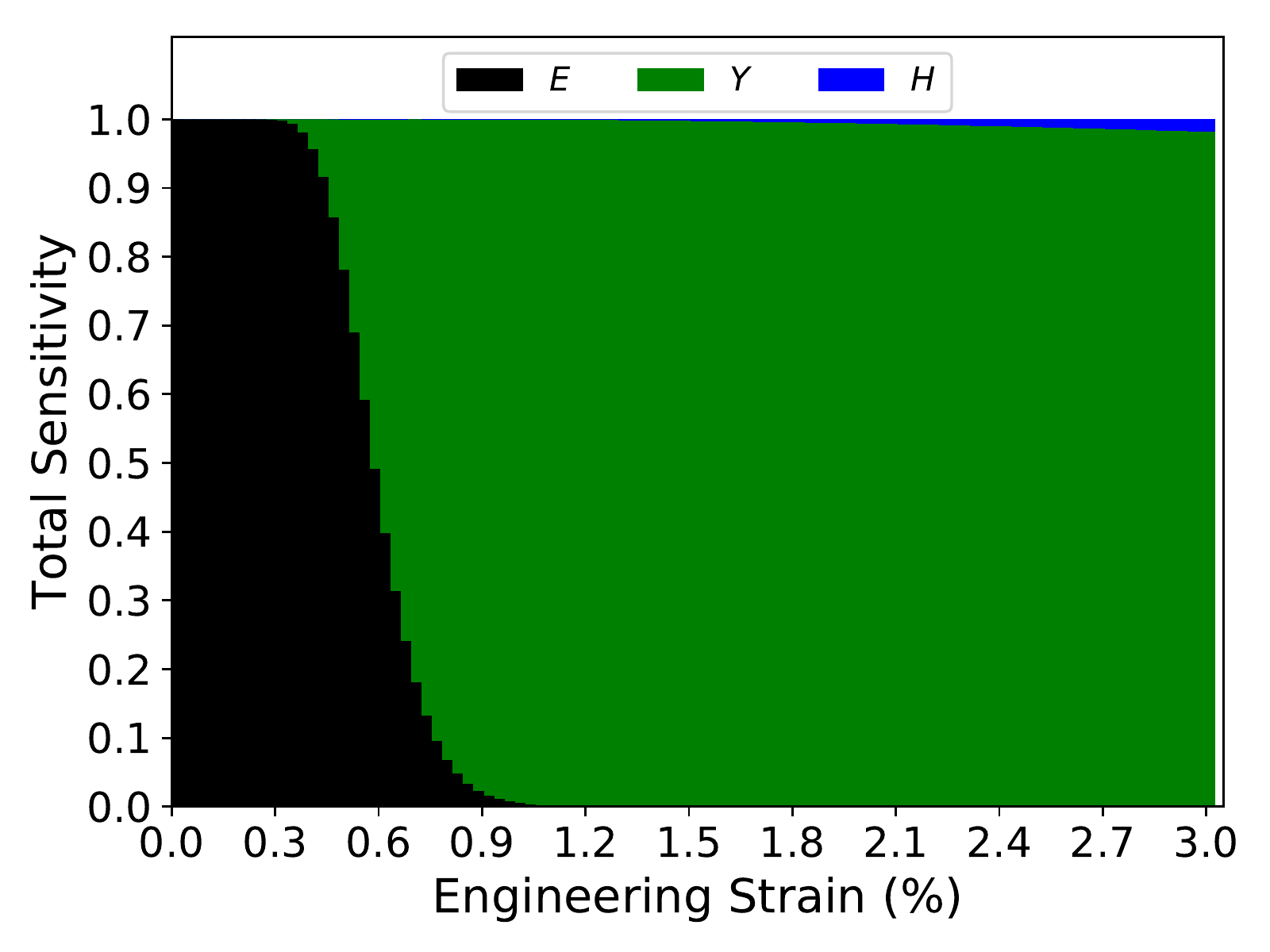}}

\subcaptionbox{}%
{\includegraphics[width=\figwidth]{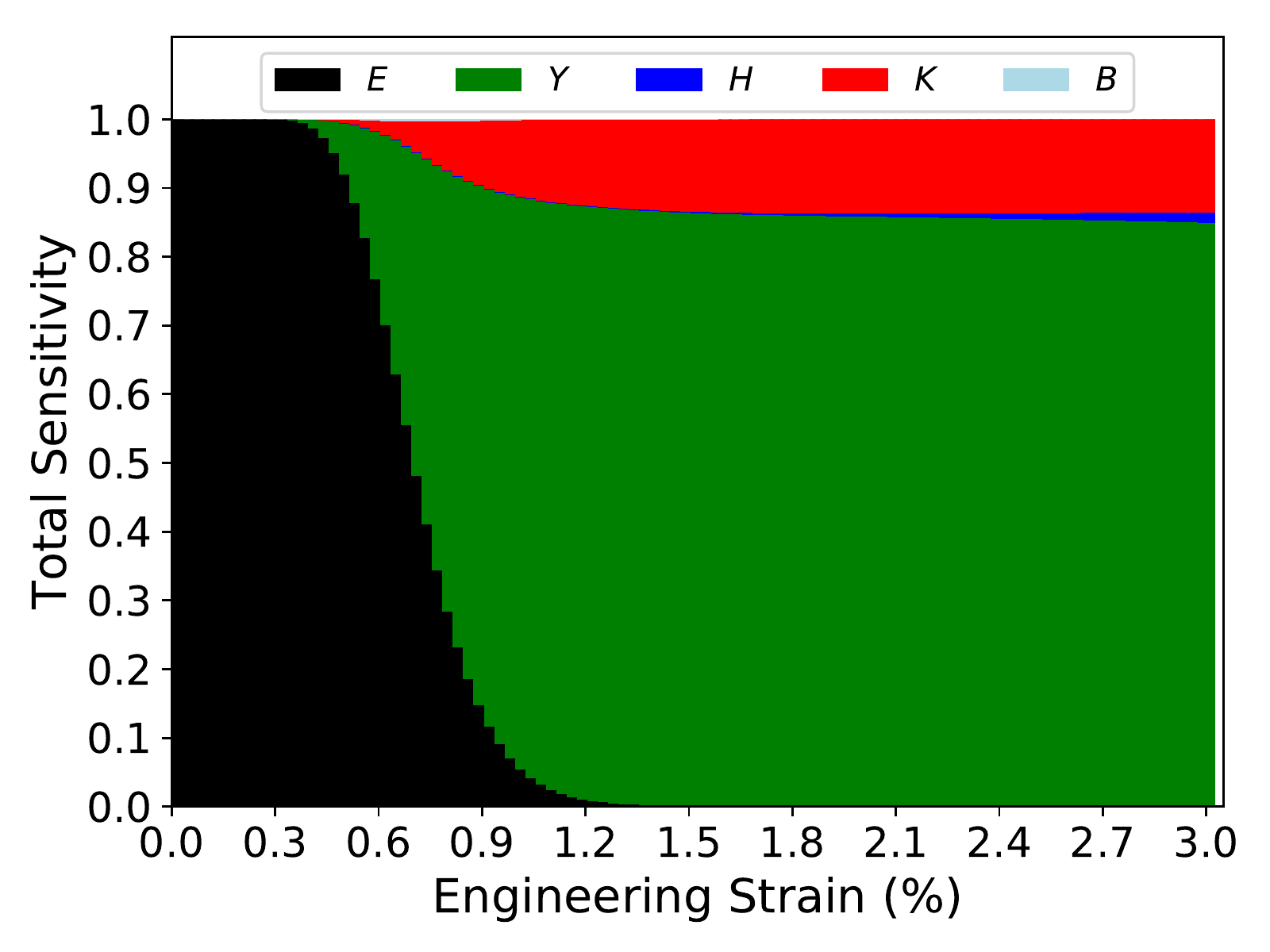}}%
\caption{Total sensitivities as a function of the strain obtained for the (a) two, (b) three, and (c) five parameter model.}
\label{fig:tot_sens}
\end{figure}

As mentioned, one advantage of building a PCe surrogate is that one can obtain global Sobol sensitivities of a target quantity of interest with respect to the input parameters \cite{SUDRET2008964}. 
Here we compute the total sensitivities \cite{SUDRET2008964} of the stress using the surrogate built at each strain point. 
\fref{fig:tot_sens} shows the sensitivities obtained over the range of the surrogate for the: (a) two, (b) three, and (c) five parameter models. 
The sensitivities are influenced by the range chosen to build the surrogate model. 
The experimental stress-strain curves show little hardening and so the surrogate was constructed with over a narrow range of hardening parameters, \eref{eq:inputPCE}, which appropriately minimizes their importance.
Also, it is apparent that the relative importance of the parameters evolves with strain.
As expected, \fref{fig:tot_sens} shows that the Young's modulus $\young$ is the dominant parameter within the elastic regime. 
At larger strains, in the transition between the elastic and plastic regimes, $\young$ becomes gradually less important and the yield $\yield$ starts to dominate. 
In the more complex models, the hardening $\hard$ and the saturation parameters $\satmod$ and $\satexp$ play a relative minor role due to the fact that the data displays little hardening and the two parameter model is a good representation of the majority of the stress behavior.
In both the three and five parameter models the sensitivity to $\hard$ is slim and almost negligible, whereas the added $\satmod$ in the five parameter model is clearly not negligible apparently for its role in determining the ``knee'' in the stress-strain at the elastic-plastic transition.
More discussion of this behavior will be given in \sref{sec:comparison}.

\subsection{Calibration} \label{sec:calibration_results}

In this section, we discuss and contrast the results obtained from the inverse problem formulated the additive error and those obtained using the embedded formulation.
We use the three parameter, linear hardening model as a reference case, which we discuss in detail, and then show the main results for the other models. 

\subsubsection{Inversion with the additive error model} \label{sec:additive_inversion}

We assume the measurement noise to be constant, $\varsigma^2$, along the strain axis, \ie the measurement error does not depend on the strain $\strain$. 
Hence, the parameters to be inferred are $\thetab = \{E,Y,H, \varsigma^2\}$. 
As priors, we choose uniform densities with ranges coinciding with those chosen to build the surrogate model in \eref{eq:inputPCE}. 
For the variance, we choose a uniform prior over the positive axis.
This is typically appropriate because the surrogate might not be reliable outside the range where it was computed on.
The prior plays a minor role if a substantial amount of data is available, making the problem likelihood-informed rather than prior-informed in the large data limit.

We leverage this case to highlight some key features of Bayesian calibration applied to the present problem of representing material variability: first, the dependence on the type of surrogate model used, second, the batch-to-batch differences in the resulting parameters, third, the correlations between the target parameters, and finally, convergence of the results with the amount of data used in the inverse problem.  

\begin{figure}[!t]
\centering
\subcaptionbox{}%
{\includegraphics[width=\figwidthtwo]{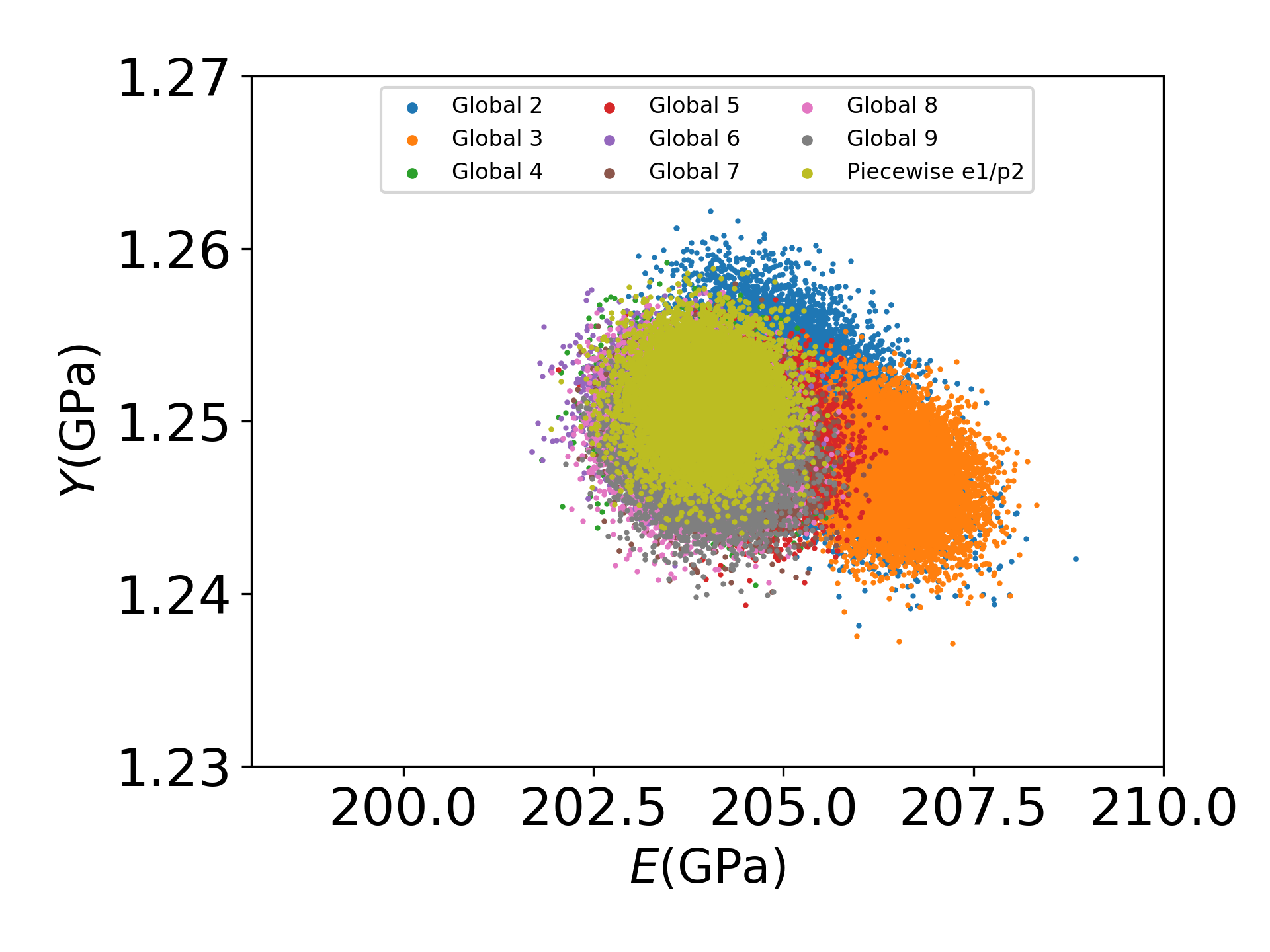}}%
\hfill%
\subcaptionbox{}%
{\includegraphics[width=\figwidthtwo]{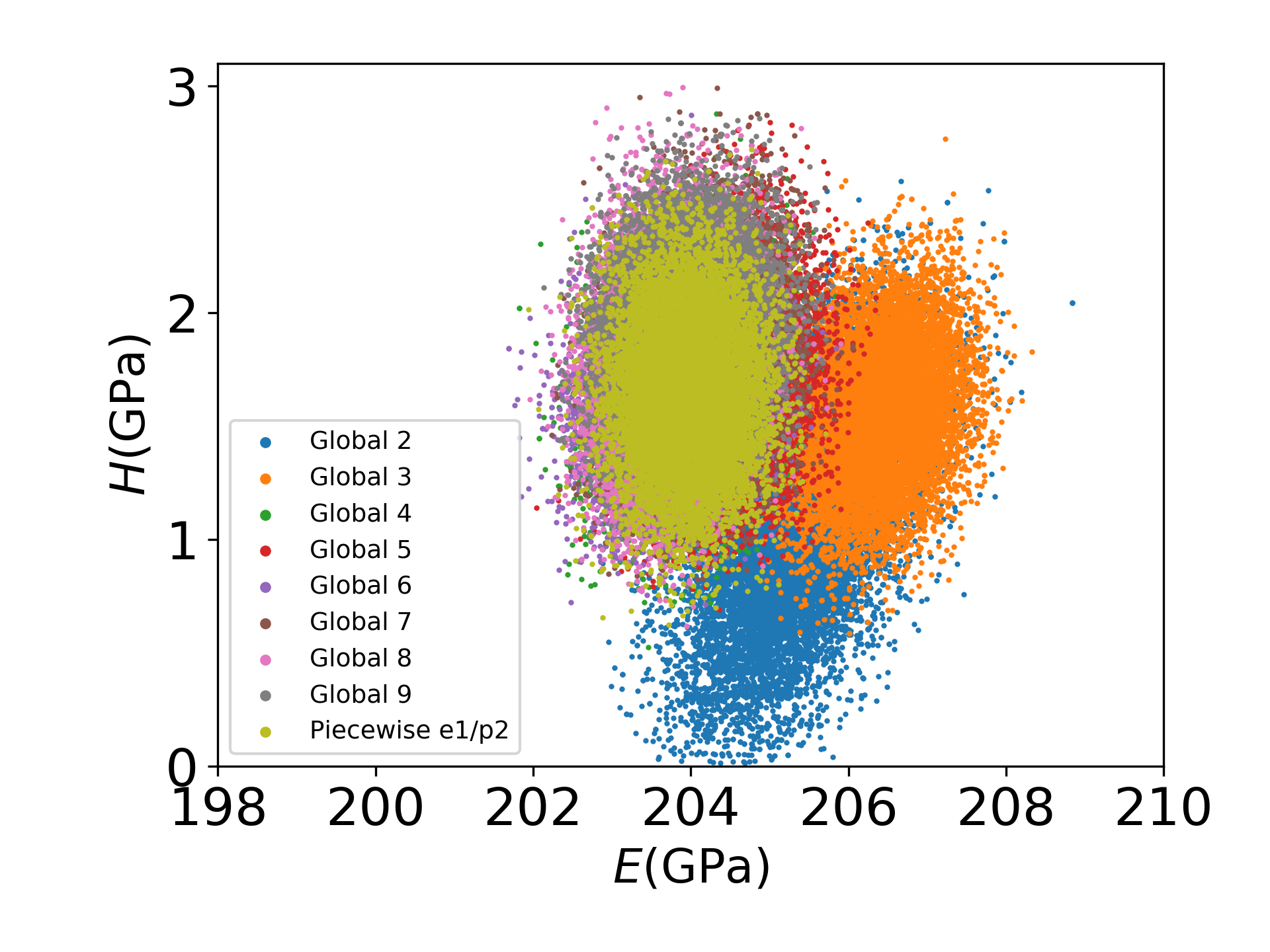}}

\subcaptionbox{}%
{\includegraphics[width=\figwidthtwo]{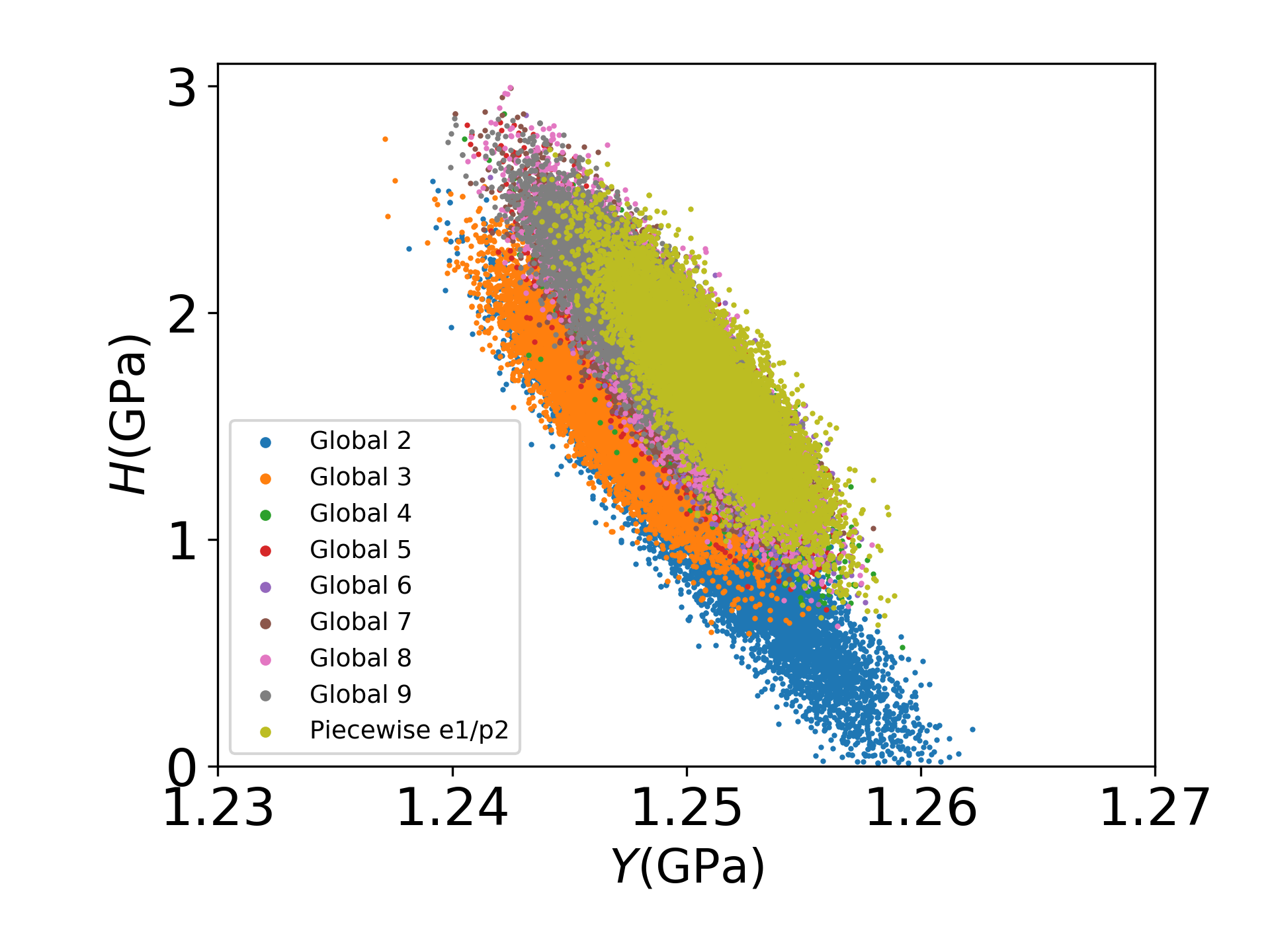}}%
\hfill%
\subcaptionbox{}%
{\includegraphics[width=\figwidthtwo]{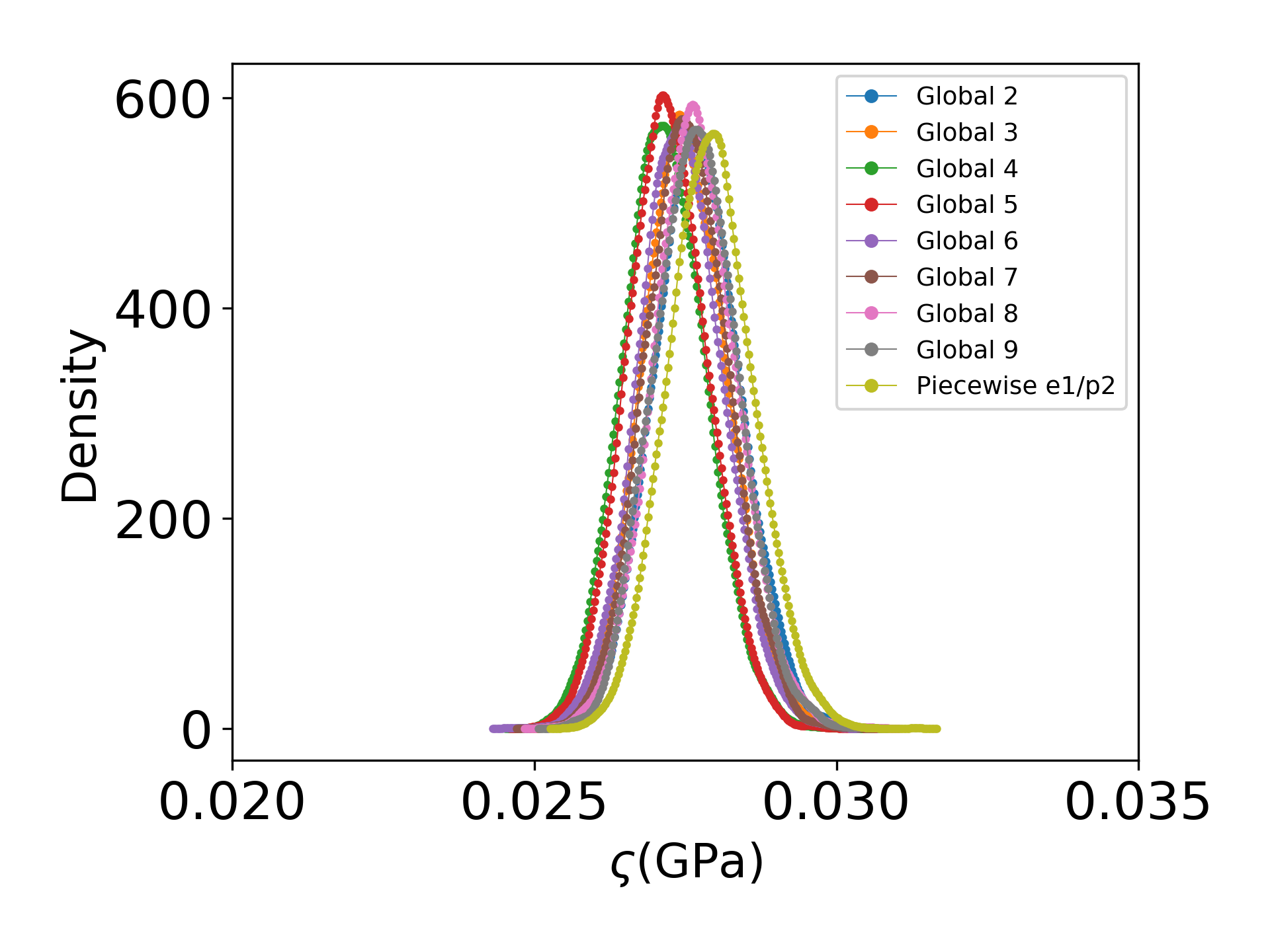}}%
\caption{Results for the additive-model-based inversion run with the three parameter model using $10$ stress-strain curves from $\Dc_3$ showing the effect of the surrogate.
We plot samples of the joint posteriors: (a) $\prob(\young, \yield)$, (b) $\prob(\young, \hard)$, and (c) $\prob(\yield,\hard)$, as well as (d) the marginalized posterior for the standard deviation of the measurement error, $\prob(\varsigma)$. }
\label{fig:3P_calibclassFig1}
\end{figure}

\fref{fig:3P_calibclassFig1} shows the joint posteriors between the physical parameters, \fref{fig:3P_calibclassFig1}(a,b,c), and the marginalized posterior for the standard deviation of the measurement error, \fref{fig:3P_calibclassFig1}(d), obtained using $10$ stress-strain curves randomly chosen from $\Dc_3$. 
Results are shown for each of the surrogates we constructed. 
The convergence of the resulting PDFs with polynomial order gives us confidence that the higher order global surrogates and the piece-wise surrogate lead to sufficiently accurate posterior parameter distributions.
Since the distributions are not skewed, it is apparent that the extracted elastic modulus $\young$ is not correlated with the yield $\yield$, or the hardening $\hard$, whereas there is a weak negative correlation between yield and hardening. 
We conjecture that the pre-yield data informs $\young$ independent of the other parameters and, likewise, the yield point informs $\yield$; however, there is a trend in the experimental data, see \fref{fig:experiment}c, for high yield points to lead to subsequent low post yield slopes and vice versa.
The maximum \aposteriori (MAP) value of the inferred standard deviation is around  0.027~GPa. 
This value is larger than the one estimated directly from the experimental data (0.020~GPa). 
This is expected because we are not accounting for model error and, therefore, the model discrepancy is lumped into the measurement error. 
Note that the range used for the plots are much smaller than those originally chosen for the surrogate construction, \eref{eq:inputPCE}.

\begin{figure}[!t]
\centering
{\includegraphics[width=0.55\textwidth]{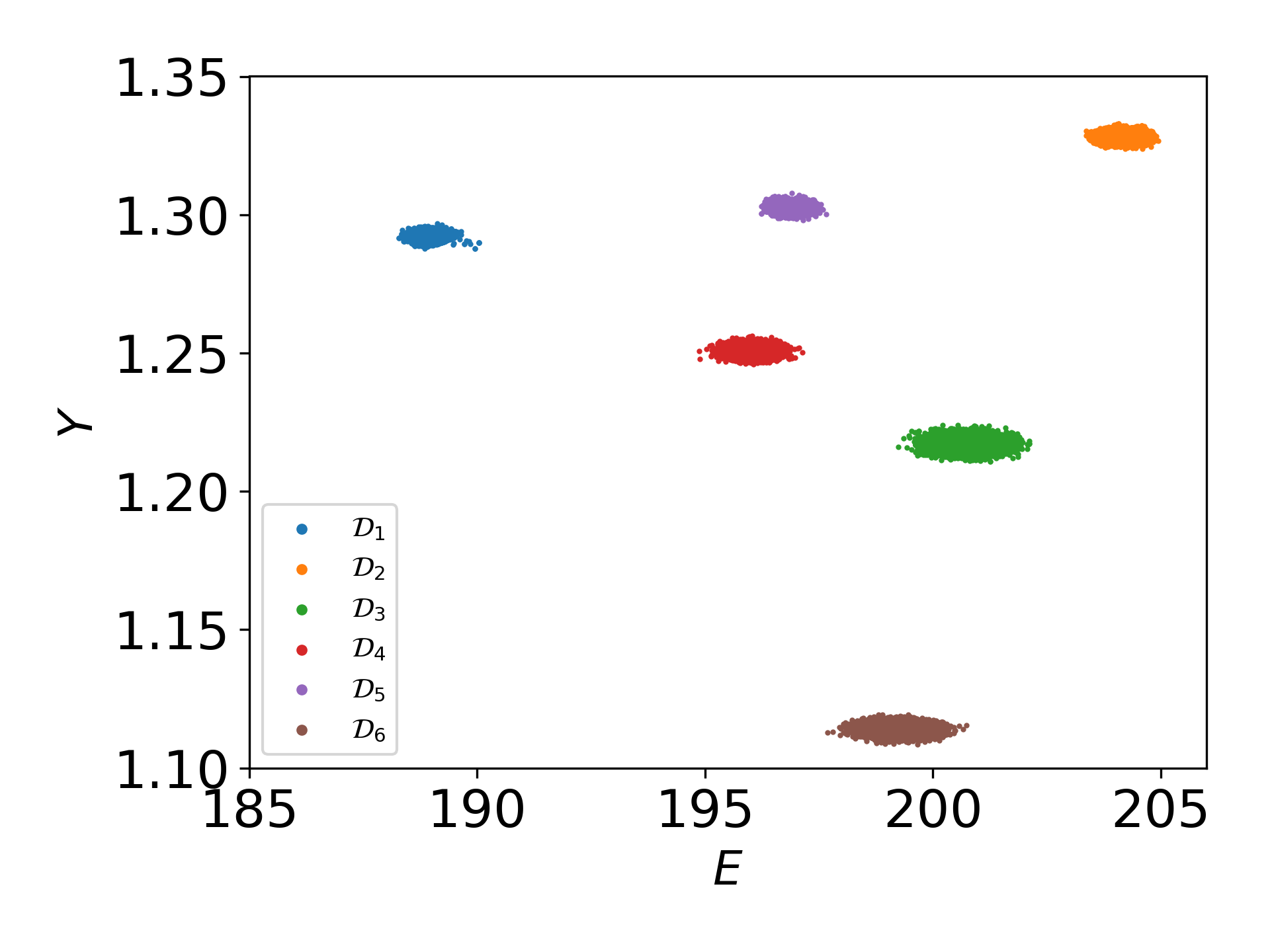}(a)}\\
{\includegraphics[width=0.55\textwidth]{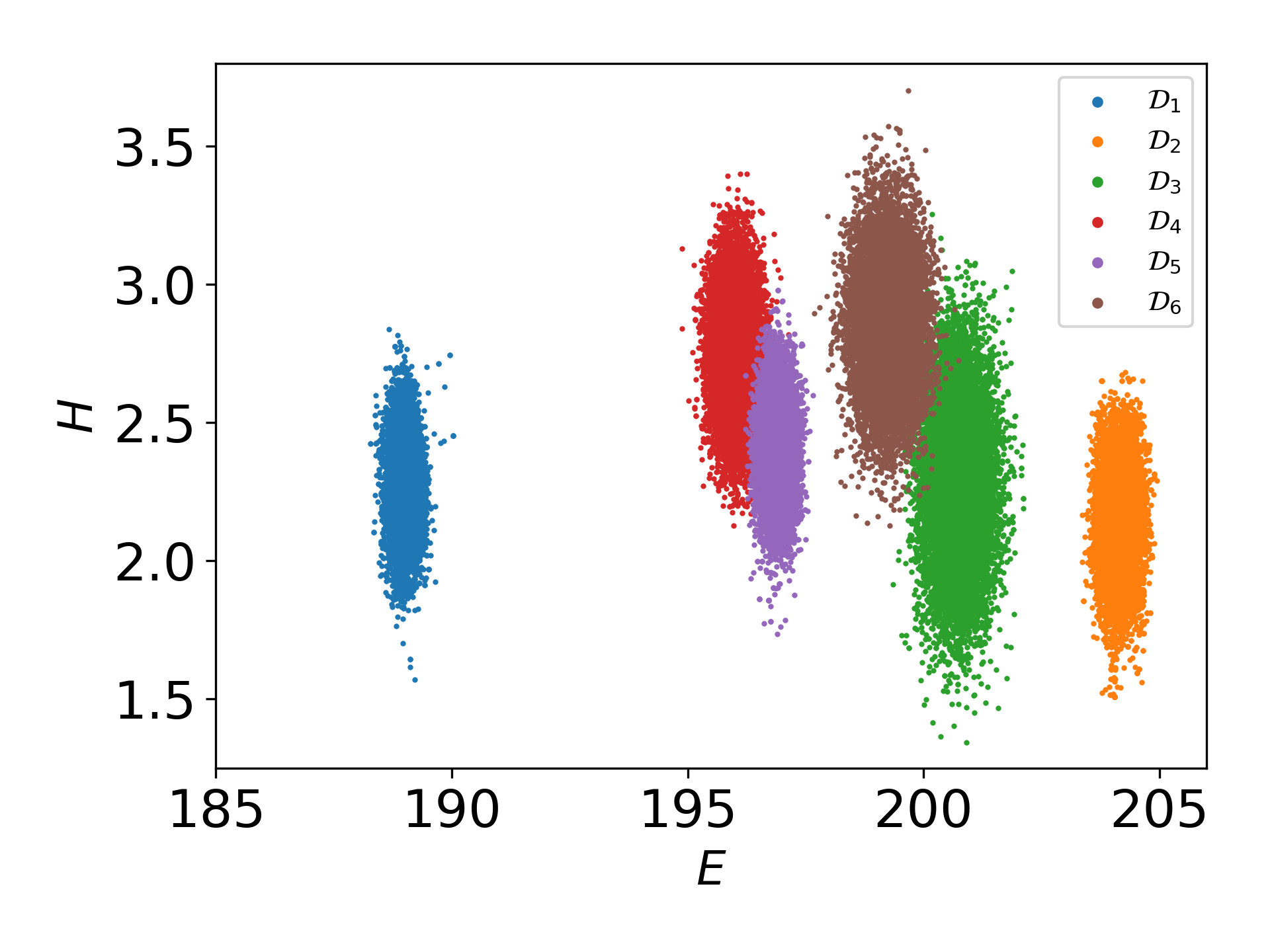}(b)}\\
{\includegraphics[width=0.55\textwidth]{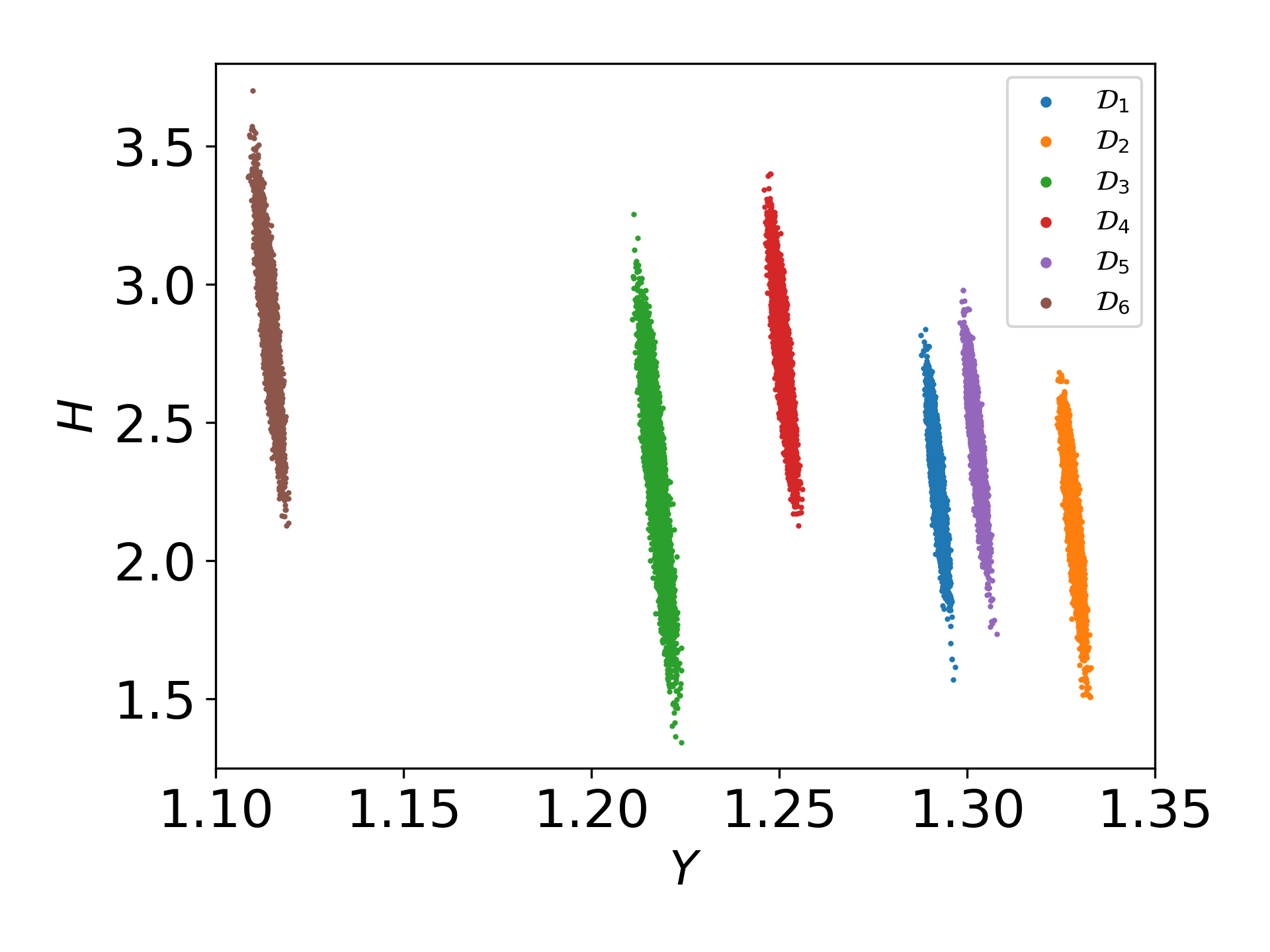}(c)}
\caption{Batch-to-batch comparison of the joint posteriors: (a) $\prob(\young, \yield)$, (b) $\prob(\young, \hard)$, and (c) $\prob(\yield,\hard)$ resulting from additive inversion with the piecewise elastic/plastic surrogate for the three parameter model using $40$ stress-strain curves for each batch.}
\label{fig:3P_calibclassFig2}
\end{figure}

\fref{fig:3P_calibclassFig2} shows the joint posteriors among the physical parameter for the separate batches obtained using $40$ randomly selected curves from each batch.
Clearly, the mean parameters of the batches are quite variable and the distributions are, for the most part, distinct and well-separated. 
However, the correlation structures are similar, suggesting that the batches behave qualitatively in the same way.

\begin{figure}[!t]
\centering
{\includegraphics[width=0.55\textwidth]{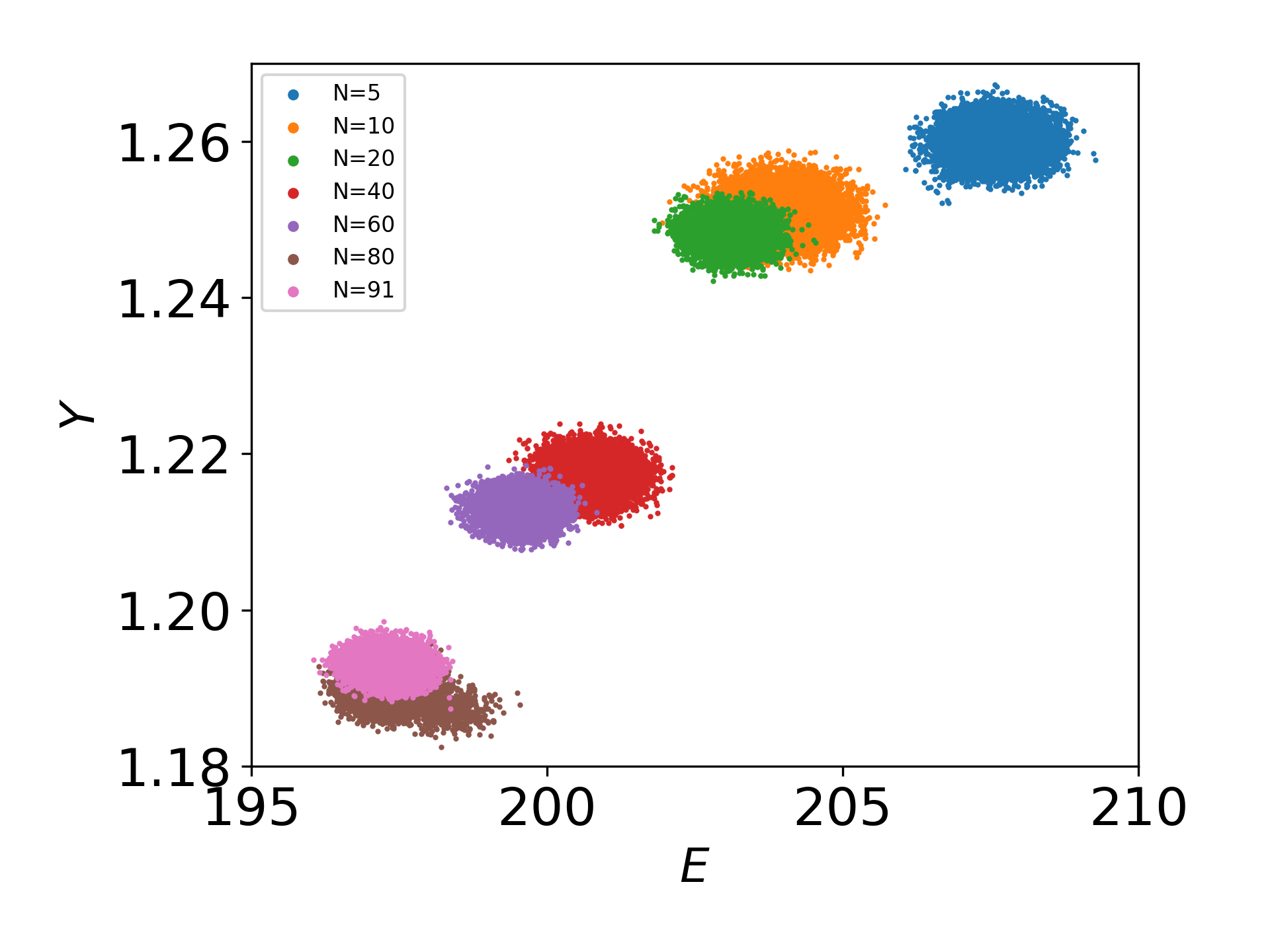}(a)}\\
{\includegraphics[width=0.55\textwidth]{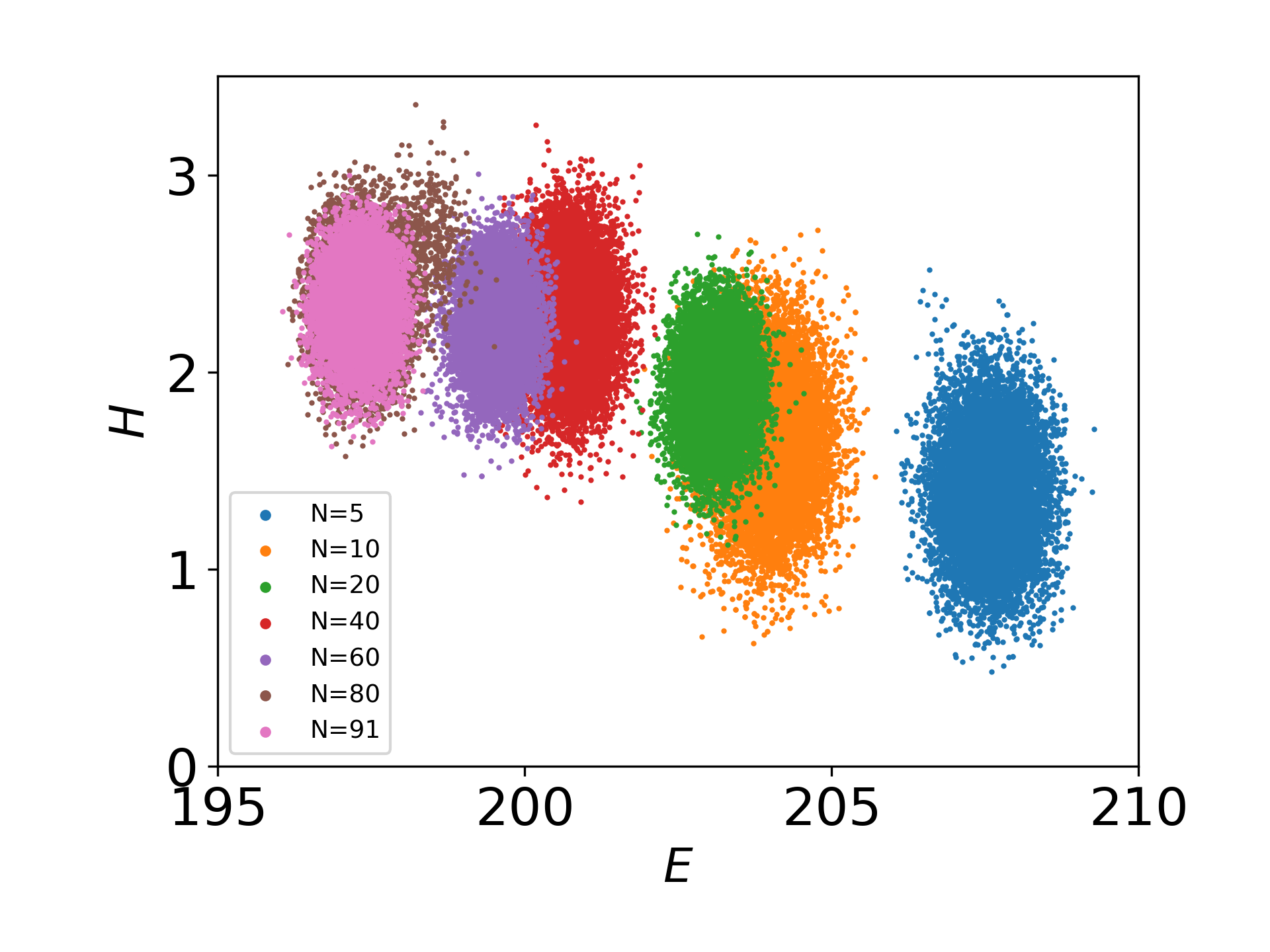}(b)}\\
{\includegraphics[width=0.55\textwidth]{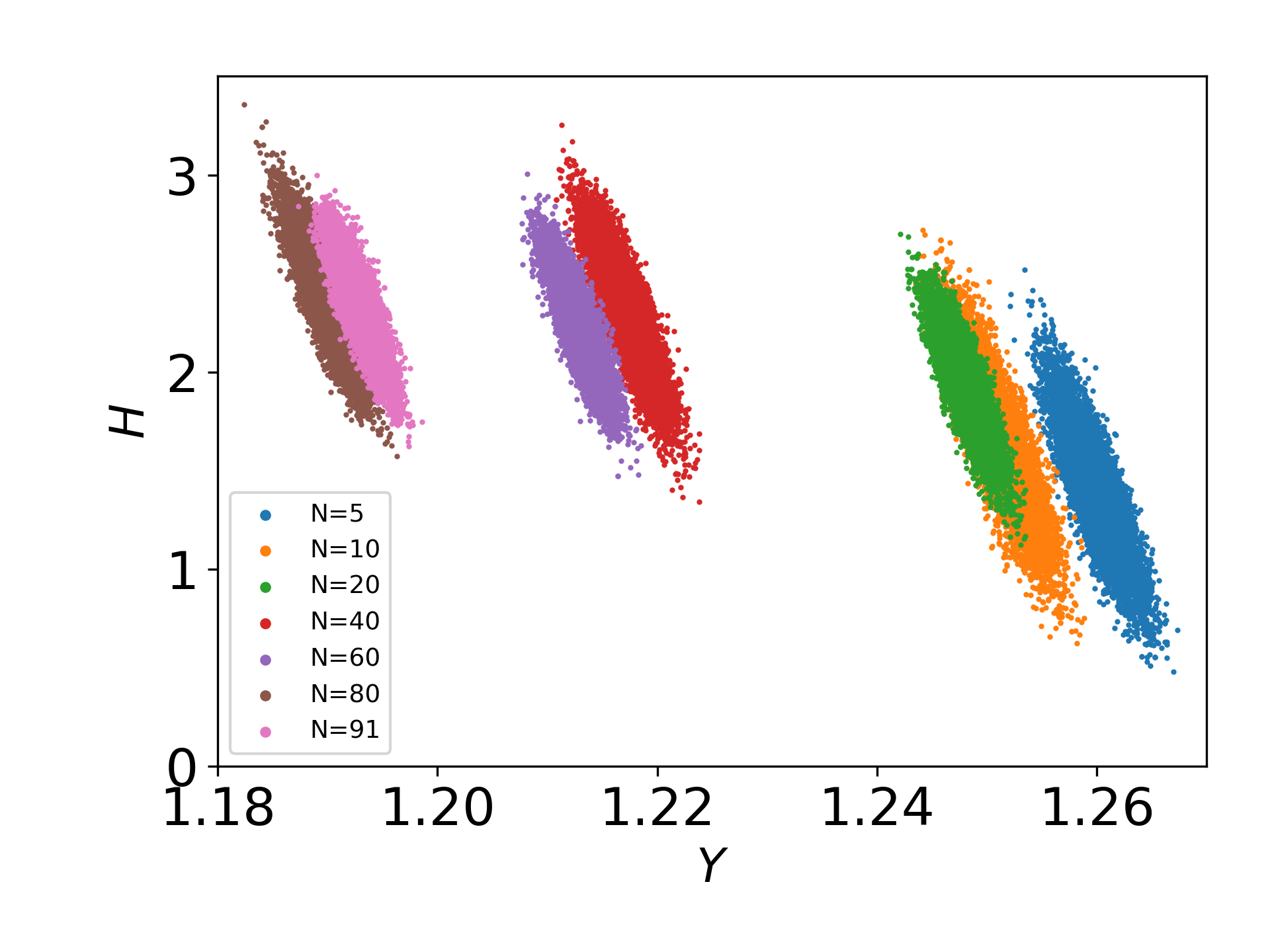}(c)}
\caption{Convergence of posterior distributions with data: (a) $\prob(\young, \yield)$, (b) $\prob(\young, \hard)$, and (c) $\prob(\yield,\hard)$.
Results are obtained using the additive inversion with the piece-wise elastic/plastic surrogate for the three parameter model $\tilde{M}^{(3)}$ and different number $N$ of curves from batch $\Dc_3$.}
\label{fig:3P_calibclassFig3}
\end{figure}

\fref{fig:3P_calibclassFig3} illustrates the convergence trend in the posterior distributions as a function of the number $N$ of curves used in the calibration.
The panels of \fref{fig:3P_calibclassFig3} show samples of the joint posteriors $\prob(\young, \yield)$, $\prob(\young, \hard)$ and $\prob(\yield,\hard)$ obtained from the third batch, as a function of the number $N$ of curves used to run the problem. 
The densities shift and narrow as more data is taken into account, and, given the range of the graphs, it appears that the densities have more-or-less converged which suggests sufficient data has been obtained to characterize the parameters. 

\begin{figure}[!t]
\centering
\includegraphics[width=\figwidth]{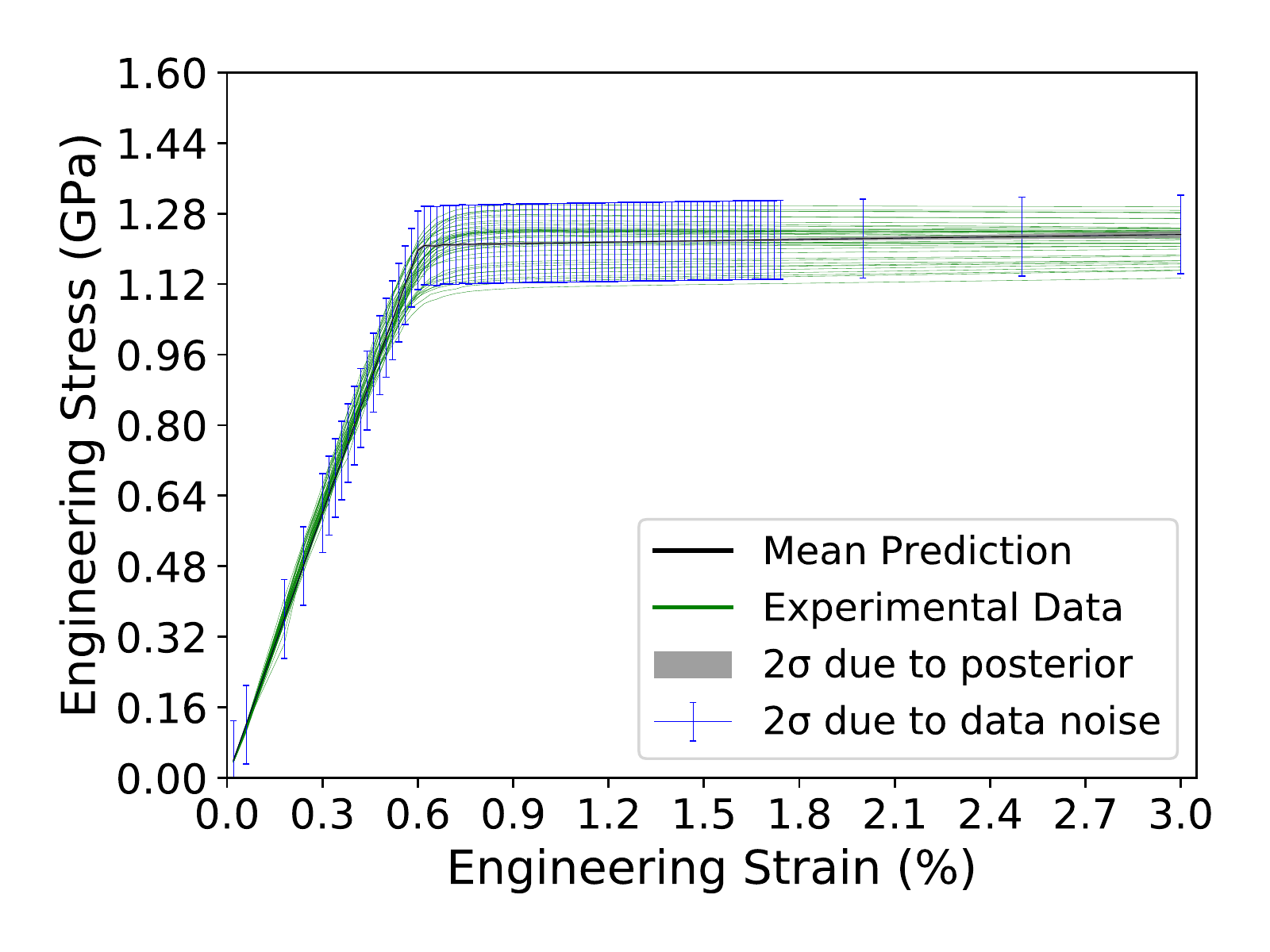}
\caption{Posterior predictive results obtained for the additive inversion using the three parameter model along with 40 curves (green) from $\Dc_3$. 
The blue bars represent the posterior uncertainty from the data noise, while the black solid line represents the mean prediction and the gray band represents the $\pm 2$ standard deviations due to posterior uncertainty.}
\label{fig:3P_calibclassFigPred}
\end{figure}

\fref{fig:3P_calibclassFigPred} shows the predictions using the posterior distribution along with an ensemble of curves from the third batch used to run the inference. 
The error bars represent the posterior predictive uncertainty stemming from the data noise, while the black solid line represents the mean prediction and the gray band represents the $\pm 2$ standard deviations due to posterior uncertainty. 
In this case, the data set leads to a narrow posterior distribution around the mean curve, which is reflected in a tight gray band. 
Overall, we see that the model prediction and the superimposed error bars cover the experimental data and, hence, provide a good representation of it.
However, the inferred value of the measurement noise is not representative of the actual noise that was estimated from the experimental data since the lack of model discrepancy term amplifies the inferred data noise. 
Hence, despite the predictive analysis showing that we are able to recover the variability of the data, our conclusion is this is an unsuitable inversion model since the full variability is represented as measurement error.  
Consequently, predictions made using this model would underestimate and inadequately account for the actual variability in the material.

\subsubsection{Inversion with embedded error model} \label{sec:embedded_inversion}

In this section, we discuss some key calibration results obtained using the embedded error model.

\begin{figure}[!t]
\centering
{\includegraphics[width=0.55\textwidth]{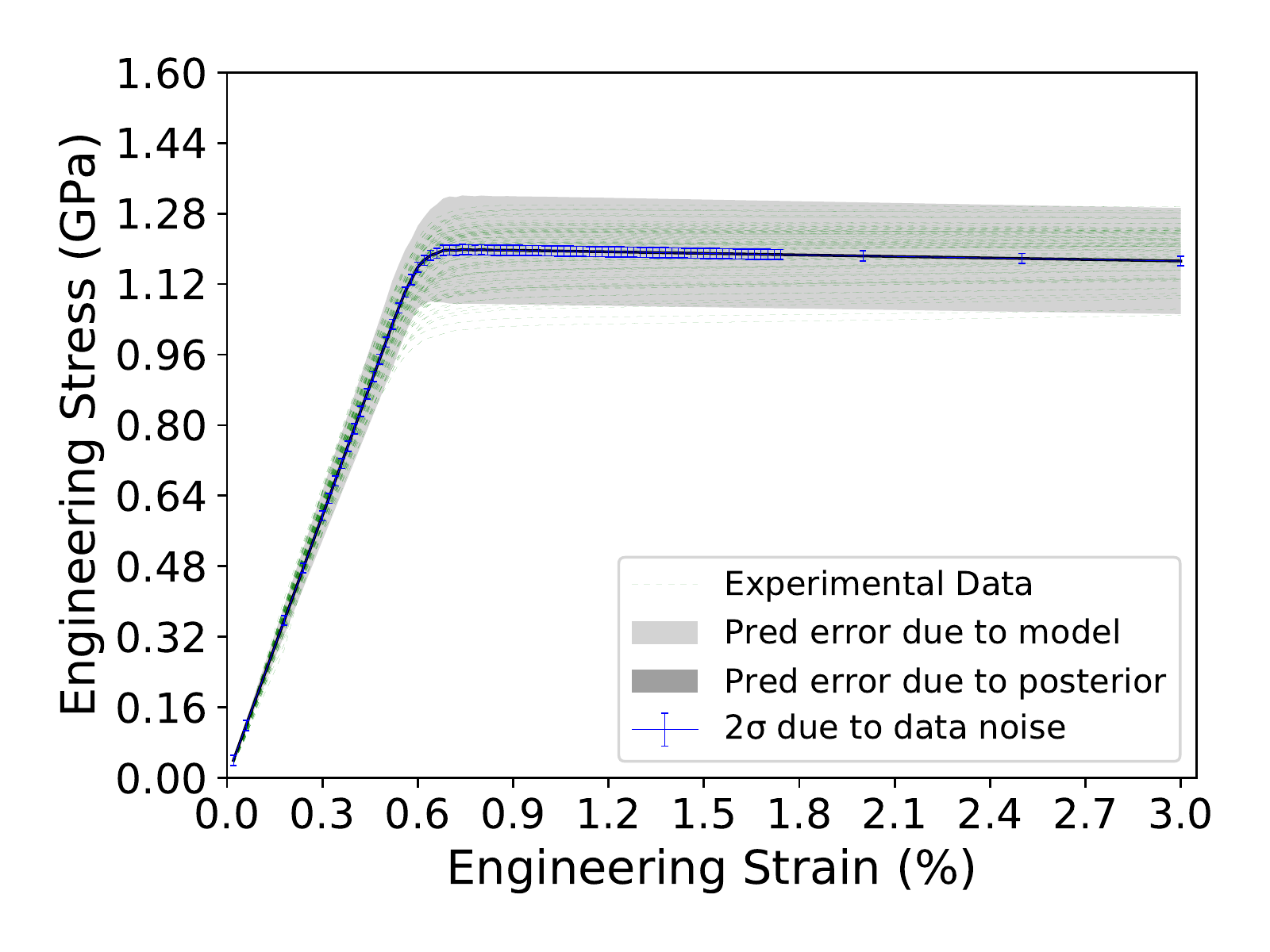}(a)}\\
{\includegraphics[width=0.55\textwidth]{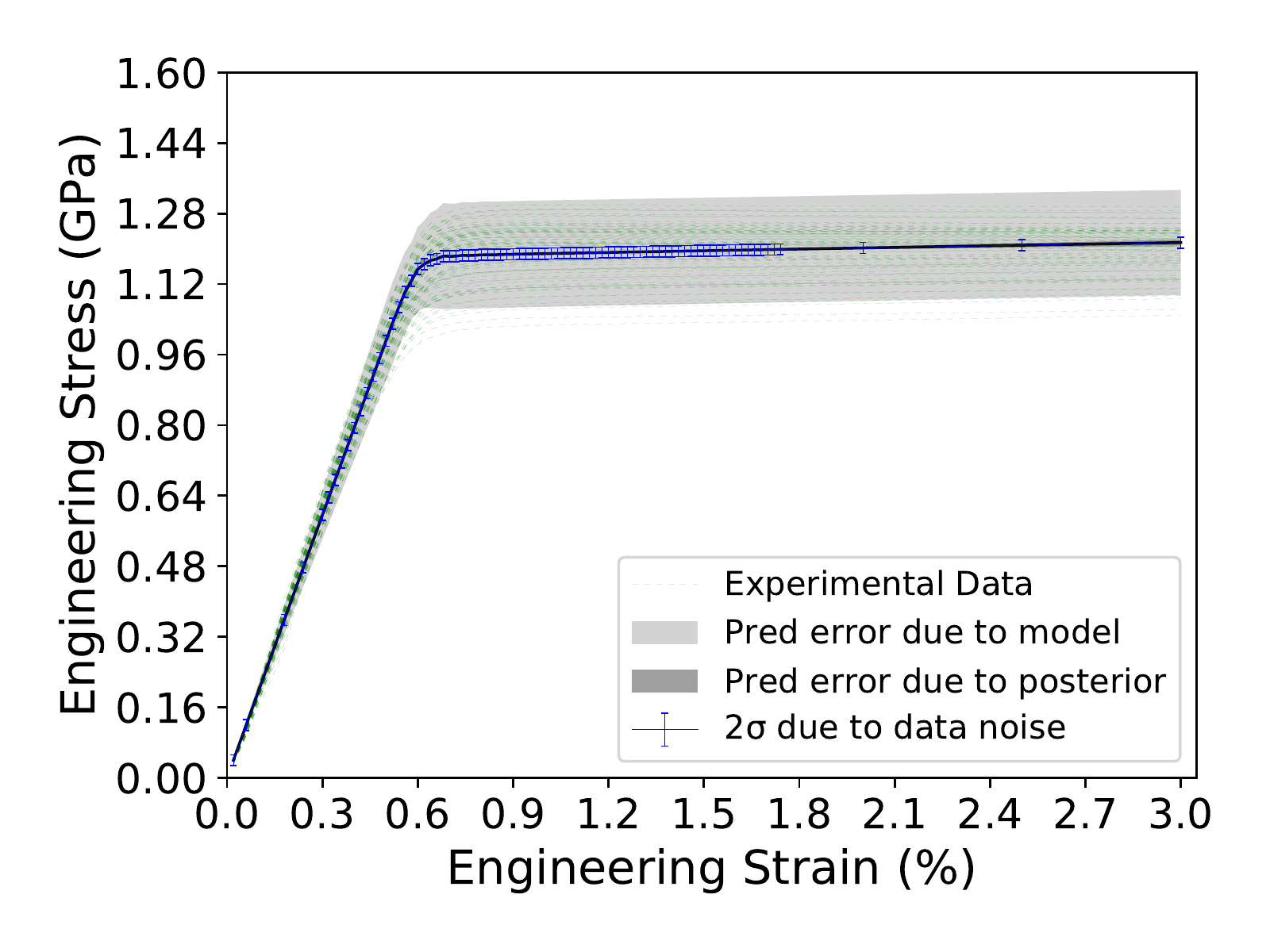}(b)}\\
{\includegraphics[width=0.55\textwidth]{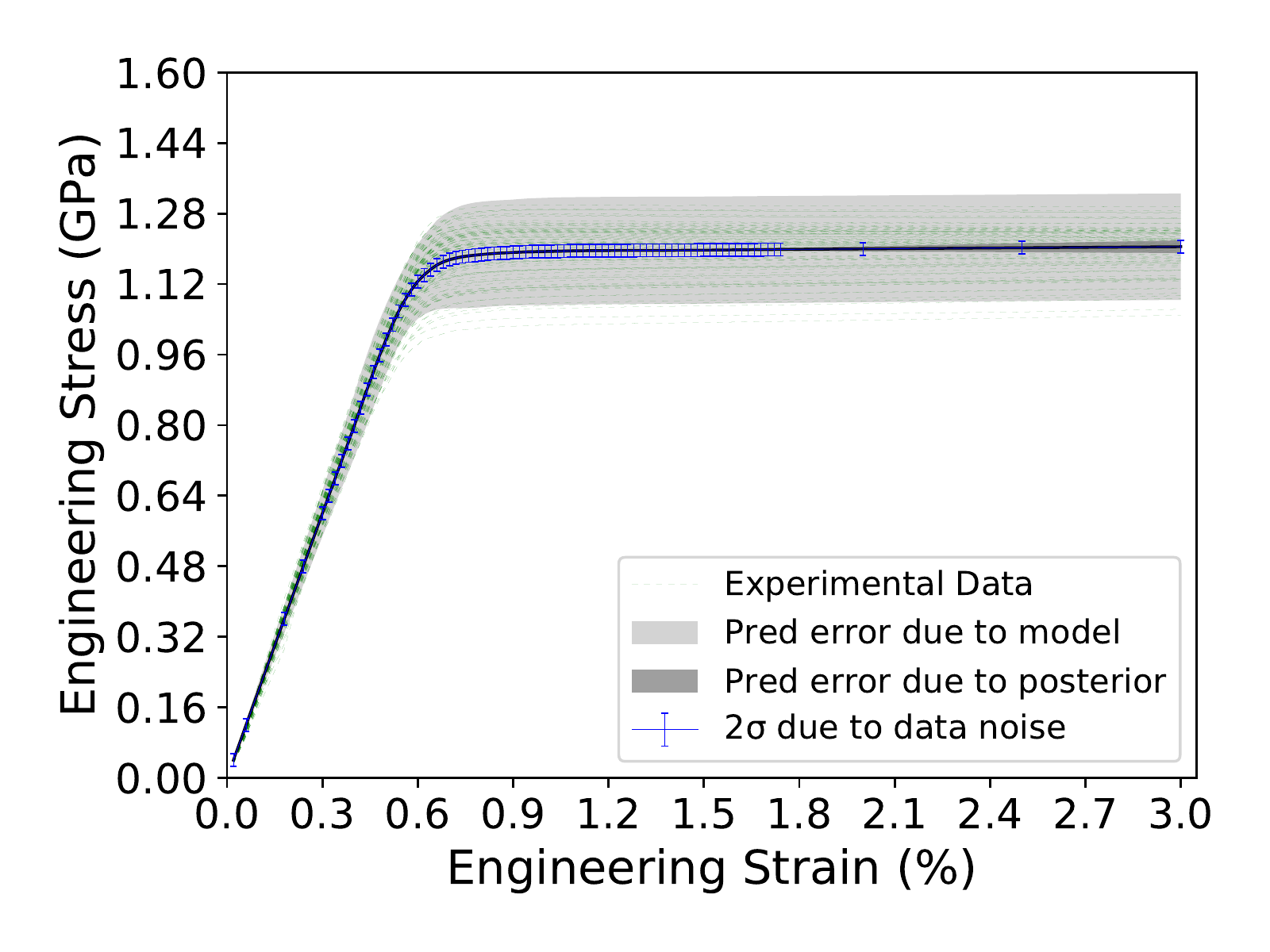}(c)}
\caption{Posterior predictive results obtained with the embedded model-based inversion using the (a) 2, (b) 3, and (c) 5-parameter models using $91$ curves (green) from $\Dc_3$. The blue bars represent the posterior uncertainty from data noise, the black solid line represents the mean prediction, the light gray band represents the $\pm 2$ standard deviations due to model inadequacy and the dark gray band is the posterior uncertainty.}
\label{fig:235P_calibEmb}
\end{figure}

\fref{fig:235P_calibEmb} shows the posterior predictive plots for the 2, 3, and 5 parameter models along with an ensemble of stress-strain curves from $\Dc_3$.
(For brevity we only show the results for one representative batch, since the others yield similar results.)
For each plot, the blue bars represent the posterior uncertainty from the data noise, the black solid line represents the mean prediction, the dark gray band represents the $\pm 2$ standard deviations due to posterior uncertainty, and the light gray band represents the $\pm 2$ standard deviations due to model error. 
We observe that in all cases the experimental data is captured and described well by the model predictions.
Here, the variability of the data is mostly described by the prediction uncertainty due to parameter variability, while the data noise is small and comparable to the estimate of the measurement error obtained from the data itself. 
This is the key difference with respect to the additive results shown in \fref{fig:3P_calibclassFigPred}.
By accounting for model discrepancy through the embedded error terms, we are able to characterize the variability more properly because it is not artificially lumped in the data noise.
The contribution stemming from the posterior uncertainty is again quite small, suggesting that we have accounted for sufficient amount of data. 
The results for three models show that overall they have similar predictive capabilities.

\subsubsection{Comparison of additive and embedded error models} \label{sec:comparison}

\begin{figure}[!ht]
\centering

{\includegraphics[width=0.8\figwidthtwo]{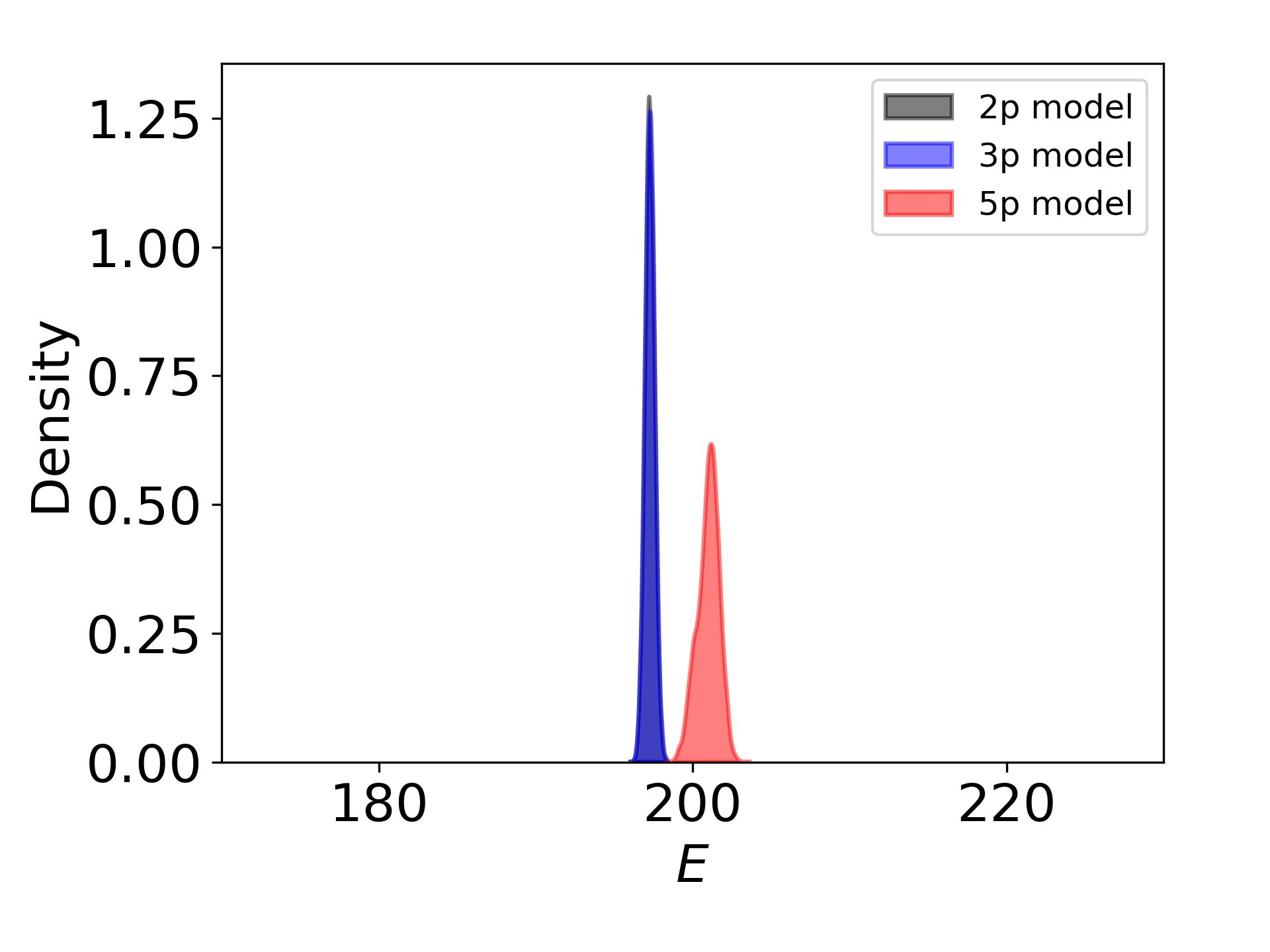}}%
{\includegraphics[width=0.8\figwidthtwo]{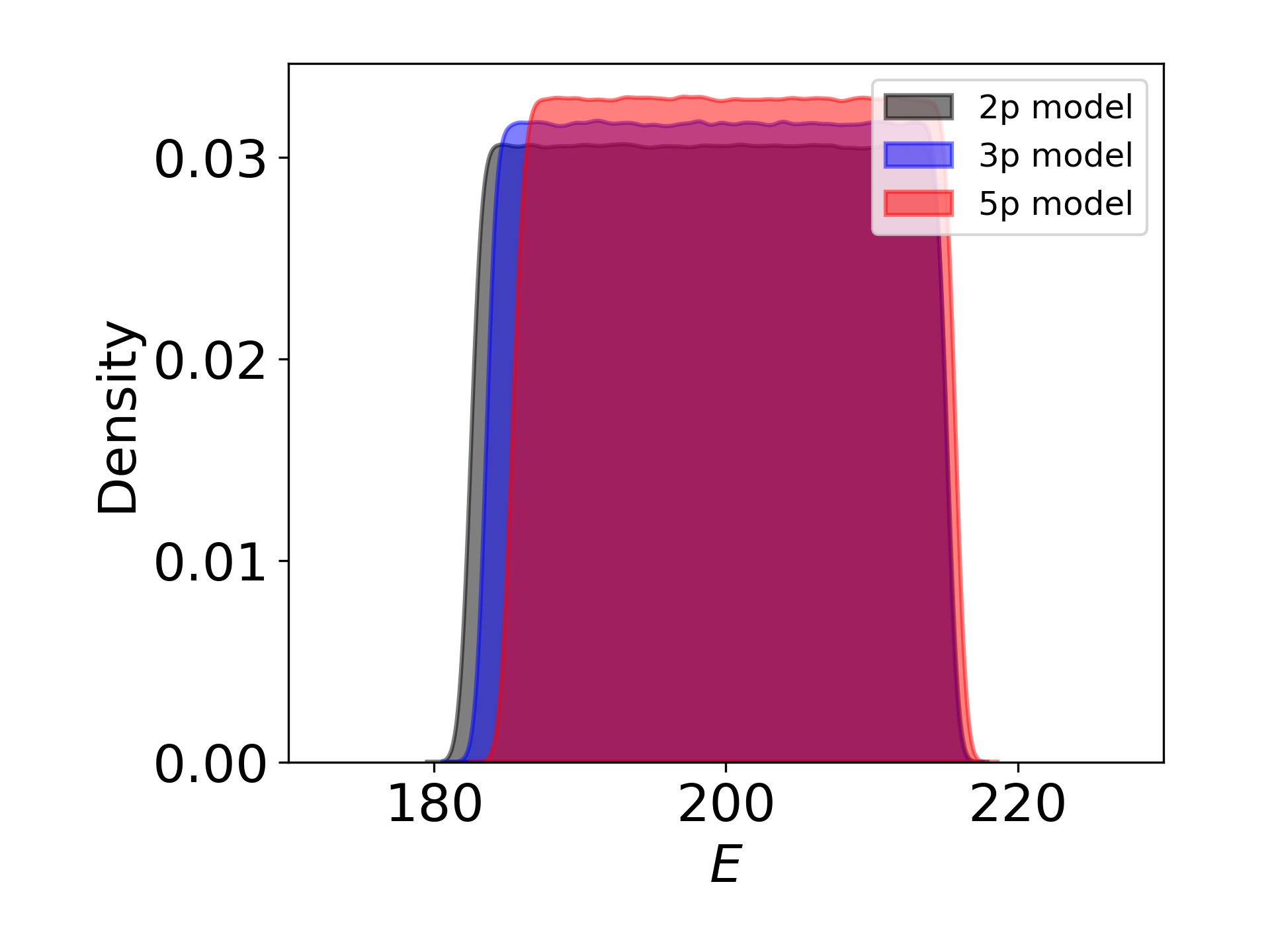}}%

{\includegraphics[width=0.8\figwidthtwo]{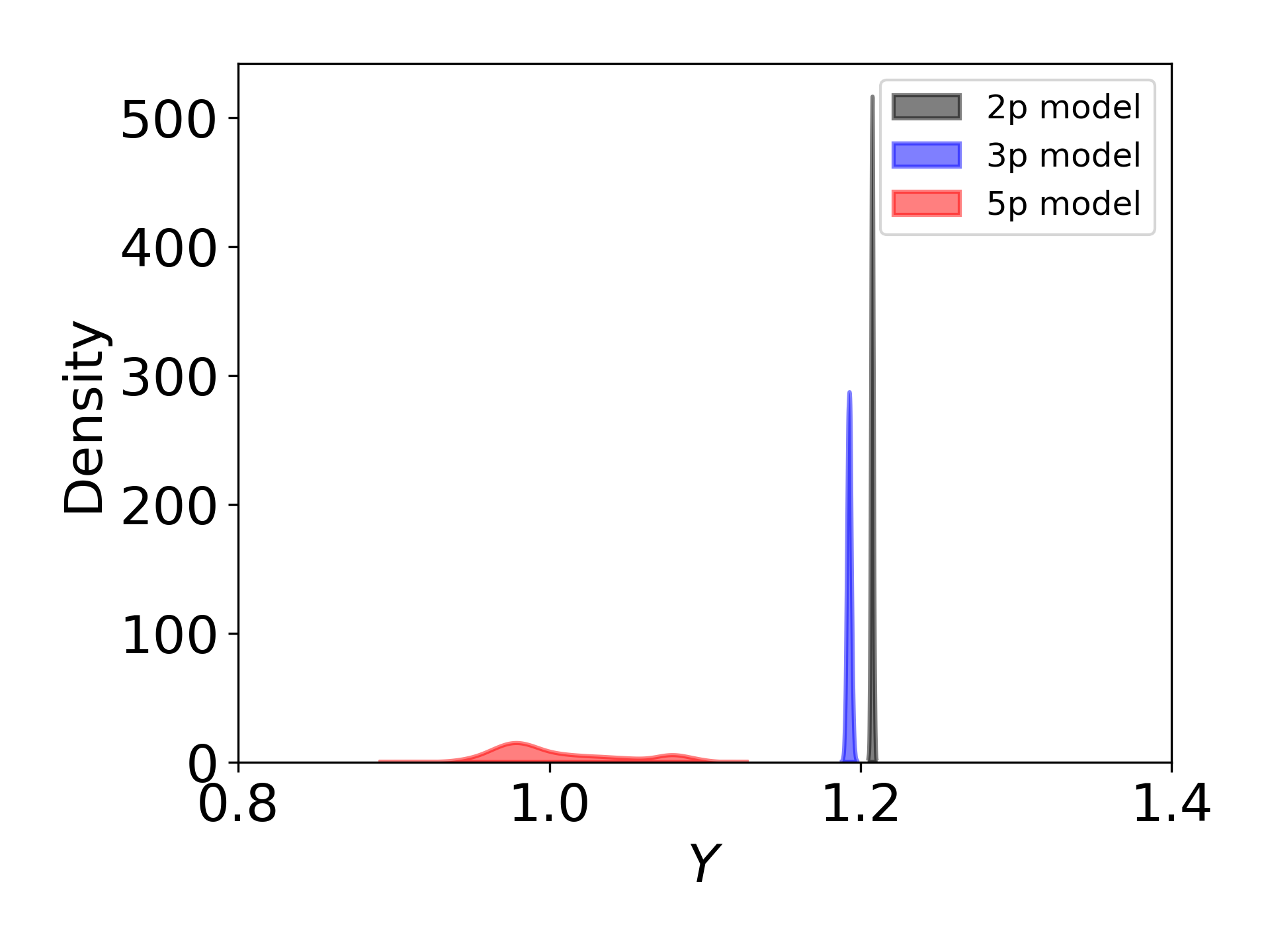}}%
{\includegraphics[width=0.8\figwidthtwo]{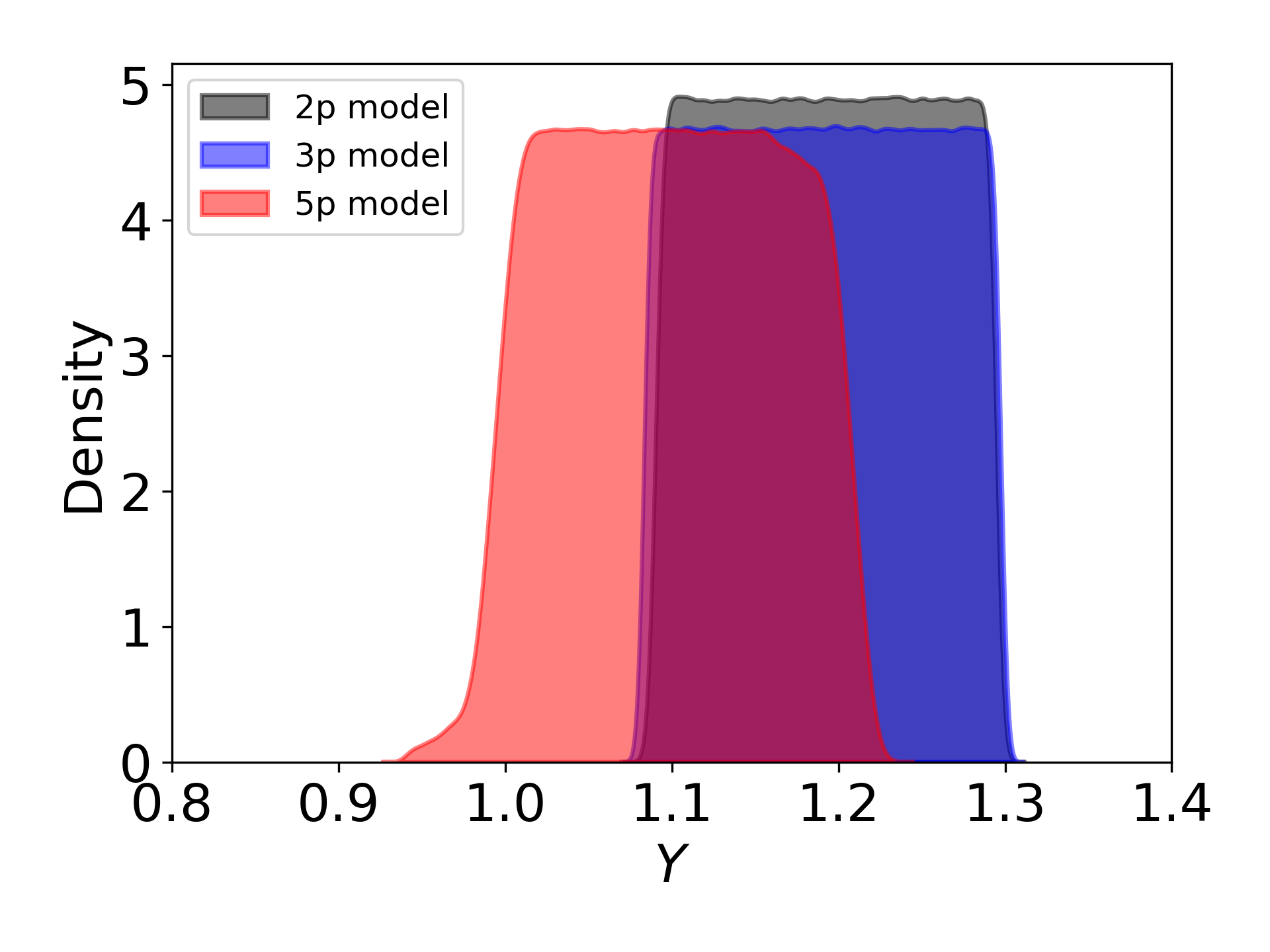}}%

{\includegraphics[width=0.8\figwidthtwo]{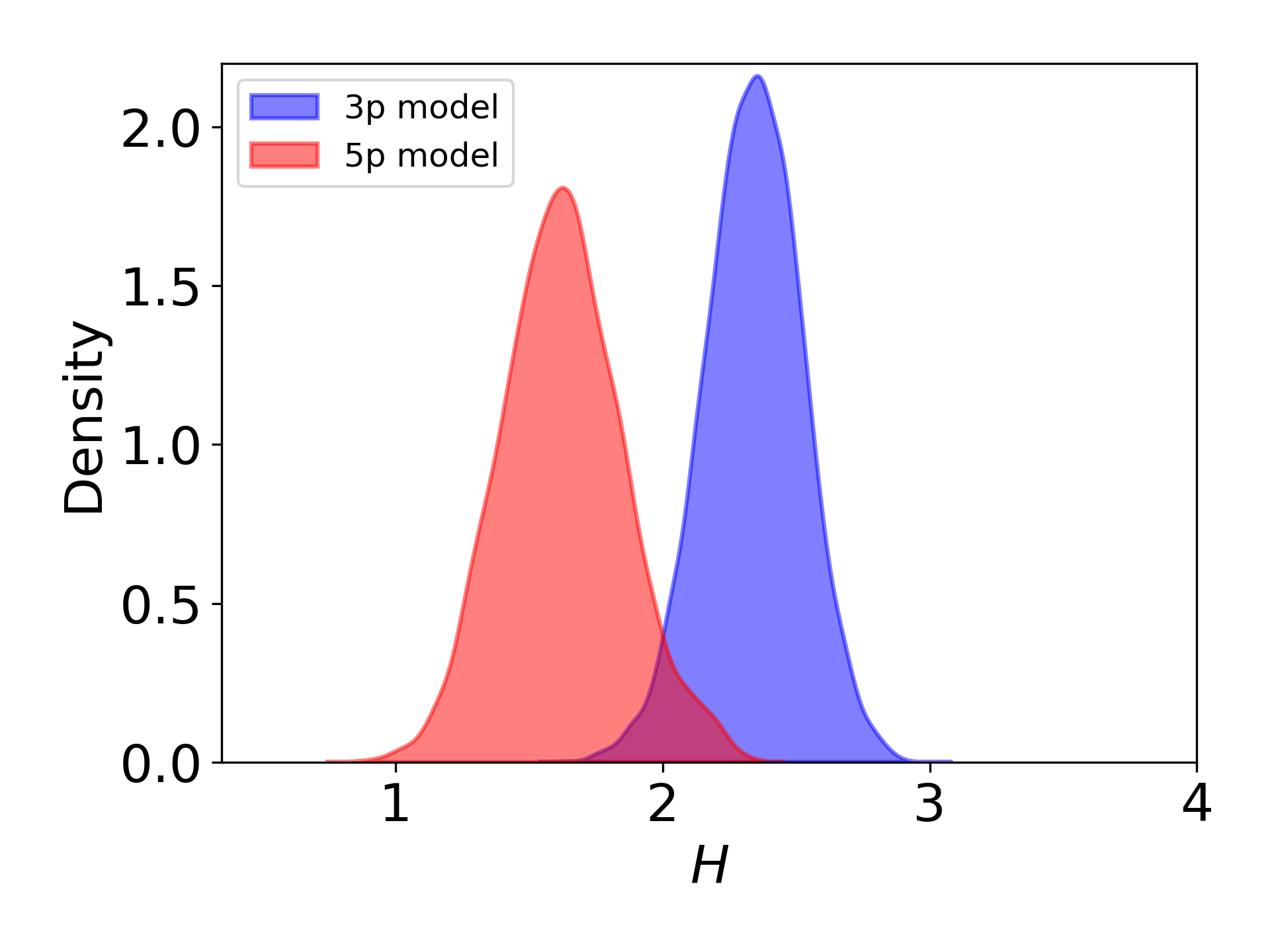}}%
{\includegraphics[width=0.8\figwidthtwo]{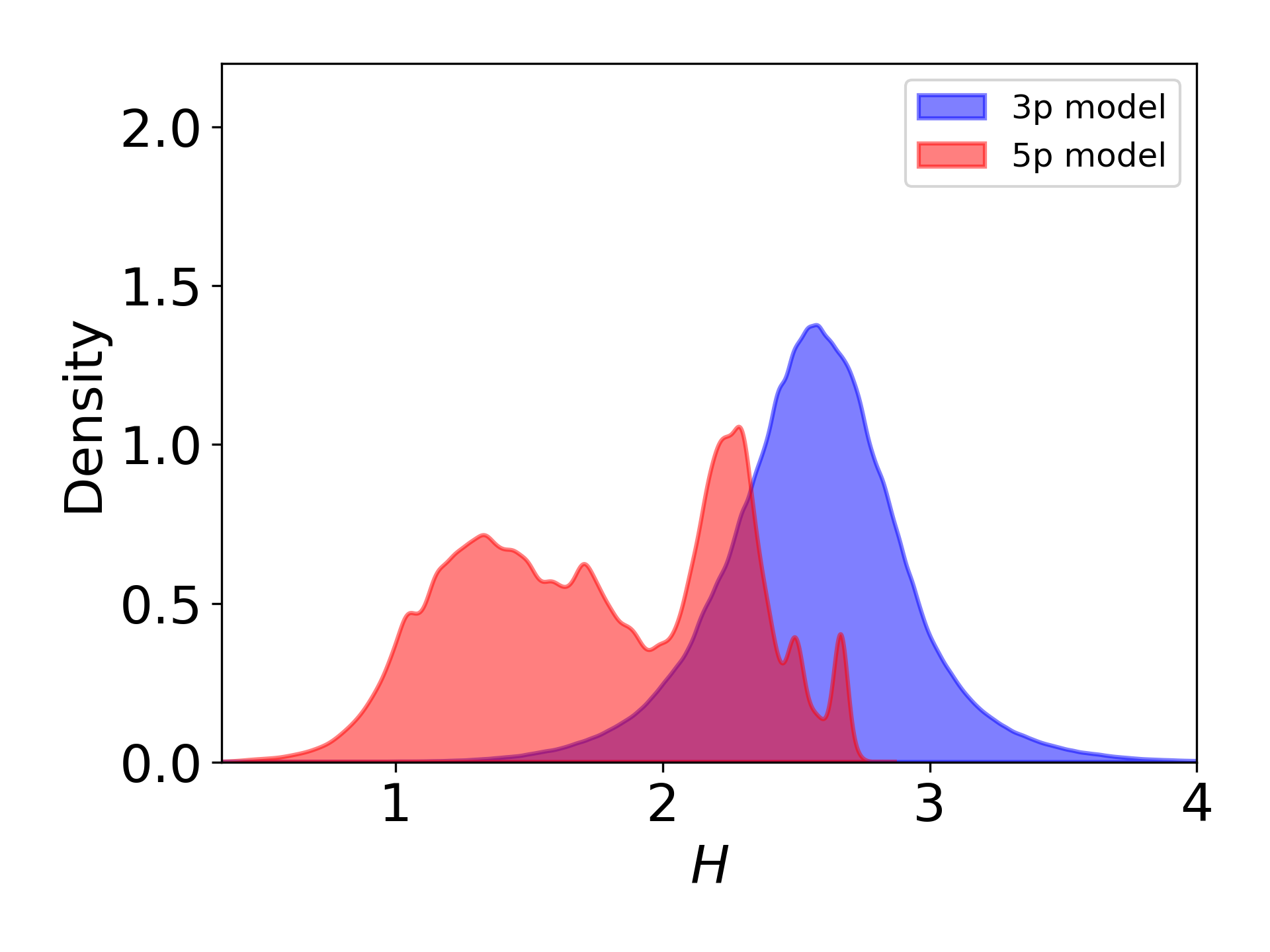}}%

{\includegraphics[width=0.8\figwidthtwo]{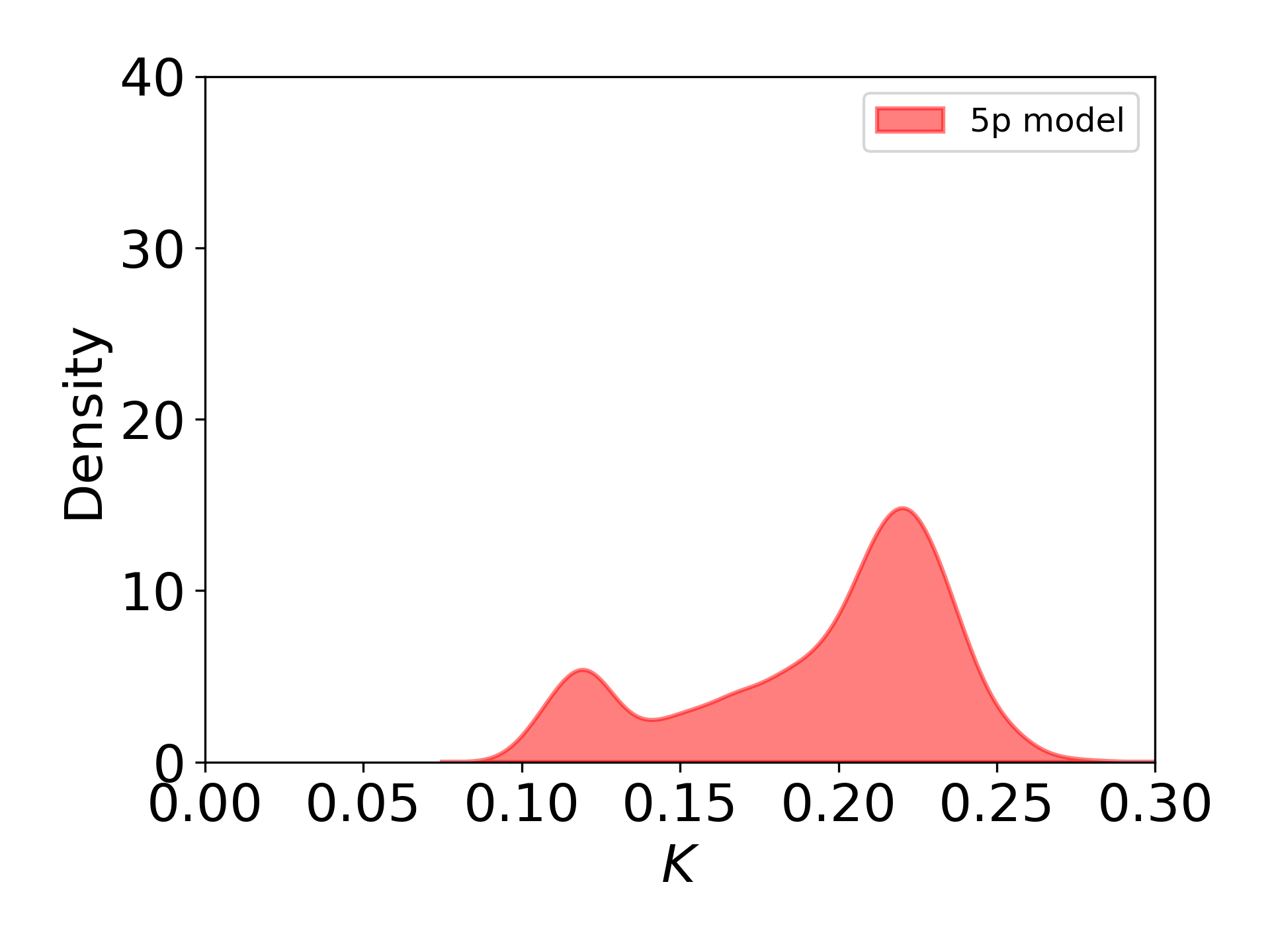}}%
{\includegraphics[width=0.8\figwidthtwo]{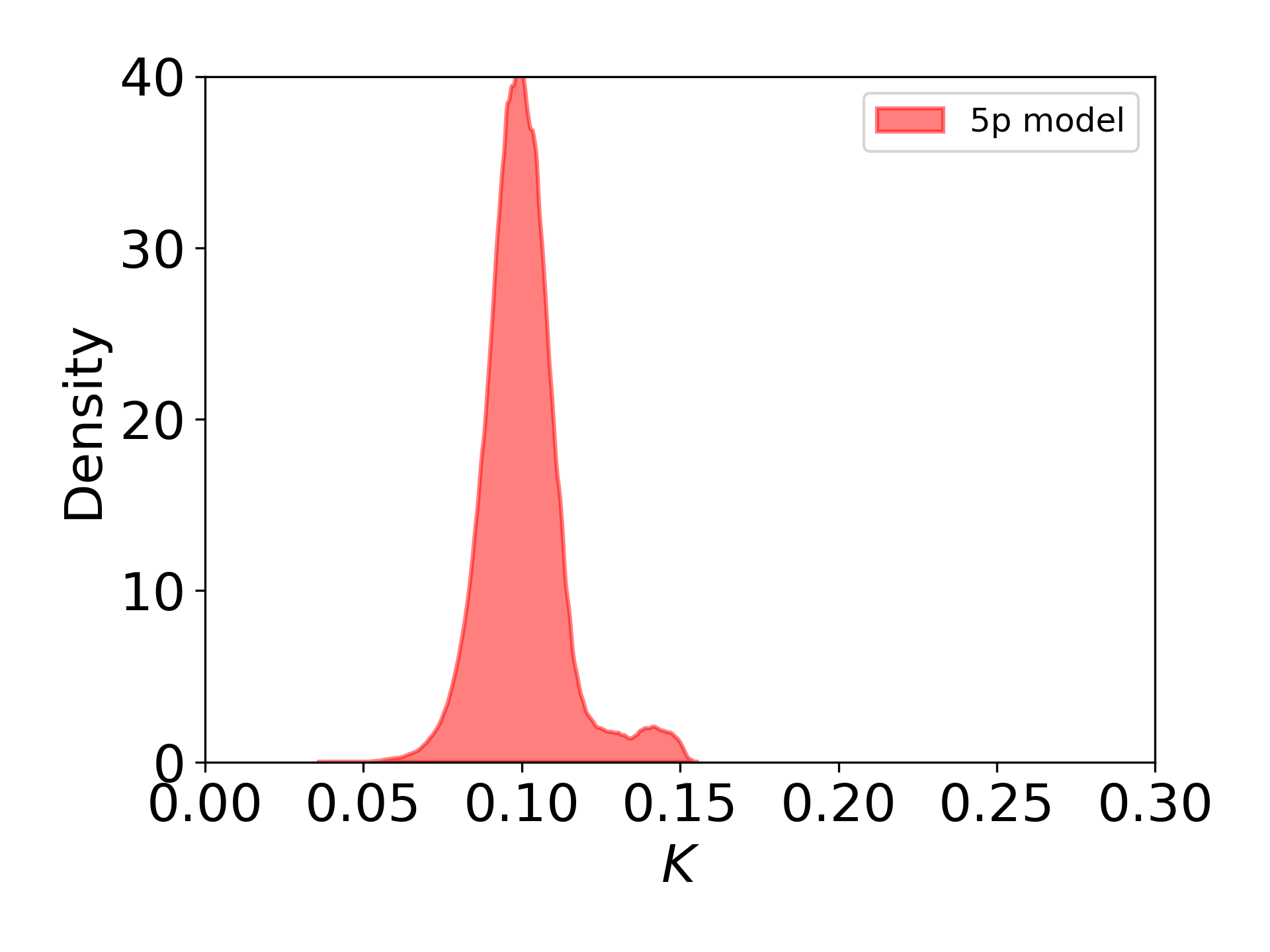}}%

{\includegraphics[width=0.8\figwidthtwo]{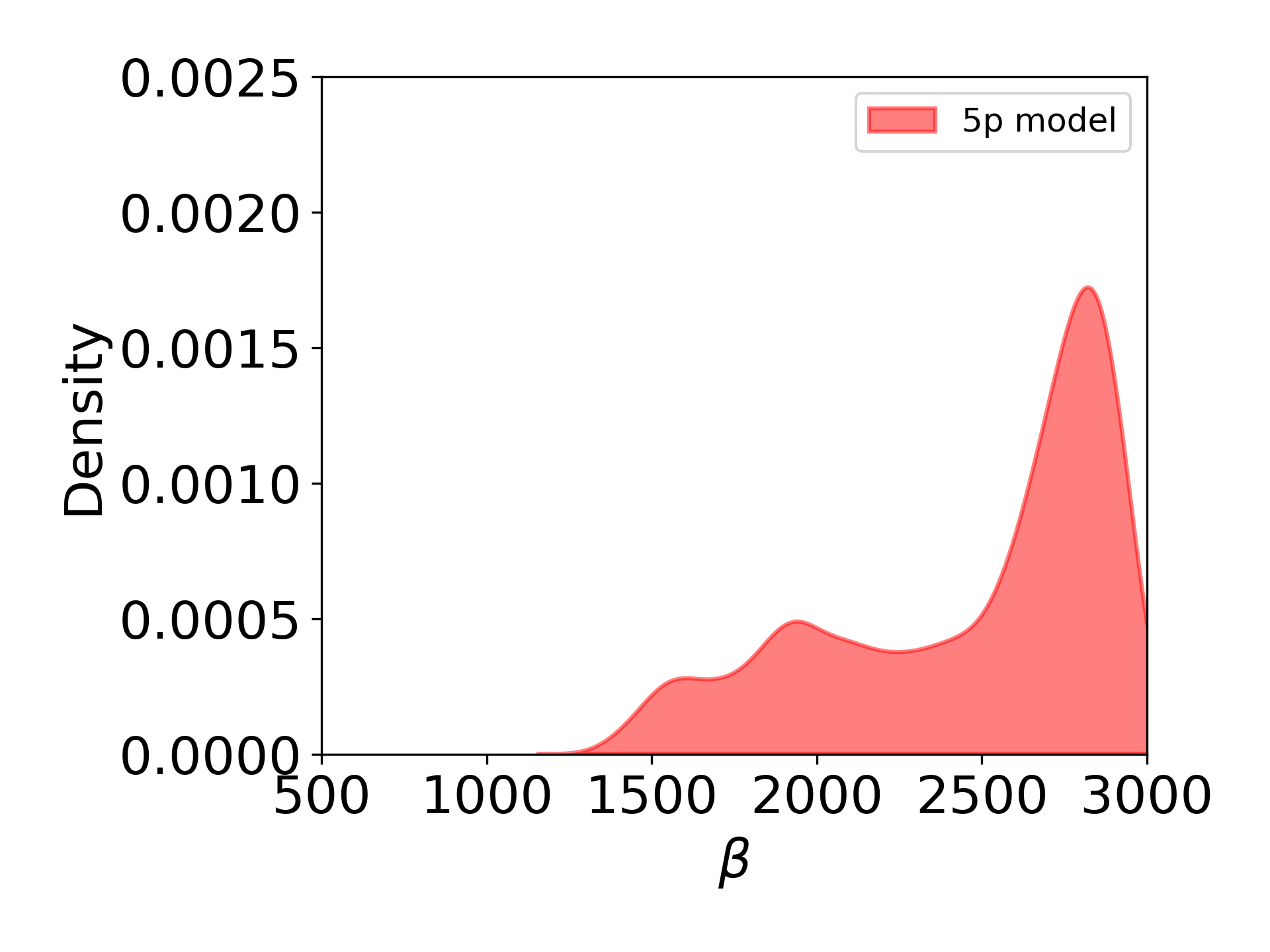}}%
{\includegraphics[width=0.8\figwidthtwo]{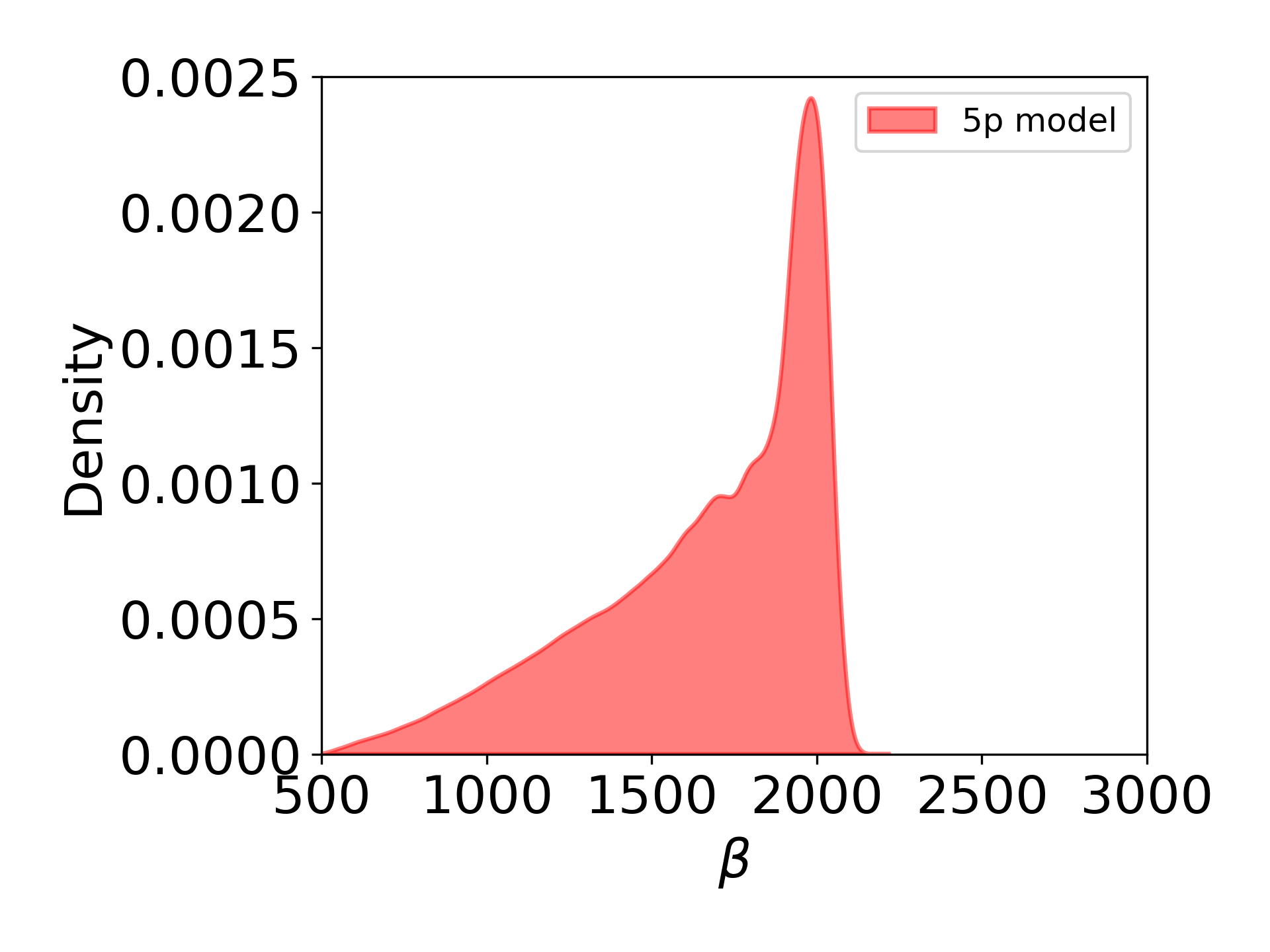}}%
\caption{Comparison of posterior PDFs of the material parameters obtained using the additive (left) and embedded (right) model inference using the data from $\Dc_3$.}
\label{fig:cmpclassEmbe}
\end{figure}

To illustrate the differences between the additive and embedded error-based inference results, \fref{fig:cmpclassEmbe} shows the posterior PDFs for each parameter using the data from the third batch obtained from the additive inference (left column) and embedded error approach (right column).
We note in both cases, the contribution stemming from the posterior uncertainty of the parameter estimates is again quite small, as shown in \figref{fig:3P_calibclassFigPred} and \figref{fig:235P_calibEmb}, suggesting that we have accounted for a sufficiently large amount of data.
The key difference in the parameter posterior distributions is the embedded term which explicitly enables the calibrated models to reflect the inherent variability in the material parameters.
In contrast, calibration with the additive error term tends to attribute variability to that component of the model which cannot be interpreted as a specific level of variability in any particular parameter.

Posterior PDFs of the two parameters common to all the models, the Young's modulus $\young$ and yield strength $\yield$, highlight this difference in uncertainty quantification strategies.
\figref{fig:cmpclassEmbe} shows that the embedded approach attributes a significant amount of variability in $\young$ based on the data while the additive approach has a small uncertainty.
In effect, the distribution in the additive approach converges to the best average value for $\young$, since $\young$ is treated as a constant rather than a distribution.
All variability arising from data variation, as opposed to measurement error and finite sample size, is attributed to the additive term where it is confused with all other uncertainties.
With the embedded approach, the distributions of the elastic modulus $\young$ for the three models are broad, flat, and fairly consistent.
Likewise, the difference in the two approaches is readily apparent in the distributions of $\yield$ shown in \figref{fig:cmpclassEmbe}.  
The distributions of the yield strength $\yield$ for the two simpler plasticity models are essentially the same, as they treat yield as a well-defined point unlike the more complex 5-parameter, saturation hardening model.
When interpreted in light of the model sensitivities shown in \figref{fig:tot_sens} and discussed in \sref{sec:sens_results}, $\yield$ is well informed by the data, resulting in the narrow distribution and small uncertainties in the additive approach and consistent broad distributions using the embedded approach.  
The low sensitivity to $\hard$ in the 3 and 5-parameter models gives rise to broad Gaussian distributions with the additive formulation indicating more informative data is needed. 
This deficiency also contributes to the qualitative differences in the densities for $\hard$ and $\yield$ with the additive formulation.
In contrast to $\hard$, the 5 parameter model is sensitive to both the $\yield$ and $\satmod$, as shown in \fref{fig:tot_sens}, and yet the data is not sufficient to fully inform each independently.
Instead, the calibration results in a broad joint PDF of the two parameters since they effect similar changes in the model's behavior near the elastic-plastic transition.
This confounding effect also likely gives rise to the bimodal posterior distributions of these parameters.
Lastly, given that the sensitivity to $\satexp$ is essentially negligible, it is not surprising that the prior, restricted by the range of the surrogate, exerts significant influence on the posterior distribution of this parameter in both formulations.

\begin{figure}[!t]
\centering

\subcaptionbox{}%
{\includegraphics[width=\figwidthtwo]{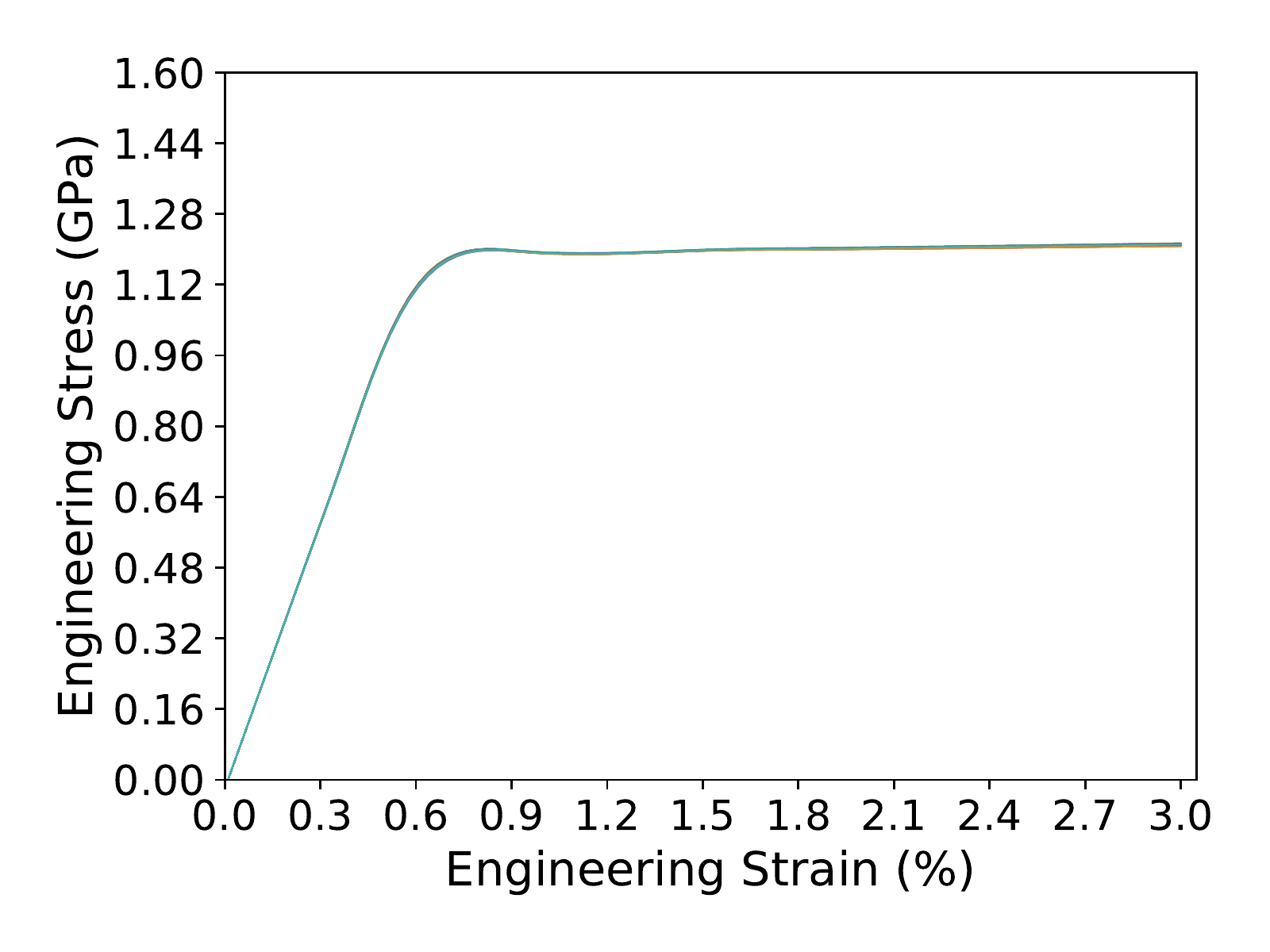}}
\hfill%
\subcaptionbox{}%
{\includegraphics[width=\figwidthtwo]{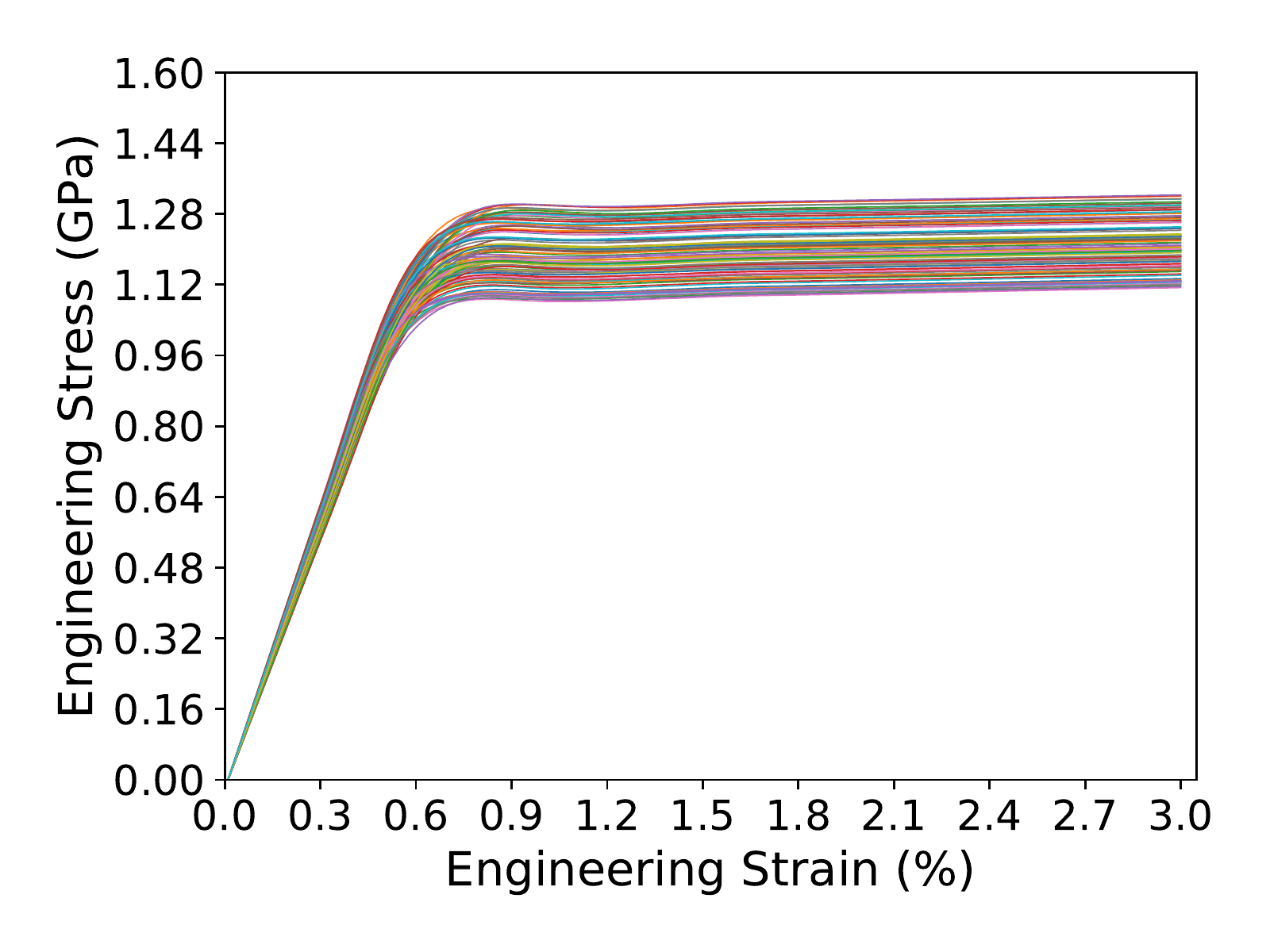}}\\

\subcaptionbox{}%
{\includegraphics[width=\figwidthtwo]{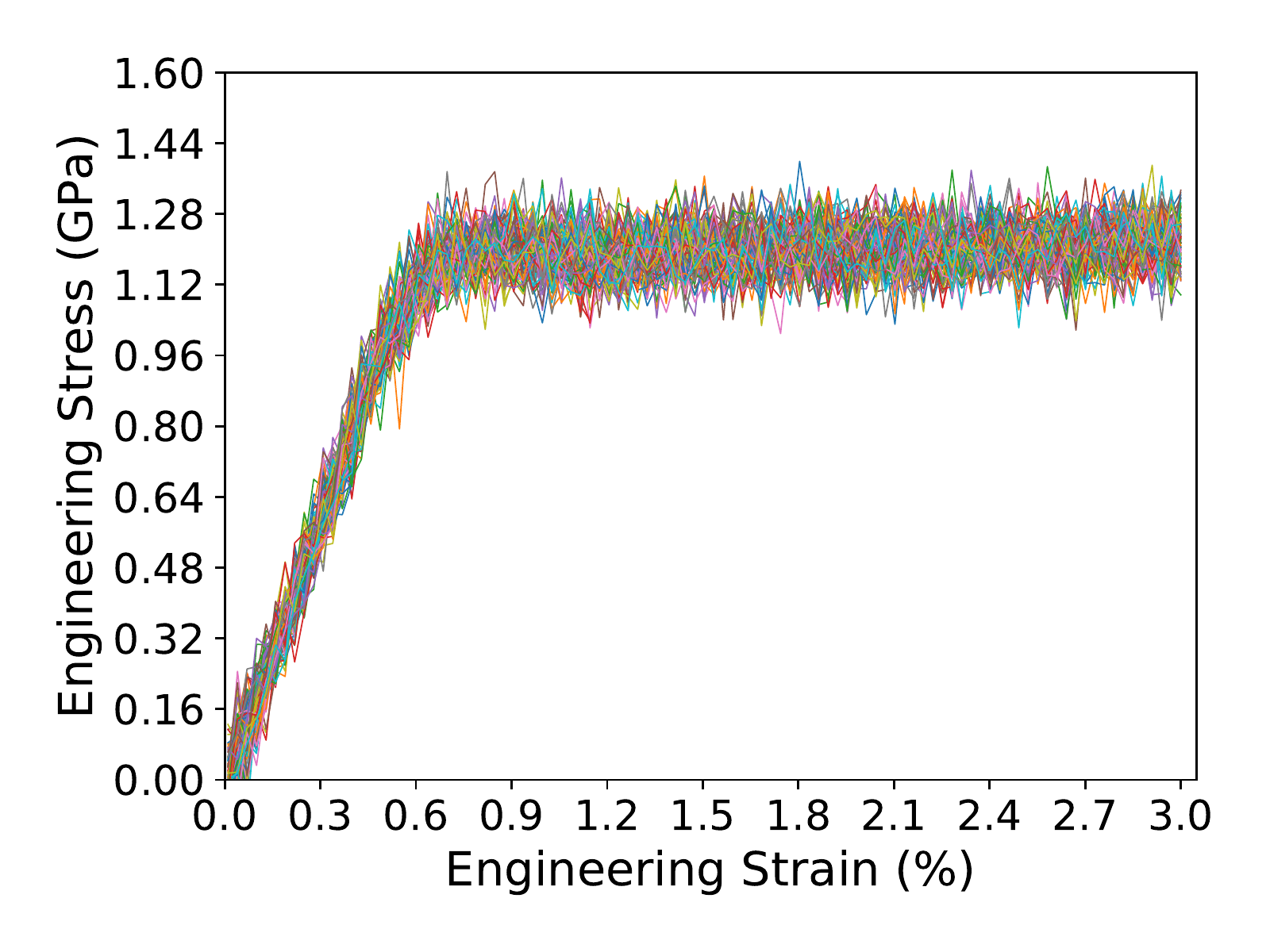}}
\hfill%
\subcaptionbox{}%
{\includegraphics[width=\figwidthtwo]{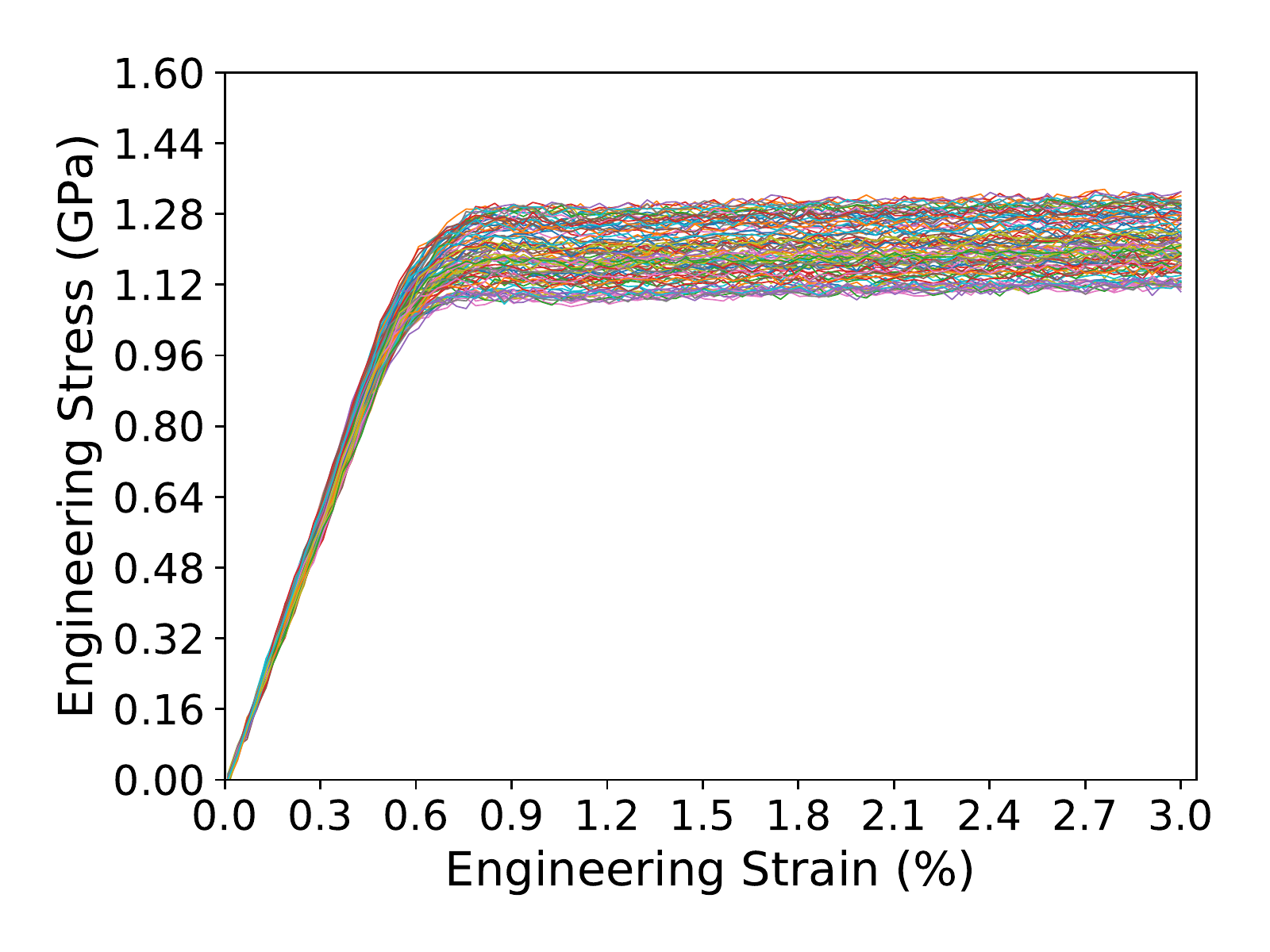}}\\
\caption{Sample posterior predictive realizations obtained using the 5 parameter model for the additive (left) and  embedded (right) approach calibrated using $\Dc_3$.
The top row shows the results obtained by sampling the joint posterior density and pushing these samples through the model only, while the bottom row shows the results with the contribution of measurement error.}
\label{fig:realizPlots}
\end{figure}

\fref{fig:realizPlots} shows 100 posterior predictive realizations obtained using the 5 parameter model for the additive approach \fref{fig:realizPlots}(a,c) and embedded approach \fref{fig:realizPlots}(b,d) calibrated using $\Dc_3$.
The top row shows the results obtained by sampling the joint posterior density of the parameters $\thetab$, and pushing these samples through the model $\tilde{M}(\strain_j;\thetab)$ only; while the bottom row shows the results with the contribution of the measurement noise $\eta_j^{(3,k)}$. 
It is evident from comparing \fref{fig:realizPlots} to the data shown in \fref{fig:experiment}b that the embedded error approach yields a suitable representation where the variability of the response is determined by the variability of the material parameters. 
On the contrary, the additive approach yields a tight envelope of predictions, and the full variability is only captured by the added (and over-estimated) measurement noise. 
Comparing individual realizations to the curves obtained experimentally, it is  clear that the embedded approach with noise yields curves that match quite well the trends observed in the experiment, the classical method deviates considerably and is highly dominated by the high frequency, uncorrelated noise.

\section{Discussion} \label{sec:discussion}

Following the discussion begun in \sref{sec:embedded_discrepancy}, we will use a simplification of the model $M(\strain,\thetab)$ to summarize the key concepts in this work and help generalize the intuition needed to model variability.
Let us consider only the elastic response so that the nominal model is:
\begin{equation}
\stress(\strain) = M(\strain;\young) = \young \strain \ ,
\end{equation}
and limit our attention to data for a single batch in the elastic regime. 
The embedded model of the data is
\begin{equation}
\stress_j = M(\strain_j;\young+\alpha\, \hat{\eta} ) + \eta_j = (\young+\alpha \hat{\eta} ) \strain_j + \eta_j \ ,
\end{equation}
where $\hat{\eta}$ and $\eta_j$ are mean zero random variables.
The additive model omits $\alpha\, \hat{\eta}$ which varies the slope of the stress-strain curve.
In both cases, 
\begin{equation}
\mathbb{E}\left[\stress_j \right] = (\young+\alpha\,\hat{\eta} ) \strain_j + \eta_j 
= \young \, \mathbb{E}\left[\strain_j \right]
\ ,
\end{equation}
so, in the limit of infinite informative data, both formulations recover the correct mean $\young = \frac{\mathbb{E}\left[\stress_j \right]}{\mathbb{E}\left[\strain_j \right]}$.
This is illustrated in the comparison of \fref{fig:realizPlots}a with the average of \fref{fig:realizPlots}b in the elastic regime.
The difference between the two formulations becomes clear when examining the variance at a given $\strain_i$ and covariance across all samples $\strain_i$ of the calibrated representations.
Recall that the stress-strain data for a single specimen has virtually no measurement noise and yet the stress-strain curves are essentially lines with slopes that vary across a batch.
Without the $\alpha\, \hat{\eta}$ term, the simplest additive model, where the sequence of random variables $\{\eta_j\}$ are assumed independent and identically distributed, obtains a variance around the mean that is determined by the total variance of the dataset.
Furthermore, the realizations are not smooth.
They have  a wide and constant variation around their mean trends and, consequently, have  distinct offsets in stress at zero strain, as can been seen in \fref{fig:realizPlots}c.
In contrast to this model employing only uncorrelated noise $\eta_j$, the embedded model accounts for most of the variation in the data with a distribution of slopes effected by the $\alpha\,\hat{\eta}$ term.
This is consistent with the variations in the dataset which is composed of highly correlated data for each test, \ie each test gives essentially the same linear relationship between $\stress_j$ and $\strain_j$ at every $j$.
The fan-like ensemble of realizations shown in \fref{fig:realizPlots}d clearly represents the continuity and the particular type of variation seen in the data.

This basic illustration was motivated by our data where each individual experiment is well-described by the hypothesized model.
When that is the case, the example shows that the embedded model better represents the intrinsic material variability and, by extension, the underlying physics.
If, however, external measurement noise were the primary source of uncertainty, the additive model would provide a good representation.
As in traditional constitutive modeling, a rational choice of how to formulate the representation of observed variability can only be assessed by examination of the data, and then confirming the validity of that choice by comparing synthetically replicated experiments to the observed behavior, as in \fref{fig:realizPlots}. 
This choice can be guided by examining whether or not the apparent noise is correlated with the mean behavior and the model prediction, as was done in this work.

In general, there are three kinds of uncertainty that should be considered during calibration to experimental data:  (A) external measurement error, (B) intrinsic variability, and (C) model form error.
Given that measurement noise is typically uncorrelated with the underlying physical response it is typically modeled with white noise.
Moreover, it is reducible by replicating the experiment and collecting more data in the sense that the posterior distributions of model parameters converge and narrow.
In contrast,  variability in the material properties cannot be reduced by increased data gathering, although more data will typically better inform the estimated joint distribution of material parameters.
The hallmark of inherent variability is individual experimental curves which are well explained by a model with appropriate physical parameters, but have systematic  parametric discrepancies across the set of curves.
The embedded error formulation is well-suited to represent this source of variability.
Finally, model form error refers to relevant physics which are unincorporated in the model, and manifests itself through discrepancies between model predictions and the actual data.
When present, a model of a single realization will display a systematic discrepancy from the data it is trying to emulate.
Since this error can confound the determination of the other errors it is crucial to perform model selection, as was done in this work albeit for a dataset that was generally well-represented by all members of the model family.
(It should be noted the embedded formulation \cite{sargsyan2015statistical} was originally developed to mitigate this type of error.)

\section{Conclusions} \label{sec:conclusion}

We have presented a method which can model the variability of a material which is well-described by existing plasticity models of mean response, but contains microstructural variability leading to different macroscopic material properties.
By leveraging the embedded error method of Sargsyan \etal \cite{sargsyan2015statistical} (which, as mentioned, was originally developed to address model discrepancy), we can  mathematically represent the material variability as material parameters drawn from a well-calibrated joint distribution.
This Bayesian approach is consistent with our understanding of microstructural variability and appropriate for UQ studies requiring forward propagation of this variability.
In particular, we developed a constitutive model of variability that is amenable to non-intrusive sampling and adaptable to direct evaluation in simulation codes that handle fields of distributions.
This enables a robust design methodology that can predict performance margins with high confidence.

Another important contribution of this work is to contrast the proposed approach with commonly used uncertainty formulations.
The standard, additive error formulation appropriately accounts for the uncertainty in the experiments arising from measurement error.
Yet, in the case of inherent variability, it characterizes all the uncertainty as measurement error which results in unwarranted confidence in the material properties and an inability to correctly understand how that variability would manifest in applications.
We have demonstrated that the embedded method accurately characterizes the aleatoric uncertainty present in the experimental observations and enables ``black-box'' engineering UQ analysis.
It gives insight into what aspects of a homogeneous, macroscale constitutive model are most strongly affected by microstructural variability and enables quantitative model selection.
It is able to distinguish the variability of different batches and therefore assess their relative performance.
It demonstrates convergence with increasing data and the convergence of common parameters in a nested hierarchy of models.
Moreover, it is capable of representing both significant external noise and inherent variability in a unified formulation.
In future work, we will extend the methodology to the post-necking failure behavior of similar materials.

\section*{Acknowledgments}
This work was supported by the LDRD program at Sandia National Laboratories, and its support is gratefully acknowledged.
B.L. Boyce would like to acknowledge the support of the Center for Integrated Nanotechnologies.  
Sandia National Laboratories is a multimission laboratory managed and operated by National Technology and Engineering Solutions of Sandia, LLC., a wholly owned subsidiary of Honeywell International, Inc., for the U.S. Department of Energy's National Nuclear Security Administration under contract DE-NA0003525.



\clearpage
\appendix
\section{Polynomial chaos expansion} \label{app:pce}

A polynomial chaos expansion (PCe) is a spectral representation of a random variable. 
Here we provide a brief description of the PCe construction and refer readers 
to \crefs{Ghanem:1991, Xiu:2002c, OlmOmk:2010} for more details. 
A PCe representation of any real-valued random variable $\lambda$ with finite variance is an expansion of the form
\begin{align}
  \lambda = \sum_{|I|=0}^{\infty} \lambda_{I}
  \Psi_{I}(\xi_1,\xi_2,\ldots),
  \label{e:PCEForm2}
\end{align}
where $\xi_I$ are independent identically distributed (i.i.d.)\ standard random variables, $\lambda_{I}$ are the coefficients, $I=(I_1,I_2,\ldots) \ \forall I_j \in \mathbb{N}_0$ is an infinite-dimensional multi-index, $|I|=I_1+I_2+\ldots$ is the $\ell_1$ norm, and $\Psi_{I}$ are multivariate normalized orthogonal polynomials written as products of univariate orthonormal polynomials:
\begin{align}
  \Psi_{I}(\xi_1,\xi_2,\ldots) = \prod_{j=1}^{\infty}
  \psi_{I_j}(\xi_j).
\end{align}
The basis functions $\psi_{I_j}$ are polynomials of order $I_j$ in the independent variable $\xi_j$ orthonormal with respect to the probability density $\prob(\xi_j)$. 
For instance, if the germ $\xi$ is a standard Gaussian random variable, then the PCe is based on Hermite polynomials.
Different choices of $\xi_j$ and $\psi_m$ are available via the generalized Askey family~\cite{Xiu:2002c}. The PCe~\eqref{e:PCEForm2} converges to the true random variable $\lambda$ in the mean-square sense~\cite{Cameron:1947}. 

For computational purposes, the infinite dimensional expansion~\eqref{e:PCEForm2} must be truncated:
\begin{align}
  \lambda = \sum_{I \in \mathcal{I}} \lambda_{I}
  \Psi_{I}(\xi_1,\xi_2,\ldots,\xi_{n_s}),
  \label{e:PCEForm3}
\end{align}
where $\mathcal{I}$ is some index set, and $n_s$ is some finite stochastic dimension that typically corresponds to the number of stochastic degrees of freedom in the system. 
For example, one popular choice for $\mathcal{I}$ is the \emph{total-order} expansion of degree $p$, where $\mathcal{I}=\{I : |I| \leq p \}$, see \eg \cref{OlmOmk:2010}.

Given the expansion for the input $\lambda(\xi)$, the PCe for a target quantity of interest $Q$ produced by the model evaluation $Q = M(\lambda)$ can be wriiten in a similar form 
\begin{align}
  Q(\xi) = M(\lambda(\xi)) = 
  \sum_{I \in \mathcal{I}} q_{I} \Psi_{i}(\xi_1,\xi_2,\ldots,\xi_{n_s}).
  \label{eq:PCEForm4}
\end{align}

Methods to compute PC coefficients are broadly divided into two groups, namely intrusive and non-intrusive \cite{OlmOmk:2010}.
The former involves substituting the expansions into the governing equations, and applying orthogonal projection to the resulting equations, resulting in a larger and modified system for the PCe coefficients. 
This approach is applicable when one has access to the full forward model and can readily modify the governing equations in the simulator. 
The other, non-intrusive approach is more generally applicable and involves finding an approximation in the subspace spanned by the basis functions by evaluating the original model many times. 

One such non-intrusive method relies on orthogonal projection of the solution 
\begin{align}
  q_{I} = \mathbb{E}[M(\lambda)
    \Psi_{I}] = \int_{\Xi}
    M(\lambda(\xi)) \Psi_{I}(\xi) \prob(\xi)
    \,d\xi.
    \label{e:NISP}
\end{align}
and is known as non-intrusive spectral projection (NISP).
In general, this integral must be estimated numerically. 
An alternative method of non-intrusively obtaining PCe coefficients is regression, which involves solving the linear system:
\begin{align}
  \underbrace{\left[
  \begin{array}{ccc} \Psi_{I^1}(\xi^{(1)}) & \cdots &
    \Psi_{I^K}(\xi^{(1)}) \\ \vdots & & \vdots \\
    \Psi_{I^1}(\xi^{(K)}) & \cdots & 
    \Psi_{I^K}(\xi^{(K)}) 
    \end{array}\right]}_{\mathsf{A}}
  \underbrace{\left[
  \begin{array}{c} 
  	   q_{I^1} \\  \vdots
    \\ q_{I^K} 
  \end{array}\right]}_{\mathsf{c}} =
  \underbrace{\left[
  \begin{array}{c} 
  M(\lambda(\xi^{(1)})) \\  \vdots
  \\ M(\lambda(\xi^{(K)})) 
  \end{array}\right]}_{\mathsf{M}},
\end{align}
where $\Psi_{I^n}$ is the $n$th basis function, $q_{I^n}$ is the coefficient corresponding to that basis, and $\xi^{(m)}$ is the $m$th regression point. 
In the regression matrix $\mathsf{A}$ each column corresponds to a basis element and each row corresponds to a regression point from the training set.

\section{Embedded discrepancy} \label{app:embedded}

As discussed in \cref{Sargsyan2015}, the embedded discrepancy likelihood often involves highly nonlinear and near-degenerate features, thus forcing one to find an alternative way to approximate it in a computationally feasible manner.
Sargsyan~\etal\cite{Sargsyan2015} suggest several options based on the assumption of conditional independence between the data points. 
In this work, we rely on the Gaussian approximation to the marginalized likelihood, which for the $i$-th batch $\Dc_i$, can be written as
\begin{align}
  \prob({\Dc_i}|\alphab, \tilde{M}) = 
  \frac{1}{(2\pi)^{\frac{N_i n_\strain}{2}}}
  \prod_{j=0}^{n_\strain-1}
  \prod_{k=1}^{N_i} 
  \frac{1}{\varsigma_{j}(\alphab)}
  \exp\left[-\frac{(\mu_j(\alphab)-\stress_j^{(i,k)})^2}
  {2\varsigma_{j}^2(\alphab)}\right],
\end{align}
where
\begin{align}
  \mu_j(\alphab) \equiv
  \mu(\strain_j;\alphab) = 
  \mathbb{E}_{\xib}[\tilde{M}(\strain_j;(\thetab+\hat{\etab})(\xib)]
\end{align}
and
\begin{align}
  \varsigma_j^2(\alphab) \equiv
  \varsigma^2(\strain_j;\alphab) = 
  \mathbb{V}_{\xib}[\tilde{M}(\strain_j;(\thetab+\hat{\etab})(\xib)]
\end{align}
are the mean and variance of the model at fixed $\alphab$ and strain point.
These moments are computed by constructing a PCe for the outputs by propagating the PCe of the input argument in \eref{eq:pcPerturbed}:
\begin{equation}
  \tilde{M}(\strain;\thetab+\hat{\etab})
  =\tilde{M}\left(\strain; \sum_{I} 
  \alpha_{I} \Psi_{I}(\xib)\right) 
  \approx 
  \sum_{I} \tilde{M}_{I}(\strain;\alphab)\Psi_{I}(\xib).
\label{e:surrogate_PCE}
\end{equation}
This can be done using NISP mentioned in \ref{app:pce} together with quadrature, and the moments can be computed from the expansion coefficients as 
\begin{align}
\mu(\strain;\alphab) \approx \tilde{M}_0(\strain;\alphab)
\qquad \textrm{ and } \qquad 
\varsigma^2(\strain;\alphab) \approx \sum_{I\neq0} 
\tilde{M}_I(\strain;\alphab).
\label{e:moments_PCE}
\end{align}

After obtaining $\alphab$ using Bayesian calibration and the likelihood just discussed the model can be used in a predictive manner.
Let $\phi(\strain;\alphab)=\tilde{M}(\strain;(\thetab+\hat{\etab})(\xib))$ be the probabilistic prediction of the model for a fixed $\alphab$. 
We remark that even if $\alphab$ is fixed, the prediction $\tilde{M}(\strain;(\thetab+\hat{\etab})(\xib))$ remains probabilistic because of the additional variability of the random variable $\thetab$ stemming from its PCe with $\xib$. 
We can then inspect the \emph{posterior predictive} random variable $\phi(\strain;\alphab)$. 
The posterior predictive random variable has the following mean 
\begin{equation}
\phi_{\textrm{mean}}(\strain)=
\mathbb{E}_{\alphab} {[\mu(\strain;\alphab)]}  
\end{equation}
and variance
\begin{equation}
\phi_{\textrm{var}}(\strain)
= 
\underbrace{
\mathbb{E}_{\alphab}{[\varsigma^2(\strain;\alphab)]}
}_{\textrm{model error}}
+
\underbrace{
\mathbb{V}_{\alphab}{[\mu(\strain;\alphab)]},
}_{\textrm{posterior uncertainty}}
\end{equation}
where $\mathbb{V}_{\alphab}$ denotes variance with respect to the posterior distribution of $\alphab$.   
Here $\varsigma^2(\alphab)$ is the forward-propagated variance of the function $\tilde{M}(\strain;(\thetab+\hat{\etab})(\xib))$ at location $\strain$ for given $\alphab$. 
The contribution of model error and posterior uncertainty are identified and separated leveraging the law of total variance. 
The uncertainty due to model error is independent of how much data we use during the inference process and, thus, it can be only improved by refining the model and/or its accuracy. 
On the other hand, the posterior uncertainty tends to shrink with more data. 
Note that, in practice, the posterior distribution is described via samples, therefore the expectation ($\mathbb{E}_{\alphab}$) and the variance ($\mathbb{V}_{\alphab}$) are computed via Monte-Carlo integration using MCMC samples.

\end{document}